\documentclass[12pt]{article}
\begin{document}

\begin{center}
{Calculation of  multi-loop
superstring amplitudes} \\
G. S. Danilov\\ Petersburg Nuclear Physics Institute,
Gatchina, Russia\\
E-mail: danilov@thd.pnpi.spb.ru
\end{center}
\begin{center}
{\bf Abstract}
\end{center}
The multi-loop interaction amplitudes in the
theory of the
closed, oriented
superstrings are obtained by the integration of local amplitudes.
The superstring local amplitude is represented by a sum
of the spinning string local amplitudes. The spinning string local amplitudes are
given explicitly through
super-Schottky group parameters and  interaction vertex
coordinates on the $(1|1)$ complex supermanifold.
The super-Schottky group is a supersymmetrical extension of the
Schottky group.   The integration is ambiguous
with respect to those replacements of the integration variables which
admix  Grassmann variables to the boson ones.  So  the calculation is
guided by a preservation of local symmetries of the superstring.  The
superstring amplitudes are free from divergences and they are
consistent with the world-sheet spinning string symmetries.  The vacuum
amplitude vanishes along with  1-, 2- and 3-point amplitudes of
massless states. The vanishing of the vacuum amplitude occurs after the
integration of the corresponding local amplitude is performed over
those modular variables that are limiting points of the super-Schottky
group. The above-mentioned massless state amplitudes are nullified
afterwards the corresponding local amplitudes are, in addition,
integrated over the interaction vertex coordinates.

\newpage
\section{Introduction}
In the
Ramond- Neveu - Schwarz  theory \cite{rnshw} the superstring
interaction  amplitudes are obtained by a summation of  the spinning
(fermion) string interaction amplitudes.  The  string world-sheet
carries the  zweibein $e_m^a$ and the 2D-gravitino  field $\phi_m$.
Due to the local 2D-symmetries, the amplitudes are  independent of
the above mentioned gauge fields $e_m^a$ and $\phi_m$.  In addition,
due to a hidden space-time symmetry of the superstring, it is expected
that the vacuum superstring amplitude is nullified along with 1-, 2-
and 3-point amplitudes of massless states.

Every $n$-loop
amplitude  is represented by an integral of a local amplitude.  The
integral is taken over $(3n-3|2n-2)$ complex moduli (if $n>1$) and over
interaction vertex coordinates on the complex $(1|1)$ supermanifold.

The supermanifold is often specified \cite{ver} as the Riemann surface
with spin structure \cite{swit}. Grassmann moduli are
assigned to the 2D-gravitino field which, therefore, is not
conformally flat. The integration over Grassmann moduli being
performed,  each spinning string amplitude  is represented by an
integral of a local function where the integral is taken over vertex
coordinates and over  Riemann moduli.  The superstring amplitude is
obtained  by  summing over spin structures.
Spin structures are not invariant under
transformations of the 2D supersymmetry that for a long time was a
source of troubles. Multi-loop amplitudes in \cite{ver} depend
on the location of the 2D-gravitino field \cite{ver,momor,as} that
means a loss of the 2D supersymmetry in these calculations.  True
two-loop amplitudes has been obtained in \cite{hok}.
Arbitrary-loop amplitudes have been obtained in \cite{dancqg}.

Alternatively, the supercovariant  gauge
\cite{bsh,fried,vec,dan93,danphr} is employed where the
zweibein and 2D-gravitino  field are conformally flat. In this case
local spinning string amplitudes have a manifest symmetry under
 $SL(2)$ and supermodular transformations. The supermodular transformations
are an extension \cite{dannph} of the modular transformations
\cite{siegal} on the Riemann surface.  The string world sheet is
specified as the $(1|1)$ , non-split supermanifold. The supermanifold
carries a ``superspin'' structure \cite{dan93,danphr,dannph,dan90}
instead of the spin structure \cite{swit}.  The superstring amplitude
is obtained by a summation over the superspin structures.  The
superspin structures are supersymmetric extensions of the spin
structures \cite{swit}.  In this case the twist about $(A,B)$-cycles
is, generally, accompanied by a supersymmetric
transformation including fermion-boson mixing.
The fermion-boson mixing
arises due to the presence of Grassmann moduli that are assigned to the
$(1|1)$ complex, non-split
supermanifold in addition to the Riemann ones.
The fermion-boson mixing differentiates the
superspin structures from the ordinary spin ones. Indeed, the ordinary
spin structures \cite{swit} imply that boson fields are single-valued
on Riemann surfaces. Only fermion fields being twisted about
$(A,B)$-cycles, may receive the sign.

The supermanifold is usually described by ``super-Schottky''
groups that are superconformal extensions \cite{dan90} of the Schottky
groups. The genus-$n>1$ super-Schottky group depends on $(3n-3)$ even
complex moduli and on $(2n-2)$  Grassmann complex moduli.
In this case  local
spinning amplitudes  with any number of loops  have  been explicitly
calculated \cite{danphr} for all the superspin structures. In doing so,
the partition functions have been computed from equations
\cite{danphr,dan90} that are nothing else than  Ward identities. These
equations  realize the requirement that the spinning amplitudes are
independent of infinitesimal local variations of both the vierbein and
the gravitino field.  Therefore, they are consistent with the gauge
invariance of the fermion string. The world-sheet gauge group is so
large that  the Ward identities fix the partition function up to a
constant multiplier.  This multiplier is determined by a factorization
condition following from the unitarity equations.
In the present paper, the integration
of the local amplitudes \cite{danphr} over moduli and over vertex
coordinates is considered.

Super-Schottky groups are, generally,
non-split that is fermion variables are mixed to the boson ones under
the super-Schottky group transformation.
In this case the integration over the
fundamental region of the Schottky group leads to the loss of the
$SL(2)$ symmetry. The modular transformations are non-split too, and
the period matrix \cite{danphr, dannph} is different from the period
matrix \cite{siegal} on the Riemann surface by terms, which are
proportional to the Grassmann moduli  (generally, these terms depend on
the superspin structure). Then the integration over the fundamental
region of the modular group leads to a loss of the modular symmetry.
To restore the symmetries, the integral over the
fundamental region must be supplemented \cite{dan04} by the integral
around its boundary (cf. with  \cite{dancqg}).

When the Riemann surface is degenerated, singularities appear in  the spinning string amplitudes.
The integration of such singular expressions is ambiguous with respect
to those replacements of the integration variables which admix
Grassmann variables to the boson ones.  The result of the integration
of the given expression can be either finite or divergent depending
on the choice of the integration variables \cite{dan04}.  Nevertheless,
the result  is the same under those replacements of the integration
variables, which do not lead to a divergence of the integral. In this
paper the superstring amplitudes are calculated through multipliers and
limiting points of the super-Schottky group.  The obtained amplitudes
are finite, and they hold both
$SL(2)$-symmetry and the supermodular one. In addition, the vacuum
amplitude is nullified together with 1-, 2- and 3-point amplitudes of
massless states.

Conventional wisdom has it that the split gauge \cite{ver} has the
essential drop on the supercovariant gauge. Indeed, in the split gauge
\cite{ver} the $(1|1)$ complex supermanifold is split  that is fermions
are not mixed to bosons under twists around non-contractible cycles.
In this case that integration measures seemingly possesses by  the
modular symmetry \cite{siegal} and are represented through
theta-functions and modular forms.
In the two-loop calculation
\cite{hok} the genus-2 integration measures are indeed modular
forms, and the GSO projections of the spinning string amplitudes with
less than four legs are  equal to zero.  The GSO projection of the
four-point amplitude does not depend on $\phi_m$.  The spinning string
amplitude ceases to depend on $\phi_m$ due to the integration over
vertex coordinates.  The papers \cite{hok} have initiated the efforts
\cite{capige} to build genus $n>2$ amplitudes assuming certain
properties of the amplitudes, the modular symmetry being among them.
This strategy meets with difficulties \cite{dunmorsl,ma2vo}, at least
for $n>3$.  In fact modular transformations on  the split
$(1|1)$ supermanifold are, generally, distinguished from the modular
transformation on the Riemann surface \cite{dancqg}.

The calculation of the arbitrary-loop
spinning  amplitudes in \cite{dancqg} exploits the
gauge symmetry on the string world sheet and uses no assumptions.
The
spinning string amplitudes are independent of local variations of the
2D gravitino field due to the integration over vertex coordinates, but
local amplitudes depend on $\phi_m$ (the last takes place even
in the two-loop case \cite{hok}). Modular transformations,
generally, change $\phi_m$.  At the same time, the integration of
the local function over moduli and over vertex coordinates is
performed at the given location of $\phi_m$.  Returning back to
the original $\phi_m$ is achieved by an extra transformation of a
local 2D supersymmetry.  Therefore, the symmetry group of the amplitude
consists of modular transformations accompanied by the relevant
supersymmetry ones.  So far as the local amplitudes depend on
$\phi_m$ the distinguishing of these
``supermodular'' transformations from the ordinary modular
transformations is important.

The period matrix on
the discussed  supermanifold
collects periods of scalar superfunctions which vanish under the
supercovariant Laplacian \cite{dan04}.  The supermodular transformation
changes \cite{dancqg,dan04} this matrix
just as the relevant modular transformation changes the period matrix
\cite{siegal} on the Riemann surface.
Since the superscalar functions depend on  $\phi_m$,
the period matrix on the $(1|1)$ complex supermanifold is, generally,
distinguished from the period matrix  on the Riemann surface by  terms
proportional to the Grassmann moduli. In this case the integration of
the local amplitude over the fundamental region of the modular group
leads to the loss of 2D supersymmetry.  As the result, the spinning
string amplitudes depend on $\phi_m$. To restore
the  supersymmetry, the integration over the fundamental
region of the modular group
must be supplemented \cite{dancqg,dan04} by the integral around the
boundary of the region just as it occurs in the supercovariant gauge.

A loss of the supersymmetry in \cite{ver} is due to
the difference between the supermodular symmetry  and the modular one was
ignored, and because of  an incomplete calculation of the
ghost zero mode contribution to the amplitude \cite{dancqg}.

If $n\leq3$, then
the periods of the superscalar functions can be taken \cite{hok}  as
the moduli set.  In this case the boundary integral does not arise.  If
$n\leq3$ and the moduli setting \cite{ver} is used, the boundary
integral is removed \cite{dancqg} by a re-definition \cite{dancqg} of the
local amplitude.  If $n>3$, then periods of
superscalar functions depend on Grassmann moduli for any choice of
moduli variables. In this case  the boundary integral is necessary
present in the expression for the amplitude. Then nullification of the
vacuum amplitude, vacuum-dilaton transition constant and of the 2- and
3- point amplitudes for the massless boson states is achieved due to
integration over the interaction vertex coordinates and (for $n>3$)
over moduli. It is akin to what occurs in the superconformal gauge.
Furthermore,  the genus-$n>2$ superghost correlator is different from
that, which is proposed in \cite{ver}. The difference is due
\cite{dancqg} an incomplete calculation in \cite{ver} of the superghost
zero mode contribution to the amplitude.  The true correlator
\cite{dancqg} is not represented through theta (and kindred to the
theta)-functions.  Besides,  the genus-$n>2$ superstring amplitudes
cease to depend on $\phi_m$ only due to integration over the
interaction vertex coordinates and (for $n>3$) over moduli. Superstring
local amplitudes depend on $\phi_m$ that is rather awkward for
applications.  So expressions for the genus-$n>2$  amplitudes in the
split gauge are much more complicated that it has been expected
naively, and it is not obvious that in this case the split gauge \cite{ver} has an advantage over
the supercovariant gauge.

The paper is organized as it follows. Section 2
contains a brief review of  super-Schottky groups
\cite{danphr,dannph}. In Sections 3
a supercovariant integration of the local amplitudes is proposed.
In Sections 4, 5 and 6
superfield vacuum correlators  and   integration measures are given
for a following study of  degenerated modular configurations. In
Section 7 an ambiguity in the integrations  of local amplitudes is
demonstrated.  The cancelation of divergences in the superstring
amplitudes is shown for those modular configurations where either one
of the Scottky multipliers goes to zero, or the genus-$n$ supermaifold
is degenerated into the sum of the genus-1 supermanifold and of the
genus-$(n-1)$ one.  In Section 8  the finiteness of the
superstring amplitudes is established.  The vanishing of the 0-, 1-, 2-
and 3-point functions is verified. The preservation of the local
symmetries of the amplitudes is argued.

\section{Superspin structures}

As it was noted above, the super-Schottky group
determines the super-spin structure on the complex
$(1|1)$ supermanifold. The supermanifold  is mapped by the
$t=(z|\vartheta)$ coordinate. The genus-$n$ super-spin structure
presents a superconformal extension of the relevant genus-$n$ spin
structure given by the set of transformations
$\Gamma^{(0)}_{a,s}$ and $\Gamma^{(0)}_{b,s}$
(where $s=1,\dots n$) corresponding to the round of
$A_s$-cycle and, respectively, of the $B_s$-cycle on the Riemann
surface. They depend on the theta function characteristics
$l_{1s}$ and $l_{2s}$ that are assigned to the given handle $s$.  A
discrimination is made only between those (super-)spin
structures, for which field vacuum correlators are distinct. So
$l_{1s}$ and $l_{2s}$ can be restricted by 0 and 1/2. In this
case $l_{1s}=0$ is assigned to the Neveu-Schwarz handle while
$l_{1s}=1/2$ is reserved for the Ramond one.
Furthermore,
\begin{eqnarray}
\Gamma_{b,s}^{(0)}\equiv\Gamma_{b,s}^{(0)}(t)
\equiv\Gamma_{b,s}^{(0)}(l_{2s};t)
=\left\{z^{(b0)}_s=g_s(z),\quad
\vartheta^{(b0)}_s=
-(-1)^{l_{2s}}\vartheta \sqrt{g_s'(z)}\right\}\,,
\nonumber\\
\Gamma_{a,s}^{(0)}\equiv\Gamma_{a,s}^{(0)}(t)
\equiv\Gamma_{a,s}^{(0)}(l_{1s};t)=\{z^{(a0)}_s=z,\quad
\vartheta^{(a0)}_s=(-1)^{2l_{1s}}\vartheta\}
\label{zgama}
\end{eqnarray}
where $g'_s(z)=\partial_z g_s(z)$ and $g_s(z)$ is the Schottky transformation $g(z)$ corresponding to
the round about $B_s$-cycle. The Schottky transformation $g(z)$ is given by
\begin{equation}
g(z)=\frac{az+b}{cz+d}\quad{\rm where}
\quad ad-bc=1.
\label{sch}
\end{equation}
The parameters in (\ref{sch}) are expressed
through two unmoved (limiting) points
$u$ and $v$ on the complex $z$-plane as well as through
the  complex multiplier $k$ (where $|k|\leq1$). In this case
\begin{equation}
a=\frac{u-kv}{\sqrt k(u-v)}\,,\quad d=\frac{ku-v}
{\sqrt k(u-v)}\quad{\rm and}\quad c=\frac{1-k}{\sqrt k(u-v)}\,.
\label{uvk}
\end{equation}
The Schottky
transformation  (\ref{sch})
turns the boundary of the Schottky  circle $C_v$ into the
boundary of the $C_u$ circle. In this case
\begin{equation}
C_v=\{z:|cz+d|^2=1\}\,,\qquad C_u=\{z:|-cz+a|^2=1\}
\label{circ}
\end{equation}
Using (\ref{uvk}), one can see that $v$ lies inside $C_v$
and outside $C_u$. Correspondingly, $u$ is inside
$C_u$ and outside $C_v$.
Since
$\Gamma^{(0)}_{a,s}$ in (\ref{zgama}) corresponds to the
round of the Schottky circle, in the Ramond case a square root
cut appears on $z$-plane between $u_s$ and $v_s$.
The
set of $n$ forming transformations (\ref{zgama}) and their
group products forms the Schottky group of the genus-$n$.  The
fundamental region of a group is that, no two of whose
points are congruent\footnote{The congruent points, or curves, or domains
are those related by a group transformation other that the identical
transformation \cite{ford}.} under the group transformations, and such
that the neighborhood of any point on the boundary contains points
congruent to the points in the region \cite{ford}.  The integral of a
conformal $(1,1)$ tensor  over a fundamental region of the group does
not depend on the choice of the fundamental region \cite{ford}.

If the Schottky circles of the forming transformations all are
separated from each other, then the exterior of them can be taken as
the fundamental region. We need, however, to consider, in addition,
those modular configurations where  the  circles corresponding
to distinct handles may overlap each
other (see eq.(\ref{inosp}) in Sec. 3).
In any case the exterior of all the Schottky circles $C_{v_g}$
and $C_{u_g}$ can be taken as the fundamental region. The $C_{v_g}$
and $C_{u_g}$ circles correspond to the transformation $g\neq I$ of
the Schottky group. One can replace \cite{ford}  any part of the
fundamental region by a congruent part and still have a fundamental
region. Therefore, $(-cz+a)$ in (\ref{circ}) can be
replaced   by relevant function $\ell_g(z)$.
Simultaneously, $(cz+d)$ is replaced by $\ell_g(g(z))$.
To bound the integration with respect to $(z,\bar z)$ by the fundamental
region,
the integrated expression can  multiply by a factor
$B^{(n)}(z,\bar z;\{q,\bar q\})$ given as
\begin{equation}
B^{(n)}(z,\bar
z;\{q,\bar q\})= \prod_{g\neq I}\theta(|\ell_g(z)|^2-1)
\theta(1-|\ell_g(g(z))|^2)
\label{zbound}
\end{equation}
where
$\theta(x)$ is the step function that is $\theta(x)=1$ at $x>0$ and
$\theta(x)=0$ at $x<0$.

In the superstring theory the transformations
(\ref{zgama}) are replaced by $SL(2)$ transformations
$\Gamma_{a,s}\equiv\Gamma_{a,s}(l_{2s})
\equiv\Gamma_{a,s}(t)\equiv\Gamma_{a,s}(l_{1s};t)$
and $\Gamma_{b,s}\equiv\Gamma_{b,s}(l_{2s})
\equiv\Gamma_{b,s}(t)\equiv\Gamma_{b,s}(l_{2s};t)$ with
\cite{dan93,danphr,dannph}
\begin{equation}
\Gamma_{a,s}=\tilde\Gamma^{-1}_s
\Gamma^{(0)}_{a,s}\tilde\Gamma_s\,,
\qquad \Gamma_{b,s}=
\widetilde\Gamma^{-1}_s\Gamma^{(0)}_{b,s}
\widetilde\Gamma_s
\label{gamab}
\end{equation}
where $\Gamma^{(0)}_{b,s}$ and
$\Gamma^{(0)}_{a,s}$ are given by (\ref{zgama}) while
$\widetilde\Gamma_s$ depends, among other things, on two
Grassmann parameters $(\mu_s,\nu_s)$ as  follows
\begin{equation}
\tilde\Gamma_s\equiv\tilde\Gamma_s(t)=\left\{
z=z^{(s)} +\vartheta^{(s)}\varepsilon_s(z^{(s)})\,,\quad
\vartheta=\vartheta^{(s)}\left(1+
\frac{\varepsilon_s\varepsilon'_s}2\right)+
\varepsilon_s(z^{(s)})
\right\}\,,
\label{tgam}
\end{equation}
\begin{equation}
\varepsilon_s(z)=\frac{\mu_s(z-v_s)-\nu_s(z-u_s)}
{u_s-v_s}\,,\qquad
\varepsilon'_s=\partial_z\varepsilon_s(z)
\label{teps}
\end{equation}
where $(u_s|\mu_s)$ and $(v_s|\nu_s)$ are limiting points of
transformations (\ref{gamab}). The set of the transformations
(\ref{gamab}) for $s=1,\dots,n$ together with their group
products forms the genus-$n$ super-Schottky group. In the explicit form the
$\Gamma_{b,s}$ transformation
$(z|\vartheta)\to(z^{(b)}_s|\vartheta^{(b)}_s)$ is as follows
\begin{eqnarray}
\vartheta^{(b)}_s=\frac{(-1)^{2l_{2s}+1}}{
(c_sz+d_s)}\biggl[(1+\varepsilon_s\varepsilon_s')\vartheta+
\epsilon_s(z)\biggl]-g'(z)\vartheta\varepsilon_s\varepsilon_s'\,,
\nonumber\\
z^{(b)}_s=g_s(z)+g_s'(z)\vartheta\epsilon_s(z)-
g_s'(z)\varepsilon_s(z)\epsilon_s(z)\,.
\label{sbtr}
\end{eqnarray}
In this case
\begin{equation}
\epsilon_s(z)=(-1)^{2l_{2s}+1}(c_sz+d_s)
\varepsilon_s(g_s(z))-\varepsilon_s(z)
\label{eps}
\end{equation}
If $l_{1s}=0$, then
$\Gamma_{a,s}$ is the identical transformation. If
$l_{1s}=1/2$, then the $\Gamma_{a,s}$ transformation
$(z|\vartheta)\to(z^{(a)}_s|\vartheta^{(a)}_s)$ is
not identical transformation being
\begin{equation}
\vartheta^{(a)}_s=-
(1+2\varepsilon_s\varepsilon_s')\vartheta+
2\varepsilon_s(z)\,,
\qquad
z^{(a)}_s=z-2\vartheta\varepsilon_s(z)\,.
\label{satr}
\end{equation}
Therefore, in this case both $\Gamma_{a,s}$
and $\Gamma_{b,s}$
are non-split transformations.
The superconformal $p$-tensor $T_p(t)$ is changed under the $SL(2)$
transformation $\Gamma(t)=\{t\rightarrow
t_\Gamma=(z_\Gamma(t)|\vartheta_\Gamma(t))\}$ as follows
\begin{equation}
T_p(t_\Gamma)=T_p(t)Q_\Gamma^p(t)
\label{stens}
\end{equation}
where $Q_\Gamma(t)$ is that factor, which the spinor left derivative
$D(t)$ receives under the $\Gamma(t)$ transformation. In this case
\begin{equation}
Q_\Gamma^{-1}(t)=D(t)\vartheta_\Gamma(t)\,;\qquad
D(t_\Gamma)= Q_\Gamma(t)D(t)\,,\qquad
D(t)=\vartheta\partial_z+\partial_\vartheta
\label{supder}
\end{equation}
It follows from (\ref{supder}) that
\begin{equation}
Q_{\Gamma_1\Gamma_2}(t)=Q_{\Gamma_1}(\Gamma_2(t))Q_{\Gamma_2}(t)
\label{suprel}
\end{equation}
Furthermore, $\Gamma_{b,s}$ turns the ``boundary" of the
$\hat C_{u_s}$ "circle" to the ``boundary" of $\hat C_{v_s}$
where\footnote{Through the paper
the overline denotes the complex conjugation.}
\begin{eqnarray}
\hat C_{v_s}=\{t:(-1)^{2l_{2s}+2l_{2s}'}
Q_{\Gamma_{b,s}(l_{2s})}(t)\overline{Q_{\Gamma_{b,s}(l_{2s}')}(t)}=1\}\,,
\nonumber\\
\hat C_{u_s}=\{t:(-1)^{2l_{2s}+2l_{2s}'}
Q_{\Gamma_{b,s}^{-1}(l_{2s})}(t)
\overline{Q_{\Gamma_{b,s}^{-1}(l_{2s}')}(t)}=1\}
\label{hcirc1}
\end{eqnarray}
where $l_{2s}$ is assigned to the right movers and $l_{2s}'$ is
assosiated with the left ones.  Pairs of the congruent ``circles''
($\hat C_{v_{\Gamma_b}},\hat C_{u_{\Gamma_b}}$)
can be built for any group product
of the $\Gamma_{b,s}$ transformations.

As for the Schottky group, one can replace   ``circles'' (\ref{hcirc1})
by a pair of any relevant ``curves".
In particular,
the ``circles'' (\ref{hcirc1}) can be replaced by a pair of the
congruent ``circles'' ($\tilde C_{v_s},\tilde C_{u_s}$) where
\begin{equation}
\tilde C_{v_s}=\{t:|c_sz^{(s)}+d_s|^2=1\}\quad{\rm and}\quad
\tilde C_{u_s}=\{t:|-c_sz^{(s)}+a_s|^2=1\}\,,
\label{hcirc}
\end{equation}
$z^{(s)}$ being defined by (\ref{tgam}).
"Circles" (\ref{hcirc1}) and (\ref{hcirc}) are
different from each other and from (\ref{circ}) by terms that
are proportional to Grassmann quantities.

All the congruent ``curves'' are distinguished from the
corresponding  congruent curves for the Schottky group   by terms that
are proportional to Grassmann quantities.
Therefore, the integration of the
conformal $(1/2,1/2)$-tensor
over the Schottky group fundamental region
destroys the $SL(2)$ symmetry.  The symmetry is restored due to
supplementing of the integral by an
integral over the boundary of the integration region.
To derive this
boundary integral, we extend of the step function $\theta(x)$ to the
case when $x=x_b+x_s$ contains the ``soul'' part $x_s$ that is the part
proportional to the Grassmann parameters.  Then  $\theta(x)$ is
understood in the sense that it is the Taylor series in $x_s$. In the
calculation of the Taylor series one employs the known relation
$d\,\theta(x_b)/d\,x_b=\delta(x_b)$ where $\delta(x)$ is the Dirac
delta-function. Under this condition,  the fundamental
region can be bounded just as it was done for the Schottky group
fundamental region. In doing so,  the integrand is multiplied by a step
function factor.  Besides $(t,\bar t)$ and the $\{q,\bar q\}$ set of
the super-Schottky group parameters, this $B_{L,L'}^{(n)}(t,\bar
t;\{q,\bar q\})$ factor may depend on the superspin structure
$L=\{l_{1s},l_{2s}\}$ of the right movers and on the superspin
structure $L'$ of the left movers.
Excepting the unity transformation, the group transformations are
divided into transformations $G$ and inverse $G^{-1}$ to them. Then in
the general case
\begin{equation}
B_{L,L'}^{(n)}(t,\bar t;\{q,\bar
q\})= \prod_G\theta(\ell_G^L(t)\overline{\ell_G^{L'}(t)}-1)
\theta(1-\ell_G^L(\Gamma_{b,G^{-1}}(L;t))
\overline{\ell_G^{L'}(\Gamma_{b,G^{-1}}(L';t))})
\label{zsbound}
\end{equation}
where  $\ell_G^L(t)$ is assigned to the super-Schottky
group transformation $\Gamma_{b,G}(L;t)$ for the superspin structure
$L$.  The step function $\theta(x)$ is understood as the Taylor series
in those Grassmann quantities, which are contained in its argument.
The above expansion originates $\delta$-functions and their
derivatives, which give rise to boundary terms in the integral. The
$(1/2,1/2)$ supertensor being integrated,  the integral is independent
of $\ell_G^L(t)$ that is directly verified for infinitesimal variations
of $\ell_G^L(t)$. The integration variables being replaced, the
arguments of the step functions are  correspondingly replaced. As the
result, the integral is independent of the choice of the integration
variables.  For every given modular configuration only a finite number
of the multipliers in (\ref{zsbound}) are different from the unity.

\section{Integration of the local amplitude}

The superstring amplitude  is calculated by the
integration of the local amplitude. The integration  is performed with
respect to the interaction vertex coordinates $t_j=(z_j|\vartheta_j)$
and over super-Schottky group parameters $k_j$,
$U_j=(u_j|\mu_j)$ and $V_j=(v_j|\nu_j)$ exceping those
$(3|2)$ parameters  among
$U_j$ and $V_j$, which are
fixed due to
$SL(2)$ symmetry.
If $U_a$, $V_a$ and $u_b$ are fixed, then
the local amplitude is multiplied by a factor $|\widehat
H(U_a,V_a,U_b)|^2$ where
\cite{danphr}
\begin{equation}
\widehat H(U_a,V_a,U_b)=(u_a-u_b)(v_a-u_b)
\left[1-\frac{\mu_a\mu_b}{2(u_a-u_b)}-
\frac{\nu_a\mu_b}{2(v_a-u_b)}\right]\,.
\label{fixm}
\end{equation}
Due to this factor, the amplitude does not depend on  a choice of the
$\{N_0\}$ set of fixed parameters. Besides the fixed parameters, the
discussed factor (\ref{fixm}) depends on $\mu_b$  which is the
Grassmann partner of $u_b$.

The integration region over $t_j=(z_j|\vartheta_j)$ is
restricted by the  product
\begin{equation}
\widehat B_{L,L'}^{(n)}(\{t_j,\bar
t_j\};\{q,\bar q\})=\prod_{j=1}^mB_{L,L'}^{(n)}(t_j,\bar
t_j;\{q,\bar q\})
\label{prodb}
\end{equation}
of the step functions
(\ref{zsbound}). In line with Section 2, the step function  $\theta(x)$ is
understood as the Taylor series in those Grassmann quantities which
are contained in its argument $x$.

The integration over the super-Schottky group
parameters is performed in such a way that the period matrix  is kept
in the fundamental region of the modular group.  The modular
transformation on the supermanifold is
a globally defined, holomorphic superconformal
transition from
the supermanifold cordinate $t$ to $\hat t=\hat t(t,\{ q\};L)$. The
transition is accompanied by the holomorphic  $q\to\hat q(\{q\};L)$
change of the super-Schottky group parameters $q$ and, generally, by
the change  of the super-spin structure $L\to\hat L(L)$.  In this case
$\hat t$ and $\hat q$  depend on
the superspin structure by terms proportional to Grassmann parameters
\cite{dannph}.  Like to the modular transformation on the Riemann surface
\cite{siegal}, the modular transformation on the supermanifold realizes
the transition to a new basis of non-contractable cycles. Therefore,
the period matrix $\omega(\{q\},L)$ is changed  just as the period
matrix $\omega(\{q\}_0)$  on the Riemann surface
\cite{siegal}, that is
\begin{equation}
\omega(\{q\},L)=[A\omega(\{\hat q\},\hat
L)+B] [C\omega(\{\hat q\},\hat L)+D]^{-1}
\label{modtr}
\end{equation}
where integer $A$, $B$, $C$ and $D$ matrices satisfy \cite{siegal} the
following relations\footnote{ The right-top "$T$" symbol labels the
transposing.}
\begin{equation}
C^TA=A^TC\,,\qquad D^TB=B^TD\,,\qquad
D^TA-B^TC=I\,.
\label{mrel}
\end{equation}
All the Grassmann parameters
being equal to zero, the period matrix on the supermanifold
is reduced to the period
matrix $\omega(\{q\}_0)$ on the Riemann surface. In this case
entries
$\omega_{sp}(\{q\}_0)$ of $\omega(\{q\}_0)$ are  given through
$\{q\}_0$ as follows \cite{fried,vec,danphr,mart}
\begin{eqnarray}
2\pi i\omega_{sp}(\{q\}_0)=\ln\frac{(u_s-u_p)
(v_s-v_p)}{(u_s-v_p)(v_s-u_p)}+
{\sum_{\Gamma\neq I}}^{''}\ln\frac{[u_s-g_\Gamma(u_p)]
[v_s-g_\Gamma(v_p)]}{[u_s-g_\Gamma(v_p)][v_s-g_\Gamma(u_p)]}\,,
\, (s\neq p);
\label{om0sp}\\
2\pi i\omega_{ss}(\{q\}_0)=\ln k_s+
{\sum_{\Gamma\neq I}}'\ln\frac{[u_s-g_\Gamma(u_s)]
[v_s-g_\Gamma(v_s)]}{[u_s-g_\Gamma(v_s)][v_s-g_\Gamma(u_s)]}
\label{om0ss}
\end{eqnarray}
where $I$ is the identical transformation.
The summation in
(\ref{om0sp}) is performed over all the Schottky group transformations
$\Gamma$ whose leftmosts are not group
powers of $g_s$, or whose rightmosts are not group powers of
$g_p$. In (\ref{om0ss}) the summation is performed over all those
$g_\Gamma$, whose leftmosts and  rightmosts are not group
powers of $g_s$. Jumping
from the given branch of the logarithmic function to another branch
presents a certain transformation (\ref{modtr}) with $C=0$ and $A=D=I$.
We fix
branches of  logarithmic functions in
(\ref{om0sp}) and in (\ref{om0ss})
by cuts which are drawn between  singular points
of the logarithmic functions
on the $u_s$ and $v_s$ on complex planes. In so doing we suppose
that
\begin{eqnarray}
|\arg
k_s|\leq\pi\,,\quad|\arg(u_s-u_r)|\leq\pi\,,
\,\,|\arg(u_s-v_r)|\leq\pi\,,
\nonumber\\
|\arg(v_s-v_r)|\leq\pi\,,
\quad(s\neq r)\,.
\label{argum}
\end{eqnarray}
Adding of $\pm1$ to $\omega_{sp}(\{q\}_0)$ in (\ref{om0sp}) is
achieved by a suitable round of the limiting points $(u_s,v_s)$ over
limiting points of the handle $p$. In this case the sum on the right
side of (\ref{om0sp}) is not changed.  So  constraints (\ref{argum}) on
the arguments of the limiting point differences discriminate between
$(l_{1s}=l_{1r}=1/2, l_{2s}=l_{2r}=0)$ and $(l_{1s}=l_{1r}=1/2,
l_{2s}=l_{2r}=1/2)$.  The $|\arg k_s|\leq\pi$ condition discriminates
between $(l_{1s}=0,l_{2s}=0)$ and $(l_{1s}=0,l_{2s}=1/2)$. Hence one
can re-define spin structures  using  relations (\ref{argum}) instead
of the usual constraint $|Re\,\omega_{sp}(\{q\}_0)|\leq1/2$.  In the
both cases the sum over the spin structures includes all the distinct
spin structures without a double counting.

Conditions (\ref{argum}) can be also used for a definition of superspin
structures at non-zero Grassmann parameters. If the functions in
(\ref{argum}) have  "soul" (i.e.  proportional to Grassmann parameters)
parts, the corresponding relation
(\ref{argum}) is written down  for the "body" of the function.

The modular transformation (\ref{modtr}) with
\begin{equation}
B=C=0\,,\quad A=F\,,\quad D^{-1}=
F^T\,,\quad \det  F=\pm1\,,
\label{parttr}
\end{equation}
corresponds to a redefining of  the forming transformation set. If
$(3|2)$ limiting points   are fixed by (\ref{fixm}), then the redefining
touches only $(n-2)$ handles among $n$ ones.  In particular, the
interchange between  the $s_1$ and $s_2$ handles corresponds to the $F$
matrix having  non-zero non-diagonal entries only to be
$F_{s_1s_2}=F_{s_2s_1}=1$ and diagonal entries $F_{s_1s_1}=F_{s_2s_2}
=0$, the rest diagonal entries being equal to the unity.
To avoid
a multi-counting of the handles in calculation of the $n$-loop
amplitude, one can either  order the handles in any manner, or multiply
the local amplitude by $1/(n-2)!$.

Diagonal $F\neq I$ matrices correspond to the replacement of
the given group transformation by its inverse.  In this case the
diagonal $F$ matrix with $F_{ss}=-1$ corresponds to the interchange
between $(u_s|\mu_s)$ and $(v_s|\nu_s)$. For zero Grassmann parameters
it is seen directly from (\ref{om0ss}).
To avoid double-counting of the handles in calculation of
the $n$-loop amplitude, one can
multiply the local amplitude by $2^{-(n-2)}$.

Remaining $F\neq I$ matrices  are
associated with   transitions to other sets $\{G_j\}$
of the  {\it kleinian} group basic transformations. The resulting basic
transformations are group products of the former basic transformations.
The period matrix $\omega(\{q\}_0)_{G_j,G_l}=
(F\omega(\{q\}_0)F^T)_{jl}$ in the $\{G_j\}$ basis is obtained from
(\ref{om0sp}) and (\ref{om0ss}) by the $\{g_j\}\to\{G_j\}$ replacement.
The basic group
transformations are usually \cite{siegal}
chosen in such a way  that entries
$Im\omega_{jl}(\{q\}_0)\equiv y_{jl}(\{q,\bar q\}_0)$ of the
matrix $Im\omega(\{q\}_0)\equiv y(\{q,\bar q\}_0)$
obey relations, as follows
\begin{equation}
[Fy(\{q,\bar q\}_0) F^T]_{jj}\geq y_{jj}(\{q,\bar q\}_0)
\label{imagin}
\end{equation}
where $F$ is the  $n$-dimensional integer, unimodular  matrix
whose last $n-j+1$
entries of every column $f_j$ are relatively prime.
Conditions (\ref{imagin}) restrict the fundamental region in the
period matrix space. In addition,
the fundamental region  is
bounded by conditions
\cite{siegal}
\begin{equation} |\det
[C(\omega(\{q\}_0)+\widetilde B)+D]|^2\geq1\,,
\label{dbound}
\end{equation}
$C$ and $D$ being any  of
the integer matrices in (\ref{modtr}). The integer
$\widetilde B$ matrix in (\ref{dbound})
is chosen from the condition that
$|Re\omega(\{q\}_0)+ \widetilde B|\leq1/2$. If the
usual constraint
$|Re\,\omega_{sp}(\{q\}_0)|\leq1/2$ is employed instead of
(\ref{argum}), then $\widetilde B=0$.

Restrictions (\ref{imagin})-(\ref{dbound}) do not forbid an overlap of
Schottky circles corresponding  to distinct forming
transformations.  Indeed, from (\ref{imagin}) it follows \cite{siegal}
that
\begin{equation}
min[y_{jj}(\{q,\bar q\}_0),\,y_{rr}(\{q,\bar
q\}_0)]\geq2|y_{rs}(\{q,\bar
q\}_0)|.
\label{imnd}
\end{equation}
Assuming  the Schottky multipliers to be small, one obtains  that $y_{rr}(\{q,\bar
q\}_0)\approx(\ln k_r)/2\pi i$. Then from (\ref{imnd}), it is follows that
$y_{jl}(\{q,\bar q\}_0)=Im\omega_{jl}(\{q\}_0)$  is approximately bounded as follows
\begin{equation}
\sqrt{|k|}\leq\exp[2\pi
y_{jl}(\{q,\bar q\}_0)]\leq1/\sqrt{|k|}\,,
\qquad |k|= max(|k_r|,|k_s|)\,.
\label{inosp}
\end{equation}
If, in addition, distances between Schottky circle centers are nothing much as between their radii, then
$\omega_{sp}(\{q\}_0)$ is approximated by the first term on the right side of (\ref{om0sp}) so that
\begin{equation}
\exp[2\pi
i\omega_{sp}(\{q\}_0)]\approx
\frac{(u_s-u_p)(v_s-v_p)}{(u_s-v_p)(v_s-u_p)}\,.
\label{aposp}
\end{equation}
If, for instance, $2\pi i\omega_{sp}(\{q\}_0)\to\ln k_p$, then from
(\ref{aposp}), the $u_s$ point being near $u_p$, may be either inside,
or outside $C_{u_p}$-circle, against the magnitude of
$|(v_s-v_p)/(v_s-u_p)|$.

Grassmann moduli  being equal to  zero, the integration
region over moduli can be restricted through the
multiplying   of
the local amplitude
by the product of the $\theta({\cal G}_j(\omega,\overline{\omega}))$
step functions where $\omega\equiv\omega(\{q\}_0)$ and
equations
${\cal G}_i(\omega,\overline\omega)=0$ determine the boundary
of the integration region  in line with (\ref{imagin}) and (\ref{dbound}).
If Grassman parameters are present, then
the local amplitude
is multiplied by
\begin{equation}
{\cal
O}(\omega_L,\overline{\omega_{L'}})=
\prod_j\theta({\cal G}_j(\omega_L,\overline{\omega_{L'}}))\,,
\quad\omega_L\equiv\omega(\{q\},L)
\label{bound}
\end{equation}
where the step function  $\theta(x)$ is
understood as the Taylor series in those Grassmann quantities, which
are contained in $x$.

The period matrix does not
changed under redefinitions
\begin{equation}
\Gamma_{a,s}(l_{2s})\to
\Gamma_{G_s}\Gamma_{a,s}(l_{1s})\Gamma_{G_s}^{-1}\,,
\quad \Gamma_{b,s}(l_{2s})\to
\Gamma_{G_s}\Gamma_{b,s}(l_{2s})\Gamma_{G_s}^{-1}
\label{morph}
\end{equation}
of the forming transformations (\ref{gamab}) where
$\Gamma_{G_s}$ is  any super-Schottky group transformation.
Generically, $G_s$ depends on $s$.
The limiting points
$U_s=(u_s|\mu_s)$ and $V_s=(v_s|\nu_s)$ are replaced by
$\Gamma_{G_s}U_s$ and $\Gamma_{G_s}V_s$. So the step function
factor (\ref{zsbound}) does not changed.   So the integration
region over $(u_s,v_s)$  must be more bounded, for instance, by the
requirement that either $u_s$, or $v_s$ is exterior of all the Schottky
circles $C_{u_g}$ and $C_{v_g}$ with the exception of $C_{u_s}$ or,
respectively, of $C_{v_s}$ (see Sec. II for definitions).
As in Sec. II,
the Schottky circles can be replaced by
other relevant curves. The required boundation
is realized
via multiplying the local amplitude by a relevant
step function factor. If $(u_1,v_1,u_2,\mu_1,\nu_1)$ are fixed due to
$SL(2)$-symmetry, then the step function factor can be chosen, for
instance, as $\widetilde
B_{L,L'}^{(n)}(\{V_s,\bar V_s\};\{q,\bar q\})$ given by
\begin{equation}
\widetilde B_{L,L'}^{(n)}(\{V_s,\bar
V_s\};\{q,\bar q\})=\prod_{s=3}^n{\cal B}_{L,L'}^{(n,s)}(V_s,\bar
V_s;\{q,\bar q\})
\label{uprodb}
\end{equation}
where
\begin{equation}
{\cal B}_{L,L'}^{(n,s)}(V_s,\bar V_s;\{q,\bar q\})= \prod_{G\neq
G_s}\theta\biggl(\ell_G^L(V_s)
\overline{\ell_G^{L'}(V_s)}-1\biggl)
\theta\biggl(1-\ell_G^L(\Gamma_{b,G^{-1}}(L,V_s))
\overline{\ell_G^{L'}(\Gamma_{b,G^{-1}}(L',V_s))}\biggl)
\label{zsbounq}
\end{equation}
and $\ell_G^L(t)$  can be  the same as in (\ref{zsbound}).
Like (\ref{zsbound}), the group transformations with exception of
the unity transformation are divided into transformations $G$ and
inverse $G^{-1}$ to them.  The product is calculated over all the group
elements $G$ except $G=I$ and  $G=G_s$ where $G_s$
corresponds to $\Gamma_{b,s}(l_{2s})$ transformation.
The step functions are understood as series in Grassmann
quantities.

Henceforth,
the $n$-loop, $m$-point amplitude
$A_{n,m}(\{p_j,\zeta^{(j)}\})$ for the interaction states
carrying 10-dimensional momenta $\{p_j\}$ and the polarization
tensors $\zeta^{(j)}$, is represented as follows
\begin{eqnarray}
A_m^{(n)}(\{p_j,\zeta^{(j)}\})=\frac{g^{2n+m-2}}{2^{n-2}(n-2)!}\int
\sum_{L,L'}|\widehat H(U_1,V_1,U_2)|^2
Z_{L,L'}(\{q,\overline q\})
{\cal O}^{(n)}(\omega_L,\overline{\omega_{L'}})
\nonumber\\
\times
\widehat B_{L,L'}^{(n)}(\{t_j,\bar
t_j\};\{q,\bar q\})\widetilde B_{L,L'}^{(n)}(\{V_s,\bar
V_s\};\{q,\bar q\})<\prod_{r=1}^mV(t_r,\overline t_r;p_r;\zeta^{(r)})>
(dqd\overline q)'(dtd\overline t)
\label{ampl}
\end{eqnarray}
where $Z_{L,L'}(\{q,\overline q\})$ is the
integration measure (the partition function),
$<...>$ denotes the vacuum expectation of the
interaction vertex product,
$g$ is the coupling constant and $L$
($L'$) labels the super-spin structures of right (left) movers.
The $\widehat H(U_1,V_1,U_2)$ factor (\ref{fixm}) arises because  $(3|2)$
super-Schottky group parametres are fixed due to $SL(2)$-symmetry.
The integration is
performed over vertex coordinates $\{t_j,\overline t_j\}$ on the
$(1|1)$ complex supermanifold and over the
$\{k_s,u_s,v_s,\mu_s,\nu_s\}$ set of those  $(3n-3|2n-2)$ super-Schottky
group complex parameters, which are not fixed.
For any Grassmann variable $\eta$
we define $\int d\eta\eta=1$. For any boson variable $x$ we define $dxd\overline
x=d(Re\,x)d(Im\,x)/(4\pi)$.  In this case the amplitude normalization
corresponds to the normalization in \cite{gsw}.
Step
function factors are given by (\ref{prodb}), (\ref{bound})
and by (\ref{uprodb}).
They are
treated as Taylor series with respect to Grassmann quantities in the
argument that originates $\delta$-functions and their
derivatives, which give rise to the integral over the boundary of the
integration region.

Choosing relevant $C$ and $D$ in (\ref{dbound}), one obtains
that
\begin{equation}
|(F\omega(\{q\}_0)F^T+\widetilde B)_{jj}|^2\geq1
\label{ddig}
\end{equation}
where the $F$ matrix is just one as in (\ref{parttr}). From
(\ref{om0ss}), the boundary of inequality (\ref{ddig}) is achieved
for all  $k_G$  to be small, $k_G$ being the multiplier of
the Schottky group transformation $G$. Indeed,  in this case
$|\omega(\{q\}_0)_{G,G})|^2\approx|(2\pi)^{-1}\ln k_G|^2$ that
corresponds to $|k_G|\leq e^{-3\pi/2}$.  Seemingly, the
fundamental region (\ref{imagin})-(\ref{dbound})
contains only small $k_G$, but this matter still needs
an additional study.  In any case, it is accepted in the paper that
all the Schottky transformations are loxodromic. Among other things, in
this case the $|k_G|\approx1$ multipliers do not contained in the
fundamental region (\ref{imagin})-(\ref{dbound}).

Furthermore, if all the
$k_G$ multipliers are small and (\ref{aposp})
takes place, then conditions
(\ref{bound})-(\ref{uprodb}) uniquely determine the integration region
over Schottky moduli with exception of the genus-$n=3$ case. In the
genus-3 case there are two sets of the $(u_3,v_3)$ moduli corresponding
to the same period matrix (it is supposed that $u_1$, $v_1$ and $u_2$
are fixed due to $L2$-symmetry). Indeed,   $v_2$
is uniquely expressed from (\ref{aposp}) through $\omega_{12}(\{q\}_0)$
and through $(u_1,v_1,u_2)$ which are fixed due to
$L2$-symmetry.  Then $(u_s,v_s)$ for
every $s>2$ are determined from the pair of equations (\ref{aposp})
with $s$ to be given and
$p\leq2$. These two
equations are reduced to an equation set containing
a quadratic equation and  a linear one.
These two equations calculate two  $(u_s,v_s)$-sets
for $s>2$. They are given in terms of
$\omega_{s1}(\{q\}_0)$, $\omega_{s2}(\{q\}_0)$,
$\omega_{12}(\{q\}_0)$, $u_1\,,u_2$ and $v_1$.

So in the  genus $n=3$ case
two  $(u_3,v_3)$-sets are assigned
to the same
period matrix
that may be taken into account multiplying the amplitude
(\ref{ampl}) by the 1/2 factor.

If $n>3$, then non-diagonal entries
$\omega_{sp}(\{q\}_0)$  with  $s>p>2$ are distinguished  for
various sets of the limiting points, and one-to-one
correspondence arises between the period matrix and the Schottky group
moduli.

The period matrix being  in the fundamental region,
the limiting points  are  found to be in the domain where
approximation(\ref{aposp}) is valid. It is  plausible in this case
that  step functions in (\ref{ampl}) wholly determine the integration
region over the Scottky moduli.

The vacuum
expectation and  integration measures in (\ref{ampl})
are calculated through vacuum correlators
\begin{equation}
<X^N(t_1,\overline t_1)X^M(t_2,\overline t_2)>=-\delta^{NM}
\hat X_{L,L'}(t_1,\overline t_1;t_2,\overline t_2;\{q\})
\label{scorr}
\end{equation}
of the superfields  $X^N(t,\overline t)$ of the matter
($N=0,\dots  9$) and through the
correlator
\begin{equation}
<C(t_1,\overline t_1)B(t_2,\overline t_2)>=-G_{gh}(t_1,t_2;\{q\})
\label{ghcor}
\end{equation}
of the $(C,B)$ ghosts. In this case $C$ is the vector
supermultiplet and $B$ is the 3/2-tensor one. In (\ref{scorr})
the ``mostly plus'' metric is implied.  The
normalization of the fields corresponds to the action $S$ given below
as follows
\begin{equation}
S=\int\,
\frac{d^2z}{\pi}\,d\vartheta\,d\overline\vartheta
\biggl[
B\overline D C
-\overline
BD
\overline C
-2
\overline DX_{\cal N}D
X^{\cal N}\biggl]\,.
\label{lagr}
\end{equation}

We consider the massless boson interaction amplitudes. Thus, for the
Lagrangian (\ref{lagr}), the expression \cite{fried}
for the interaction vertex $V(t,\overline
t;p;\zeta) $ is  written down  through
superfields $X^N(t,\overline t)$ of the matter
as follows
\begin{equation}
V(t,\overline
t;p;\zeta)=4\zeta_{MN} [D(t)X^M(t,\overline
t)][\overline{D(t)}X^N(t,\overline t)]
\exp[ip_RX^R(t,\overline
t)]
\label{vert}
\end{equation}
where $p=\{p^M\}$ is 10-momentum of the interacting boson while
$\zeta_{MN}$ is its polarization tensor,
$p^M\zeta_{MN}=p^N\zeta_{MN}=0$, and $p^2=0$. The spinor
derivative $D(t)$ is defined in (\ref{supder}). The summation
over twice repeated indexes is implied.  We use the "mostly
plus" metric.  The dilaton   $\zeta_{MN}$ tensor is
given by the transverse Kronecker symbol $\delta_{MN}^\perp$.

\section{Vacuum correlator of the matter superfields}

The vacuum correlator (\ref{scorr})
is calculated using
a super-holomorphic Green function.
The super-holomorphic Green function is specified except for a scalar zero mode. As in \cite{danphr},
we use the super-holomorphic Green function
$R_L(t,t';\{q\})$ that is changed under
transformations (\ref{stens}) as follows
\begin{eqnarray}
R_L(t_r^b,t';\{q\})=R_L(t,t';\{q\})+
J_r(t';\{q\};L)\,,
\nonumber\\
R_L(t_r^a,t';\{q\})=R_L(t,t';\{q\})
\label{rtrans}
\end{eqnarray}
where $t_s^a$ and $t_s^b$ are the same as in (\ref{stens})
and
super-holomorphic scalar functions $J_r(t;\{q\};L)$ have
periods as follows
\begin{eqnarray}
J_r(t_s^b;\{q\};L) = J_r(t;\{q\};L)+2\pi i
\omega_{sr}(\{q\},L)\,,
\nonumber\\
J_r(t_s^a;\{q\};L)=J_r^{(s)}(t;\{q\};L)+
2\pi i\delta_{rs}\,.
\label{trjs}
\end{eqnarray}
The $R_L(t,t';\{q\})$ function is normalized by the condition that
\begin{equation}
R_L(t,t';\{q\})=\ln(z-z'-\vartheta\vartheta')+\widetilde R_L(t,t';\{q\})
\label{lim}
\end{equation}
where $\widetilde R_L(t,t';\{q\})$ has no a singularity at
$z\to z'$. Moreover, $\widetilde R_L(t,t';\{q\})$ is defined apart from  a constant which is excluded by a condition that
$\widetilde R_L(t,t';\{q\})$ decreases at $z\to\infty$ or at $z'\to\infty$.
Along with $R_L(t,t';\{q\})$, we define
a   function $K_L(t,t';\{q\})$ as follows
\begin{equation}
K_L(t,t';\{q\})=D(t')R_L(t,t';\{q\})
\label{kr}
\end{equation}
where the $D(t)$ super-derivative is defined by eq. (\ref{supder}).
The  vacuum correlator (\ref{scorr}) of the scalar superfields is given
by
\begin{equation}
4\hat X_{L,L'}(t,\overline t;t',\overline t';\{q\})=
R_L(t,t';\{q\})+\overline{R_{L'}(t,t';\{q\})}
+I_{LL'}(t,\overline
t;t',\overline t';\{q\})\,,
\label{corr}
\end{equation}
\begin{eqnarray}
I_{LL'}(t,\overline
t;t',\overline t';\{q,\bar q\})= [J_s(t;\{q\};L) +
\overline{J_s(t;\{q\};L')}]
\nonumber\\
\times
[\Omega_{L,L'}(\{q,\overline q \})]_{sr}^{-1}
[J_r(t';\{q\};L)+\overline{J_r(t';\{q\};L')}]
\label{illin}
\end{eqnarray}
where $\Omega_{L,L'}(\{q,\overline q\})$  is
\begin{equation}
\Omega_{L,L'}(\{q,\overline q\})=
2\pi
i[\overline{\omega(\{q\},L')}-
\omega(\{q\},L)].
\label{grom}
\end{equation}
As usual, the Green function at  $t=t'$ is
defined as $R_L(t,t;\{q\})-\ln(z-z'-\vartheta\vartheta')$ at $t=t'$.
The
dilaton emission amplitude includes the vacuum pairing
$\tilde I_{L,L'}(t,\bar t;\{q\})$ of the superfields in
front of the exponential in (\ref{vert}) as follows
\begin{equation}
\tilde
I_{L,L'}(t,\bar t;\{q,\bar q\})=
2D(t)D(\bar t')I_{L,L'}
(t,\overline t;t',\overline t';\{q\})|_{t=t'}
\label{ngcor}
\end{equation}
where  definitions are given in (\ref{corr}) and in
(\ref{illin}). The dilaton-vacuum transition constant  is found
by the integration of (\ref{ngcor}) with respect to $t$ taken
over the fundamental region of the super-Schottky group. Integrating by parts and using  the
relations (\ref{trjs}), one obtains that the integral is equal to $n$.
Hence the dilaton-vacuum transition genus-$n$ constant
is equal to  the multiplied by $n$ vacuum amplitude.

At zeroth $(\mu_s,\nu_s)$-parameters, the
Green functions
$R_L(t,t';\{q\})$ and $K_L(t,t';\{q\})$
are reduced to  $R_L(t,t';\{q\}_0)$ and,
respectively, to $K_L(t,t';\{q\}_0)$ that are written down in the
terms of  the boson Green function  $R_b(z,z';\{q\}_0)$ and  of the
$R_{fL}(z,z';\{q\}_0)$ fermion Green function  as follows
\begin{eqnarray}
R_L(t,t';\{q\}_0)=R_b(z,z';\{q\}_0)-\vartheta\vartheta'
R_{fL}(z,z';\{q\}_0)\,, \nonumber\\
K_L(t,t';\{q\}_0)=\partial_{z'}R_b(z,z';\{q\}_0)\vartheta'
+\vartheta R_f(z,z';\{q\}_0)\,.
\label{rrk}
\end{eqnarray}
The boson Green function \cite{fried,vec,danphr,mart}
can be represented as follows
\begin{equation}
R_b(z,z';\{q\}_0)=\ln(z-z')+\sum_\Gamma\ln\biggl(\frac{[z-g_\Gamma(z')]}{[z-g_\Gamma(\infty)]}\biggl)\,.
\label{rbos}
\end{equation}
Therefore, $R_b(z,z';\{q\}_0)=
R_b(z',z;\{q\}_0)\pm2\pi i$.
Furthermore, $R_b(z,z';\{q\}_0)$ differs only in scalar zero mode from
the usual  $\ln E(z,z')$ expression \cite{ver}
where $E(z,z')$  is the prime form.
So,
\begin{equation}
\partial_z\partial_{z'}R_b(z,z';\{q\}_0)=
\partial_z\partial_{z'}E(z,z')\,.
\label{prfgr}
\end{equation}
The fermion Green function $R_{fL}(z,z';\{q\}_0)$ for the even spin
structure $(l_1,l_2)$ is  as follows \cite{danphr}
\begin{equation}
R_{fL}(z,z';\{q\}_0)=\exp\{\frac{1}{2}[R_b(z,z;\{q\}_0)
+R_b(z',z';\{q\}_0)]-
R_b(z,z';\{q\}_0)\}
\frac{\Theta[L]({\bf z}-{\bf
z'}|\omega)}{\Theta[L](0|\omega)}
\label{fgrfr}
\end{equation}
where $\Theta[L]({\bf z}-{\bf
z'}|\omega)$ is the theta function,
$L=(l_1,l_2)$ being its characteristics, and  ${\bf z}$ is related to
$z$ by the Jacobi mapping. The function
$R_b(z,z;\{q\}_0)$ at $z'=z$ is defined as
the limit of  $R_b(z,z';\{q\}_0)-\ln(z-z')$  at $z\rightarrow z'$.
The Green function (\ref{fgrfr}) coincides with the fermion Green
function \cite{ver} given by
\begin{equation}
R_{fL}(z,z';\{q\}_0)=
\frac{\Theta[L]({\bf z}-{\bf
z'}|\omega)}{E(z,z')
\Theta[L](0|\omega)}\,.
\label{fgrfe}
\end{equation}

If $(\mu_s,\nu_s)$-parameters are present, then  vacuum correlators
in the Neveu-Schwarz sector  are obtained by a
"naive" supersymmetrization \cite{vec,dan89,mart} of
the corresponding expressions at zeroth $(\mu_s,\nu_s)$-parameters.
In the Ramond sector the
fermions are
agitated with bosons   under twists as about  $B_s$-cycles,
so about $A_s$-ones. Then vacuum  correlators are obtained by the
"naive" supersymmetrization \cite{danphr} only for the genus-1 surface
when Grassmann parameters $(\mu_s,\nu_s)$ are not  moduli.  In this
case $(\mu_s,\nu_s)$ can be reduced to  zero by going to
$(z_s|\vartheta_s)$ variables (\ref{tgam}). The
genus-1
functions  $R_{Ls}^{(1)}(t_1,t_2)$  and $K_{Ls}^{(1)}(t_1,t_2)$ is
expressed through
the  genus-1 functions $R_b(z_1^{(s)},z_2^{(s)};\{q_s\}_0)$
and $R_{fL_s}(z,z';\{q_s\}_0)$ at $\mu_s=\nu_s=0$ as follows
\begin{eqnarray}
R_{Ls}^{(1)}(t_1,t_2)
=R_b(z_1^{(s)},z_2^{(s)};\{q_s\}_0)-\vartheta_1^{(s)}\vartheta_2^{(s)}
R_{fL_s}(z_1^{(s)},z_2^{(s)};\{q_s\}_0)
\nonumber\\
-\varepsilon_s'\vartheta_2^{(s)}\Xi(z_2;\{q_s\}_0)+
\vartheta_1^{(s)}\varepsilon_s'\Xi(z_1;\{q_s\}_0)\,,
\label{gir}\\
K_{Ls}^{(1)}(t_1,t_2)=D(t_2)R_{Ls}^{(1)}(t_1,t_2)
\label{g1k}
\end{eqnarray}
where
$z_i^{(s)}$ and $\vartheta_i^{(s)}$ for $i=(1,\,2)$ are related
with $z$ and $\vartheta$
by (\ref{tgam}), and
\begin{equation}
\Xi(z';\{q_s\}_0)=(z-z')R_{fL_s}(z,z';\{q\}_0)|_{z\to\infty}\,.
\label{xi}
\end{equation}
Two  last terms on the right side of (\ref{gir})  hold
decreasing $K_{L_s}(t_1,t_2;\{q_s\})$ at
$z_1\rightarrow\infty$ and at $z_2\rightarrow\infty$.
Otherwise $R_{Ls^{(1)}}(t_1,t_2)$  is obtained by
the  transformation (\ref{tgam}) of
$R_{L_s}(t_1,t_2;\{q_s\}_0)$.
For the
even spin structures, both
$R_b(z_1,z_2;\{q_s\}_0)$  and
$R_{fL_s}(z_1^{(s)},z_2^{(s)};\{q_s\}_0)$  are given by (\ref{rbos})
and, respectively, by (\ref{fgrfr}) taken in the genus-1 case. The odd
spin-structure
$R_{fL_s}(z_1^{(s)},z_2^{(s)};\{q_s\}_0)$ function is given in \cite{danphr}.

The calculation of the vacuum
correlators  on the higher genus supermanifolds  has been given in
\cite{danphr} where  the proportional to $(\mu_s,\nu_s)$
terms of the  correlators have been represented through
genus-1 functions. Now we calculate these  terms
through  genus-$n$ functions  at
zeroth Grassmann $(\mu_s,\nu_s)$-parameters.

To do so,
we derive equations calculating $R_L(t,t';\{q\})$
in terms of $R_b(z,z';\{q\}_0)$ and of $R_{fL}(z,z';\{q\}_0)$. For the
sake of simplicity, we suppose that Schottky circles do not overlap
each other.

We start with the  identity\footnote{Throughout
the paper $\int d\vartheta\vartheta=1$}
\begin{equation}
K_L(t,t';\{q\})=
-\oint D(t_1)R(t,t_1;\{q\}_0)\frac{dz_1\,d\vartheta_1}{2\pi i}
K_L(t_1,t';\{q\})
\label{ointk}
\end{equation}
where $D(t_1)$ is defined in (\ref{supder}) and the integration is performed in the positive direction
along the infinitesimal contour surrounding  $z_1$-point.
The contour is deformed into the sum of the  $C_s$ contours. The $C_s$
contour surrounds the Schottky circles $C_{v_s}$ and $C_{u_s}$ as
well as the cut  between  $u_s$ and $v_s$ (for the Ramond
handle). Then, by the Schottky transformations (\ref{zgama}),
the integral over every $C_s$ contour is reduced to to the integral
along the $C_{v_s}$ circle and along the cut.  It is
convenient to represent $K_L(t,t';\{q\})$ as follows
\begin{equation}
K_L(t,t';\{q\})=K_{bL}(z,t';\{q\})+\vartheta K_{fL}(z,t';\{q\})\,.
\label{klbf}
\end{equation}
Transforming (\ref{ointk}) to the integration over $C_s$ contours
and reducing every integral over $C_s$ to the integral over  $C_{v_s}$-contour (\ref{circ})
and over the cut (when $l_{1s}=1/2$), one is verified that
$K_{bL}(z,t';\{q\})$ and
$K_{fL}(z,t';\{q\})$ are written down as follows
\begin{eqnarray}
K_{bL}(z,t';\{q\})=\partial_{z'}R_b(z,z';\{q\}_0)\vartheta'+
\sum_s \int_{C_{v_s}}\partial_{z_1}R_b(z,z_1;\{q\}_0)
\frac{dz_1}{2\pi i}\Delta_{bL}^{(s)}(z_1,t';\{q\})
\nonumber\\
+\int_{z_{v_s}}^{z_{u_s}}\partial_{z_1}R_b(z,z_1;\{q\}_0)
\frac{dz_1}{2\pi i}\Delta_{bL}^{(s-)}(z_1,t';\{q\})\,,
\nonumber\\
K_{fL}(z,t';\{q\})=R_{fL}(z,z'\{q\}_0)-
\sum_s \int_{C_{v_s}}R_{fL}(z,z_1;\{q\}_0)\frac{dz_1}{2\pi i}
\Delta_{fL}^{(s)}(z_1,t';\{q\})
\nonumber\\
-\int_{z_{v_s}}^{z_{u_s}}R_f(z,z_1;\{q\}_0)\frac{dz_1}{2\pi i}
\Delta_{fL}^{(s-)}(z_1,t';\{q\})
\label{oink}
\end{eqnarray}
where $z_{v_s}$ lays on the $C_{v_s}$ circle and $z_{u_s}=
g_s(z_{v_s})$.
The integration around $C_{v_s}$ is performed in the positive direction
starting from $z_{v_s}$.
The integrand contains
the change of  $K_L(t,t';\{q\})$  under the
$\Gamma_{b ,s}^{(0)}(l_{2s})$ transformation (\ref{zgama})
which is represented as follows
\begin{eqnarray}
K_{bL}(g_s(z),t';\{q\})=K_{bL}(z,t';\{q\})
+D(t')J_s(t')+\Delta_{bL}^{(s)}(z,t';\{q\})\,,
\nonumber\\
-(-1)^{2l_{2s}}\sqrt{g_s'(z)}K_f(g_s(z),t';\{q\})
=K_{fL}(z,t';\{q\})+\Delta_{fL}^{(s)}(z,t';\{q\})\,.
\label{kbf}
\end{eqnarray}
In the Ramond case ($l_{1s}=1/2$)
the integrand contains
also
the change of  $K_L(t,t';\{q\})$  under the
$\Gamma_{a,s}^{(0)}(l_{1s})$-transformation  (\ref{zgama}).
Then
the $K_{bL}(z,t';\{q\})$ and $K_{bL}(z,t';\{q\})$
functions turn out into
$K_{bL}^{(s-)}(z,t';\{q\})$ and
$K_{fL}^{(s-)}(z,t';\{q\})$
where
\begin{eqnarray}
K_{bL}^{(s-)}(z,t';\{q\})=K_{bL}(z,t';\{q\})
+\Delta_{bL}^{(s-)}(z,t';\{q\})\,,
\nonumber\\
-K_{fL}^{(s-)}(z,t';\{q\})
=K_{fL}(z,t';\{q\})+\Delta_{fL}^{(s-)}(z,t';\{q\})\,.
\label{kbfa}
\end{eqnarray}
In (\ref{oink}), there is taken into account  that
\begin{equation}
\int_{C_{v_s}}\partial_{z_1}R_b(z,z_1;\{q\}_0)
\frac{dz_1}{2\pi i}=0\,.
\label{rbint}
\end{equation}
To calculate the right side of (\ref{kbf}),
it is useful to represent $K(t^{(b0)}_s,t';\{q\})$ for as follows
\begin{eqnarray}
K(t^{(b0)}_s,t';\{q\})=K(t^{(b)}_s,t';\{q\})
-(z_s^{(b)}-z_s^{(b0)})D^2(t_s^{(b)})K(t^{(b)}_s,t';\{q\})
\nonumber\\
-(\vartheta_s^{(b)}-\vartheta_s^{(b0)})
[D(t_s^{(b)})K(t^{(b)}_s,t';\{q\})
-\vartheta_s^{(b)}D^2(t_s^{(b)})K(t^{(b)}_s,t';\{q\})]
\label{calde}
\end{eqnarray}
where it is employed that $D^2(t)=\partial_z$. Due to (\ref{stens}),
eq.  (\ref{calde}) is represented as follows
\begin{eqnarray}
K(t^{(b0)}_s,t';\{q\})
=K(t,t';\{q\})+D(t')J_s(t';\{q\};L)
\nonumber\\
-(z_s^{(b)}-z_s^{(b0)})
Q_{\Gamma_{b,s}}(t)D(t)
Q_{\Gamma_{b,s}}(t)D(t)K(t,t';\{q\})
\nonumber\\
-(\vartheta_s^{(b)}-\vartheta_s^{(b0)})[Q_{\Gamma_{b,s}}(t)D(t)K(t,t';\{q\})
\nonumber\\
-\vartheta_s^{(b)}Q_{\Gamma_{b,s}}(t)D(t)
Q_{\Gamma_{b,s}}(t)D(t)Q_{\Gamma_{b,s}}(t))K(t,t';\{q\})]
\label{caldea}
\end{eqnarray}
where (\ref{rtrans}) and (\ref{kr}) are also employed.
The right side of (\ref{kbf}) is calculated directly from
(\ref{caldea}) as follows
\begin{eqnarray}
\Delta_{bL}^{(s)}(z,t';\{q\})=\varepsilon_s(z)\epsilon_s(z)
\partial_zK_b(z,t';\{q\})-\epsilon_s(z)K_f(z,t'\{q\})
\nonumber\\
\Delta_{fL}^{(s)}(z,t';\{q\})=
-\epsilon_s(z)\partial_zK_b(z,t';\{q\}) +
\nonumber\\
\varepsilon_s(z)\epsilon_s(z)
\partial_zK_f(z,t';\{q\})
+[(-1)^{2l_{2s}}(c_sz+d_s)+1]\varepsilon\varepsilon'K_f(z,t';\{q\})\,.
\label{strkbf}
\end{eqnarray}
The right side of (\ref{kbfa}) is calculated in
the kindred manner, it being found as follows
\begin{eqnarray}
\Delta_{fL}^{(-s)}(z,t';\{q\})=2\varepsilon_s(z)\partial_z
K_{bL}(z,t';\{q\}) +2\varepsilon_s\varepsilon_s'K_{fL}(z,t';\{q\})\,,
\nonumber\\
\Delta_{bL}^{(s-)}(z,t';\{q\})=2\varepsilon_s(z)K_f(z,t';\{q\})\,.
\label{strkbfa}
\end{eqnarray}
It is useful to note that (\ref{strkbfa}) can be derived from
(\ref{strkbf}) by the  $k_s=1$, $l_{2s}=0$ setting.

Using (\ref{strkbf}) and (\ref{strkbfa}),
one can prove that (\ref{oink})
is equivalent to the following equations
\begin{eqnarray}
K_{bL}(z,t';\{q\})=\partial_{z'}R_b(z,z';\{q\}_0)\vartheta'+
\sum_s \int_{C_s}\partial_{z_1}R_b(z,z_1;\{q\}_0)\varepsilon_s(z_1)
\frac{dz_1}{2\pi i}K_{fL}(z_1,t';\{q\})\,,
\nonumber\\
K_{fL}(z,t';\{q\})=R_{fL}(z,z'\{q\}_0)-
\sum_s\int_{C_s} R_{fL}(z,z_1;\{q\}_0)\varepsilon_s(z_1)
\frac{dz_1}{2\pi i}\partial_{z_1}K_{bL}(z_1,t';\{q\})
\label{eqsk}
\end{eqnarray}
where the integration is performed in the positive direction along
the $C_s$ contour, which surrounds
the Schottky circles  $C_{u_s}$ and $C_{v_s}$ as well as
the cut (if $l_{1s}\neq0$) between $u_s$ and $v_s$.
Eqs. (\ref{eqsk}) gives $K_{bL}(z,t';\{q\})$
and $K_{fL}(z,t';\{q\})$ in the form of series over
$(\mu_s,\nu_s)$.

In line
with (\ref{klbf}),  the superscalar Green function $R_L(t,t';\{q\})$
are represented as follows
\begin{equation}
R_L(t,t';\{q\})=R_{bL}(z,t';\{q\})-\vartheta R_{fL}(z,t';\{q\})\,.
\label{rlbf}
\end{equation}
Equations
for $R_{bL}(t,t';\{q\})$ and for $R_{fL}(t,t';\{q\})$ are
derived directly from (\ref{eqsk}) to be
\begin{eqnarray}
R_{bL}(z,t';\{q\})=R_b(z,z';\{q\}_0)-
\sum_s \int_{C_s}\partial_{z_1}R_b(z,z_1;\{q\}_0)\varepsilon_s(z_1)
\frac{dz_1}{2\pi i}R_{fL}(z_1,t';\{q\})\,,
\nonumber\\
R_{fL}(z,t';\{q\})=R_{fL}(z,z'\{q\}_0)\vartheta+
\sum_s\int_{C_s} R_{fL}(z,z_1;\{q\}_0)\varepsilon_s(z_1)
\frac{dz_1}{2\pi i}\partial_{z_1}R_{bL}(z_1,t';\{q\})\,.
\label{eqsr}
\end{eqnarray}
The equations for the superscalar function $J_r(t;\{q\};L)$ in (\ref{rtrans})
are obtained in the quite kindred  way
starting with the representation of $J_r(t;\{q\};L)$
by the integral along the contour that surrounds the $z$ point.
The integral is obtained from (\ref{ointk}) by
the $K_l(t,t';\{q\})\to J_r(t;\{q\};L)$ replacement. Further,
$J_r(t;\{q\};L)$ is represented as follows
\begin{equation}
J_r(t;\{q\};L)=J_{br}(z;\{q\};L)+\vartheta J_{fr}(z;\{q\};L)\,.
\label{jbf}
\end{equation}
Under the transformation (\ref{zgama}) the superscalar function is
changed as follows
\begin{eqnarray}
J_{br}(g_s(z);\{q\};L)=J_{br}(z,;\{q\};L)
+2\pi i\omega_{mr}(\{q\};L)+\varepsilon_s(z)\epsilon_s(z)
\partial_zJ_{br}(z,t;\{q\};L)
\nonumber\\
-\epsilon_s(z)J_{fr}(z,t\{q\};L)\,,
\nonumber\\
J_{fr}(g_s(z),t';\{q\})
=J_{fr}(z;\{q\};L)-\epsilon_s(z)\partial_zJ_{br}(z;\{q\};L) +
\varepsilon_s(z)\epsilon_s(z)
\partial_zJ_{fr}(z;\{q\};L)
\nonumber\\
+[(-1)^{2l_{2s}}(c_sz+d_s)+1]
\varepsilon\varepsilon'J_{fr}(z;\{q\};L)
\label{jbfb}
\end{eqnarray}
and under the
$\Gamma_{a,s}(l_{1s})$ transformation  (\ref{zgama}),
in the Ramond case,
\begin{eqnarray}
J_{fr}^{(s-)}(z;\{q\};L)
=J_{fr}(z;\{q\};L)+2\varepsilon_s(z)\partial_z
J_{br}(z;\{q\};L) +2\varepsilon_s\varepsilon_s'J_{fr}(z;\{q\};L)\,,
\nonumber\\
J_{br}^{(s-)}(z;\{q\};L)=J_{br}(z;\{q\};L)+2\pi i\delta_{rs}
+2\varepsilon_s(z)J_{fr}(z;\{q\};L)\,.
\label{jbfa}
\end{eqnarray}
The desired equations   are derived like (\ref{eqsk})
They are found as follows
\begin{eqnarray}
J_{br}(z;\{q\};L)=J_r(z;\{q\}_0)+
\sum_s \int_{C_s}\partial_{z_1}R_b(z,z_1;\{q\}_0)\varepsilon_s(z_1)
\frac{dz_1}{2\pi i}J_{fr}(z_1;\{q\};L)\,,
\nonumber\\
J_{fr}(z;\{q\};L)=-
\sum_s\int_{C_s} R_{fL}(z,z_1;\{q\}_0)\varepsilon_s(z_1)
\frac{dz_1}{2\pi i}\partial_{z_1}J_{br}(z_1;\{q\};L)
\label{eqj}
\end{eqnarray}
where $J_r(z;\{q\}_0)$ is the scalar function
having the periods  given by the $\omega_{mr}(\{q\}_0)$
period matrix. The $C_s$ contour
is the same as in (\ref{eqsk}).
In the explicit form
\cite{fried,vec,danphr,mart}
\begin{equation}
J_r(z;\{q\}_0)={\sum_\Gamma}^\prime\ln\frac{z-g_\Gamma(u_r)}
{z-g_\Gamma(v_r)}\quad,
\label{scf}
\end{equation}
$u_r$  $v_r$ being the limiting points of the Schottky
transformation $g_r(z)$.  The summation
is performed over all the group products $\Gamma$ except those  whose
rightmost is a power of $g_r$.
The
$\omega_{mr}(\{q\};L)$ period matrix
are obtained
by the $z\to g_m(z)$
transformation of both sides of (\ref{eqj}).
In this case $z$-point must be moved  inside the $C_m$ contour.
Due to the pole at $z=z_1$, the integrated term in (\ref{eqj})
arises, and (\ref{eqj}) appears to be
\begin{eqnarray}
J_{br}(z;\{q\};L)=J_r(z;\{q\}_0)-
\varepsilon_m(z)J_{fr}(z;\{q\};L)
\nonumber\\
+
\sum_s \int_{C_s}\partial_{z_1}R_b(z,z_1;\{q\}_0)\varepsilon_s(z_1)
\frac{dz_1}{2\pi i}J_{fr}(z_1;\{q\};L)\,,
\nonumber\\
J_{fr}(z;\{q\};L)=-\varepsilon_m(z)\partial_{z}J_{br}(z;\{q\};L)
\nonumber\\
-
\sum_s\int_{C_s} R_{fL}(z,z_1;\{q\}_0)\varepsilon_s(z_1)
\frac{dz_1}{2\pi i}\partial_{z_1}J_{br}(z_1;\{q\};L)\,,
\label{eqjj}
\end{eqnarray}
The condition that (\ref{eqjj}) is consistent with
(\ref{jbfb})
determines the period matrix as follows
\begin{equation}
2\pi i\omega_{mr}(\{q\};L)=2\pi i
\omega_{mr}(\{q\}_0)+\sum_s \int_{C_s}
\partial_z J_m(z;\{q\}_0) \frac{dz}{2\pi
i}\varepsilon_s(z)J_{fr}(z;\{q\};L)\,.
\label{ome}
\end{equation}
If  Schottky circles overlap each other, then the
vacuum correlator and associated with it quantities are obtained by an analytic continuation in the
Schottky parameters of the foregoing expressions. In the process, the singularities of
the analytic continued expressions are moved on the complex plane of the integrated variables
that causes a moving of the integration contours. The integration contours are moved in such a way that
the singularities do not cross the integration contour.
As the result,
the integration contours in the discussed
expressions are, generally, deformed. Otherwise  the foregoing expressions remain to be the same

\section{Ghost superfield correlator}

The  superfield $C$ in  (\ref{ghcor}) and in
(\ref{lagr}) has discontinuity  \cite{danphr} under twists about
non-contractible cycles on the Riemann surface. It is a peculiarity
of the scheme \cite{danphr} which, among other things,   calculates
the $B$-superfield zero-mode contribution
to the integration measure. In this case
the $G_{gh}(t,t';\{q\})$  function in (\ref{ghcor}) is changed
under $t\to t_s^{(p)}=(z_s^{(p)}|\vartheta_s^{(p)})$
transformations (\ref{sbtr}) and (\ref{satr}) as follows ($p=a,b)$
\begin{equation}
G_{gh}(t_s^{(p)},t';\{q\})=Q_{\Gamma_{p,s}}^{-2}(t)\biggl
(G_{gh}(t,t';\{q\})+
{\sum_{N_s}}' Y_{p,N_s}(t)\tilde \chi_{N_s}(t';\{q\})\biggl)
\label{ghcorr}
\end{equation}
where summation is performed over  parameters
$N_s=(k_s,u_s,v_s,\mu_s,\nu_s)$ excepting those
$(3|2)$ parameters which are fixed due to
$SL_2$-symmetry. Further, $\tilde\chi_{N_s}$ are superconformal
3/2-zero modes, and $Y_{p,N_s}$ are polynomials  of degree 2 in
$(z,\vartheta)$ as follows
\begin{equation}
Y_{p,N_s}(t)=Q_{\Gamma_{p,s}}^2\biggl[\frac{\partial g_s^p}
{\partial q_{N_s}}+\vartheta_s^p\frac{\partial \vartheta_s^p}
{\partial q_{N_s}}\biggl].
\label{poly}
\end{equation}
The $G_{gh}(t,t';\{q\})$ function can be represented
using a  Green function G(t,t';\{q\})
which is changed
under $t\to t_s^{(p)}=(z_s^{(p)}|\vartheta_s^{(p)})$
transformations (\ref{sbtr}) and (\ref{satr}) as follows ($p=a,b)$
\begin{equation}
G(t_s^{(p)},t';\{q\})=Q_{\Gamma_{p,s}}^{-2}(t)\biggl
(G(t,t';\{q\})+
\sum_{N_s} Y_{p,N_s}(t)\chi_{N_s}(t';\{q\})\biggl)
\label{gcor}
\end{equation}
where summation is performed over all
the super-Schottky group
parameters $N_s$ which are assigned to the $s$-handle.  Further,
$\chi_{N_s}(t';\{q\})$ is a superconformal 3/2-supertensor having a
singularity at $z'\to\infty$.  Then
\begin{equation}
G_{gh}(t,t';\{q\})=G(t,t';\{q\})-
\sum_{N_0,N_0'}Y_{b,N_0}(t)A_{N_0N_0'}^{-1}
\chi_{N_0'}(t')
\label{gghg}
\end{equation}
where summation is performed over those $(3|2)$ super-Scottky group
parameters that are fixed by $SL(2)$-symmetry, and the $A_{N_0,N_0'}$
matrix is determined from the condition that $G_{gh}(t,t';\{q\})$
satisfies to (\ref{ghcorr}).
The $Y_{p,N_s}(t)$ polynomials are given  in \cite{danphr}
and in Appendix A of this paper.

All the $(\mu_s,\nu_s)$ parameters being equal to zero,
$G(t,t';\{q\})$  is reduced to
$G(t,t';\{q\}_0)$  given in the terms of
the boson Green function  $G_b(z,z';\{q\}_0)$ and of
the fermion Green function $G_f(t,t';\{q\}_0)$  as follows
\begin{equation}
G(t,t';\{q\}_0)=G_b(z,z';\{q\}_0)\vartheta'+ \vartheta G_f(z,z';\{q\}_0).
\label{gzgm}
\end{equation}
The boson Green function $G_b(z,z';\{q\}_0)$ is given by \cite{danphr}
\begin{equation}
G_b(z,z';\{q\}_0)=-\sum_\Gamma\frac{1} {[z-g_\Gamma(z')][c_\Gamma
z'+d_\Gamma]^4}
\label{bgrf}
\end{equation}
where the summation is
performed over all the group products
$\Gamma$ of the  Schottky
group basic elements.
The fermion Green function $G_f(z,z';\{q\}_0)$ does not
represented by a Poincar\'e series. It
is calculated through the Green function $G_\sigma(z,z';\{q\}_0)$  given by
\begin{equation}
G_\sigma (z,z';\{q\}_0)=
\sum_\Gamma\frac{\exp
[\Omega_\Gamma(\{\sigma_s\})+\sum_s2l_{1s}
\sigma_s(J_s(z;\{q\}_0)-J_s(z';\{q\}_0))]}
{[z-g_\Gamma(z')][c_\Gamma z'+d_\Gamma]^3}
\label{gsig}
\end{equation}
where $\sigma_s=\pm1$ and
$\Omega_\Gamma(\{\sigma_s\})$
is defined as
\begin{equation}
\Omega_\Gamma(\{\sigma_s\})=-\pi i[\sum_{s,r}2l_{1s}
\sigma_s\omega_{sr}(\{q\}_0)n_r(\Gamma)+
\sum_r(2l_{2r}-1)n_r(\Gamma)]\,,
\label{omgs}
\end{equation}
$n_r(\Gamma)$ being the
number of times that the $\Gamma_r$ generators  are present
(for its inverse $n_r(\Gamma)$ is negative).  So, $G_\sigma(z,z';\{q\}_0) $
depends on a choice
of the $\{\sigma_s\}$ set.
The change   of $G_\sigma(z,z';\{q\}_0)$ under the $z\to g_r(z)$
Schottky transformation (\ref{zgama}) is as follows
\begin{equation}
G_\sigma(g_r(z),z';\{q\}_0)
=\frac{(-1)^{2l_{2r}-1}}{(c_rz+d_r)} \biggl(G_\sigma
(z,z';\{q\}_0) +\sum_{\alpha_r=\mu_r,\nu_r}\tilde Y_{\sigma,\alpha_r}(z)
\Phi_{\sigma,{\alpha_r}}(z';\{q\}_0)\biggl)
\label{trgsig}
\end{equation}
where  $\Phi_{\sigma,{\alpha_r}}(t';\{q\}_0)$  is 3/2-tensor and $\tilde
Y_{\sigma,\alpha_r}(z)$ for $\alpha_r=(\mu_r,\nu_r)$ is given by
\begin{equation}
\tilde
Y_{\sigma,\alpha_r}(z)=\exp[\pi
i\sum\nolimits_s2l_{1s}\sigma_sJ_s(z;\{q\}_0)]Y_{b,\alpha_r}^{(0)}(z)\,,
\label{tily}
\end{equation}
$Y_{b,\alpha_r}^{(0)}(z)$ being given by eq. (\ref{yb}) of Appendix A.
Moreover, $\Phi_{\sigma,{\alpha_r}}(z';\{q\}_0)$ is found to be
\begin{eqnarray}
\Phi_{\sigma,\mu_s}(z;\{q\}_0)=\grave N_{\mu_s}c_s\biggl[\frac{1}{\sqrt k}
\Phi^{(1)}_{\sigma,s}(z;\{q\}_0)+\Phi^{(2)}_{\sigma,s}(z;\{q\}_0)\biggl]
e^{-2\pi
i\sum_rl_{1r}\sigma_rJ_r(z)}\,,
\nonumber\\
\Phi_{\sigma,\nu_s}(z;\sigma)=\grave N_{\nu_s}c_s\biggl[\sqrt k
\Phi^{(1)}_{\sigma,s}(z;\{q\}_0)+\Phi^{(2)}_{\sigma,s}
(z;\{q\}_0)\biggl]e^{-2\pi
i\sum_rl_{1r}\sigma_rJ_r(z)}
\label{phi}
\end{eqnarray}
where
\begin{equation}
\grave N_{\mu_s}=\frac{(-1)^{2l_{2s}-1}\sqrt k_s}{2[1-(-1)^{2l_{2s}-1}\sqrt
k_s]}\,,\qquad
\grave N_{\nu_s}=\frac{1}{2[-1+(-1)^{2l_{2s}-1}\sqrt
k_s]}
\label{nors}
\end{equation}
and
\begin{equation}
\Phi^{(1)}_{\sigma,s}(z;\{q\}_0)=
\sum_g\frac{e^{\Omega_g(\sigma)}}{Q^2_{g_sg}(z)Q_g(z)}\,,
\qquad
\Phi^{(2)}_{\sigma,s}(z;\{q\}_0)=\sum_g
\frac{e^{\Omega_g(\sigma)}}{Q_{g_sg}(z)Q_g^2(z)}
\label{phic}
\end{equation}
where summation is performed over all the elements $g$ of the Schottky
group.

To calculate
$G_f(z,z';\{q\}_0))$, we represent it as the
integral over $z''$ along the infinitesimal contour
around $z'$, the integrand being $G_\sigma(z,z'';\{q\}_0))
G_f(z'',z';\{q\}_0))$. Running this contour away, we obtain
$G_f(z,z';\{q\}_0))$ as follows
\begin{eqnarray}
G_f(z,z';\{q\}_0))=
G_\sigma(z,z';\{q\}_0))-\sum_{s=1}^n\sum_{\alpha_s}\biggl(
\int_{C_{v_s}}
G_\sigma(z,z'';\{q\}_0)
Y_{b,\alpha_s}^{(0)}(z'')\frac{ dz''}{2\pi i}
\nonumber\\
+ \int_{z_{v_s}}^{z_{u_s}}G_\sigma(z,z'';\{q\}_0)
Y_{a,\alpha_s}^{(0)}(z'')\frac{ dz''}{2\pi i}\biggl)
\chi_{\alpha_s}(z';\{q\}_0)
\label{gfgsig}
\end{eqnarray}
where $z_{v_s}$, $z_{u_s}$ and $C_{v_s}$
are  the same as in (\ref{oink}), and
$\alpha_s=(\mu_s,\nu_s)$. The 3/2-tensor
$\chi_{\alpha_s}(z';\{q\}_0))$  is
calculated
from condition that
the change of $G_f(z,z';\{q\}_0)$ under $2\pi$-twist about
$B_s$-cycle
is given by (\ref{gcor}) at zero
values of the
odd Schottky parameters. Hence
\begin{equation}
\Phi_{\sigma,\alpha_s}(z;\{q\}_0))=\sum_{\alpha_r=\mu_r,\nu_r}
\widetilde M_{\alpha_s,\alpha_r}
(\{\sigma\})
\chi_{\alpha_r}(z;\{q\}_0))
\label{phichi}
\end{equation}
where
\begin{equation}
\widetilde M_{\alpha_s,\beta_r}(\{\sigma\})=
\int_{C_{v_r}}
\Phi_{\sigma,\alpha_s}(z;\{q\}_0))
Y_{b,\beta_r}^{(0)}(z)\frac{dz} {2\pi i}
+
\int_{z_{v_r}}^{z_{u_r}}
\Phi_{\sigma,\alpha_s}(z;\{q\}_0))
Y_{a,\beta_r}^{(0)}(z)\frac{dz} {2\pi i}.
\label{tilmma}
\end{equation}
Using eq.(\ref{ypy}) from Appendix A, one can re-written down
eq. (\ref{gfgsig})
as follows
\begin{equation}
G_f(z,z';\{q\}_0))=
G_\sigma(z,z';\{q\}_0))-\sum_{s=1}^n\sum_{\alpha_s}
\int_{C_s}
G_\sigma(z,z'';\{q\}_0)
Y_{\alpha_s}^{(0)}(z'')\frac{ dz''}{2\pi i}
\label{gfgsigc}
\end{equation}
where $Y_{N_s}^{(0)}(z)$ is given by eqs. (\ref{yu})
of Appendix A. The $C_s$
contour is the same as in (\ref{eqsk}). Correspondingly, (\ref{tilmma}) is re-written down as
follows
\begin{equation}
\widetilde M_{\alpha_s,\beta_r}(\{\sigma\})= \int_{C_r}
\Phi_{\sigma,\alpha_s}(z;\{q\}_0))
Y_{\beta_r}^{(0)}(z)\frac{dz} {2\pi i}\,.
\label{tilmmac}
\end{equation}
Calculating the integration measure in (\ref{ampl}), we shall employ
an expression for $G(t,t';\{q\})$  given through ancillary Green functions
$S_\sigma(t,t';\{q\})$
that have no discontinuity  being twisted about $A_s$-cycles, and satisfy
the conditions
\begin{equation}
S_\sigma(t_s^b,t';\{q\})=Q_{\Gamma_{b,s}}^{-2}(t)
\biggl(S_\sigma(t,t';\{q\})+\sum_{N_s}
\hat Y_{\sigma,N_s}^{(1)}(t)\Psi_{\sigma,N_s}(t';\{q\})\biggl)
\label{ssigtr}
\end{equation}
where $N_s=(k_s,u_s,v_s,\mu_s,\nu_s)$, and
$\Psi_{\sigma,N_s}(t';\{q\})$ is
the 3/2-supertensor. The $\hat
Y_{b,N_s}^{(1)}(t)$ function at $N_s=(k_s,u_s,v_s)$
are equal to $Y_{b,N_s}^{(0)}(t^{(s)})
Q_{\tilde\Gamma_s}^{-2}(t^{(s)})$ where
$t^{(s)}=(z^{(s)}|\vartheta^{(s)})$ is given by (\ref{tgam}), and
$Y_{b,N_s}^{(0)}(t)$ are defined by (\ref{yb}) of Appendix A.  If
$N_s=(\mu_s,\nu_s)$, then $\hat Y_{\sigma,N_s}^{(1)}(t)$  is
\begin{equation}
\hat
Y_{\sigma,N_s}^{(1)}(t)=\vartheta^{(s)}
\tilde Y^{(1)}_{\sigma,N_s}(z^{(s)})Q_{\tilde\Gamma_s}^{-2}(t^{(s)})
\label{hyty}
\end{equation}
where
\begin{equation}
\tilde Y^{(1)}_{\sigma,N_s}(z)
=Y_{b,N_s}^{(0)}(z)
\exp[l_{1s}\sigma_s J^{(1)}_s(z)]\,,
\qquad J^{(1)}_s(z)=\ln\frac{z-u_s}{z-v_s}\,.
\label{ysig}
\end{equation}
The  $G(t,t';\{q\})$ function in eq.(\ref{gcor}) is calculated
through $S_\sigma(t,t';\{q\})$ as follows
\begin{equation}
G(t,t';\{q\})=S_\sigma(t,t';\{q\})-\sum_{s=1}^n\sum_{N_s}
\sum_{p=a,b}\int_{C_p^{(s)}}
S_\sigma(t,t_1;\{q\})\frac{ dz_1d\vartheta_1}{2\pi i}
Y_{p,N_s}(t_1)
\chi_{N_s}(t';\{q\})
\label{gfsgsig}
\end{equation}
where $C_b^{(s)}=\hat C_{v_s}$ and the integration
along $C_a^{(s)}$ is performed between two
congruent points, one of them being on $\hat C_{v_s}$
(therefore, the second point lays on $\hat C_{u_s}$).
The $\hat C_{v_s}$ and $\hat C_{u_s}$ contours are given by (\ref{hcirc1}).
The deriving of (\ref{gfsgsig})
is similar to the deriving of (\ref{gfgsig}). In this case we begin
with  integral over $z''$ along the infinitesimal contour
around $z'$, the integrand being $S_\sigma(z,z'';\{q\}_0))
G_f(z'',z';\{q\}_0))$.
From condition that
the change of $G(t,t';\{q\})$ under $2\pi$-twist about
$B_s$-cycle
is given by (\ref{gcor}),  the relation
arises as follows
\begin{equation}
\Psi_{\sigma,N_s}(t;\{q\})=\sum_{N_r}
M_{N_s,N_r}(\{\sigma\};\{q\})\chi_{N_r}(t;\{q\})
\label{psichi}
\end{equation}
where $\Psi_{\sigma,N_s}(t;\{q\})$  is the same as in
in (\ref{ssigtr})  and
the $M_{N_s,N_r}(\{\sigma\};\{q\})$ entry  of the
$M(\{\sigma\};\{q\})$ matrix is given by
\begin{equation}
M_{N_s,N_r}(\{\sigma\};\{q\})=
\sum_{p=a,b}\int_{C_p^{(r)}}
\Psi_{\sigma,N_s}(t;\{q\}))\frac{d\vartheta dz} {2\pi i}
Y_{p,N_r}(t)
\label{mma}
\end{equation}
where $C_p^{(r)}$-contour is the same as in (\ref{gfsgsig}). In line
with Appendix A,
\begin{equation}
M(\{\sigma\};\{q\})=M^{(r)}(\{\sigma\};\{q\})\widehat T
\label{mmr}
\end{equation}
where $\widehat T$ is given by (\ref{Mmatr}). The
$M_{N_s,N_r}^{(r)}(\{\sigma\};\{q\})$ entry of the
$M^{(r)}(\{\sigma\};\{q\})$ matrix is
as follows
\begin{equation}
M_{N_s,N_r}^{(r)}(\{\sigma\};\{q\})=
\sum_{p=a,b}\int_{C_p^{(r)}}
\Psi_{\sigma,N_s}(t;\{q\}))\frac{d\vartheta dz} {2\pi i}
Y_{p,N_r}^{(r)}(t)
\label{mmrr}
\end{equation}
where
$Y_{p,N_s}^{(r)}$ is given by (\ref{yb}) and $C_p^{(r)}$-contour is defined just as in (\ref{gfsgsig}).
Due to eq. (\ref{psihy})
of Appendix A,
$M_{f,b}^{(r)}(\{\sigma\};\{q\})=0$ and
$M_{b,b'}^{(r)}(\{\sigma\};\{q\})=\delta_{bb'}$. In this case
$(b,b')=(k_s,u_s,v_s)$ and $f=(\mu_r,\nu_r)$.
Eq. (\ref{mmrr}) for$N_r=\alpha_r=(\mu_r,\nu_r)$ can be re-written down as follows
\begin{equation}
M_{N_s,\alpha_r}^{(r)}(\{\sigma\};\{q\})=
-\int_{C_r}
\Psi_{\sigma,N_s}(t;\{q\}))\frac{d\vartheta dz} {2\pi i}
Y_{\alpha_r}^{(r)}(t)
\label{mmrrr}
\end{equation}
where $Y_{\alpha_r}^{(r)}(t)$ is defined by eq. (\ref{tlzer}) and $C_r$-contour is defined as in (\ref{eqsk}).

Due to (\ref{ypy}) and (\ref{ssiggfy}) from Appendix A,
eq. (\ref{gfsgsig}) is re-written down
as follows
\begin{equation}
G(t,t';\{q\})=S_\sigma(t,t';\{q\})+\sum_{s=1}^n\sum_{\alpha_s}\sum_{N_r}
\int_{C_s}
S_\sigma(t,t_1;\{q\})\frac{ dz_1d\vartheta_1}{2\pi i}
Y_{\alpha_s}^{(r)}(t_1)\widehat T_{\alpha_sN_r}
\chi_{N_r}(t';\{q\})
\label{gfsgs}
\end{equation}
If all $(\mu_s,\nu_s)$  are nullified, then
$S_\sigma(t,t';\{q\})$ is reduced to $S_\sigma(t,t';\{q\}_0)$
as follows
\begin{equation}
S_\sigma(t,t';\{q\}_0)=G_b(z,z';\{q\}_0)\vartheta'+\vartheta
S_{(f)\sigma}(z,z';\{q\}_0)
\label{ssigz}
\end{equation}
where $G_b(z,z';\{q\}_0)$ is given by (\ref{bgrf}) and
\begin{eqnarray}
S_{(f)\sigma}(z,z';\{q\}_0)=G_{(\sigma)}
(z,z';\{q\}_0)-\sum_{r=1}^n\sum_{N_r=\mu_r,\nu_r}
\int_{C_b^{(r)}}
G_\sigma(z,z'';\{q\}_0)\frac{ dz''} {2\pi i}
\tilde Y_{\sigma,N_r}^{(1)}(z'')
\nonumber\\
\times
\Psi_{\sigma,N_r}(z';\{q\}_0)
\label{gsissi}
\end{eqnarray}
where $\tilde Y_{\sigma,N_r}^{(1)}(z)$ is defined by (\ref{ysig}) and $G_\sigma(z,z'';\{q\}_0)$ is given by
(\ref{trgsig}).
In this case
\begin{equation}
\Phi_{\sigma,N_s}(z;\{q\}_0)=\sum_{N_r=\mu_r,\nu_r}\hat
M_{N_s,N_R}(\{\sigma\}) \Psi_{\sigma,N_r}(z;\{q\}_0)
\label{hatmphipsi}
\end{equation}
where
\begin{equation}
\hat M_{N_s,N_R}(\{\sigma\})=
\int_{C_b^{(r)}}
\Phi_{\sigma,N_s}(z;\{q\}_0)\frac{dz} {2\pi i}
\tilde Y_{\sigma,N_r}^{(1)}(z)
\label{hatm}
\end{equation}
and $C_b^{(r)}$-contour is defined as in (\ref{gfgsig}).
At  $(\mu_s,\nu_s)$ being non-zeroth,
the genus-1  $S_\sigma(t,t';\{q\})$ function  is equal to
$S_{\sigma,s}^{(1)}(t_1,t_2;\{q_s\})$ given by  \cite{danphr}
\begin{eqnarray}
S_{\sigma,s}^{(1)}(t_1,t_2;\{q_s\})=Q_{\tilde\Gamma_s}^{-2}(t_1^{(s)})\biggl
[G_b(z_1^{(s)},z_2^{(s)};\{q_s\}_0)
\vartheta_2^{(s)}+\vartheta_1^{(s)}G_\sigma(z_1^{(s)},z_2^{(s)};\{q_s\}_0)
\nonumber\\
+
\varepsilon_s'\Sigma_\sigma(z_2^{(s)};\{q_s\}_0)\biggl]
Q_{\tilde\Gamma_s}^3(t_2^{(s)})
\end{eqnarray}
where $(z^{(s)}|\vartheta^{(s)})$  is defined by (\ref{tgam}),
the $Q_{\tilde\Gamma_s}$ factor is defined by (\ref{supder})
and
\begin{equation}
\Sigma_\sigma(z_2;\{q_s\}_0)=z_1
G_\sigma(z_1,z_2;\{q_s\}_0)|_{z_1\to\infty}\,.
\label{g1gh}
\end{equation}
Apart from
the last term, $S_\sigma(t_1,t_2;\{q_s\})$  is obtained by
the $\tilde\Gamma_s$ transformation (\ref{tgam}) of (\ref{gsig})
in the genus-1 case.
The last term is added for $S_\sigma(t_1,t_2;\{q_s\})$ to be decreased
at $z_1\rightarrow\infty$ or
$z_2\rightarrow\infty$.

To calculate
$S_\sigma(t,t';\{q\})$ on the higher-genus supermanifolds
one starts with the representation of
$S_\sigma(t,t';\{q\})$ by the integral  like (\ref{ointk}) as follows
\begin{equation}
S_\sigma(t,t';\{q\})=
-\oint S_\sigma(t,t_1;\{q\})_0)\frac{dz_1\,d\vartheta_1}{2\pi i}
S_\sigma(t_1,t';\{q\})\,,
\label{oints}
\end{equation}
the integration being performed in the positive direction
along the infinitesimal contour surrounding  $z_1$-point. In addition,
$S_\sigma(t,t';\{q\})$ is represented   like (\ref{klbf})
as follows
\begin{equation}
S_\sigma(t,t';\{q\})=S_\sigma^{(b)}(z,t';\{q\})+\vartheta
S_\sigma^{(f)}(z,t';\{q\})\,,
\label{defs}
\end{equation}
Like (\ref{ointk}), the contour in ({\ref{oints}) is reduced to the sum of the $C_s$ contours. The
integral over every $C_s$ contour is reduced to the integral along the $C_{v_s}$
circle and  (in the Ramond case) along the side of the cut as follows
\begin{eqnarray}
S_\sigma^{(b)}(z,t';\{q\})=S_\sigma^{(b)}(z,t';\{q\}_0)\vartheta'+
\sum_s \int_{C_{v_s}}S_\sigma^{(b)}(z,t_1;\{q\}_0)
\frac{dz_1}{2\pi i}\,\widetilde\Delta_{b\sigma}^{(s)}(z,t';\{q\})
\nonumber\\
+\int_{z_{v_s}}^{z_{u_s}}S_\sigma^{(b)}(z,t';\{q\}_0)
\frac{dz_1}{2\pi i}\,\widetilde\Delta_{b\sigma}^{(s-)}(z,t';\{q\})
\nonumber\\
S_\sigma^{(f)}(z,t';\{q\})=S_\sigma^{(f)}(z,t';\{q\}_0)-
\sum_s \int_{C_{v_s}}S_\sigma^{(f)}(z,t';\{q\}_0)\frac{dz_1}{2\pi i}\,
\widetilde\Delta_{f\sigma}^{(s)}(z,t';\{q\})
\nonumber\\
-\int_{z_{v_s}}^{z_{u_s}}S_\sigma^{(f)}(z,t';\{q\}_0)\frac{dz_1}{2\pi i}\,
\widetilde\Delta_{f\sigma}^{(s)}(z,t';\{q\})
\label{oins}
\end{eqnarray}
where $z_{v_s}$ lays on the $C_{v_s}$ circle and $z_{u_s}=
g_s(z_{v_s})$. Furthermore,
\begin{eqnarray}
g'(z)\widetilde\Delta_{b\sigma}^{(s)}(z,t';\{q\})=S_\sigma^{(b)}(g_s(z),t';\{q\})-g'(z)S_\sigma^{(b)}(z,t';\{q\})\,,
\nonumber\\
-(-1)^{2l_{2s}}\sqrt{g'(z)}\widetilde\Delta_{f\sigma}^{(s)}(z,t';\{q\})=S_\sigma^{(f)}(g_s(z),t';\{q\})+
(-1)^{2l_{2s}}\sqrt{g'(z)}S_\sigma^{(f)}(z,t';\{q\})\,,
\nonumber\\
\widetilde\Delta_{b\sigma}^{(s-)}(z,t';\{q\})=S_\sigma^{(b,s-)}(z,t';\{q\})-S_\sigma^{(b)}(z,t';\{q\})\,,
\nonumber\\
(-1)^{2l_{1s}}\widetilde\Delta_{f\sigma}^{(s-)}(z,t';\{q\})=S_\sigma^{(f,s-)}(z,t';\{q\})-(-1)^{2l_{1s}}S_\sigma^{(f)}(z,t';\{q\})
\label{ghchan}
\end{eqnarray}
where $\widetilde\Delta_{b\sigma}^{(s-)}(z,t';\{q\})$ and $\widetilde\Delta_{f\sigma}^{(s-)}(z,t';\{q\})$ coreespond to
$\Gamma_{a,s}^{(0)}(l_{1s})$-transformation  (\ref{zgama}). Moreover,
\begin{eqnarray}
S_\sigma^{(b)}(g_s(z),t';\{q\})=g'(z)\acute
S_\sigma^{(b)}(z,t';\{q\})+
\Delta_{b\sigma}^{(s)}(z,t';\{q\})\,,
\nonumber\\
-(-1)^{2l_{2s}}\sqrt{g'(z)}S_\sigma^{(f)}(g_s(z),t';\{q\})
=g'(z)\acute S_\sigma^{(f)}(z,t';\{q\})+
\Delta_{f\sigma}^{(s)}(z,t';\{q\})
\label{sigsch}
\end{eqnarray}
where $\acute S_\sigma^{(b)}(z,t';\{q\})$ and
$\acute S_\sigma^{(f)}(z,t';\{q\})$ is defined, as follows
\begin{equation}
g'_s(z)[\acute S_\sigma^{(b)}(z,t';\{q\})+\vartheta
\acute S_\sigma^{(f)}(z,t';\{q\})]=
Q_{\Gamma_{b,s}}^{-2}(t)
\biggl(S_\sigma(t,t';\{q\})+\sum_{N_s}
\hat Y_{\sigma,N_s}^{(1)}(t)\Psi_{\sigma,N_s}(t')\biggl)\,.
\label{whats}
\end{equation}
Further,
\begin{eqnarray}
\Delta_{b\sigma}^{(s)}(z,t';\{q\})=\varepsilon_s(z)\epsilon_s(z)
\partial_z[g'_s(z)\acute S_\sigma^{(b)}(z,t';\{q\})]-\epsilon_s(z)g_s'(z)\acute S_\sigma^{(f)}(z,t';\{q\})
\nonumber\\
\Delta_{f\sigma}^{(s)}(z,t';\{q\})=
-\epsilon_s(z)\partial_z[g'_s(z)\acute S_\sigma^{(b)}(z,t';\{q\})] +
\nonumber\\
\varepsilon_s(z)\epsilon_s(z)
\partial_z[g'_s(z)\acute S_\sigma^{(f)}(z,t';\{q\})]
+[(-1)^{2l_{2s}}(c_sz+d_s)+1]\varepsilon\varepsilon'g'_s(z)\acute S_\sigma^{(f)}(z,t';\{q\})\,,
\label{strkbff}
\end{eqnarray}
cf.(\ref{strkbf}). In the Ramond case the terms due to $\Gamma_{a,s}^{(0)}(l_{1s})$-transformation
(\ref{zgama}) are present in (\ref{oins}) and in (\ref{ghchan}) being found to be
\begin{eqnarray}
S_\sigma^{(b,s-)}(z,t';\{q\})=g'(z)\acute
S_\sigma^{(b,s-)}(z,t';\{q\})+
\Delta_{b\sigma}^{(s-)}(z,t';\{q\})\,,
\nonumber\\
-S_\sigma^{(f,s-)}(z,t';\{q\})
=\acute S_\sigma^{(f,s-)}(z,t';\{q\})+
\Delta_{f\sigma}^{(s-)}(z,t';\{q\})
\label{sigsh}
\end{eqnarray}
where $\acute S_\sigma^{(b,s-)}(z,t';\{q\})$ and
$\acute S_\sigma^{(f,s-)}(z,t';\{q\})$ is defined, as follows
\begin{equation}
\acute S_\sigma^{(b,s-)}(z,t';\{q\})+\vartheta
\acute S_\sigma^{(f,s-)}(z,t';\{q\})=
Q_{\Gamma_{a,s}}^{-2}(t)
S_\sigma(t,t';\{q\})\,.
\label{whatsc}
\end{equation}
where $Q_{\Gamma_{a,s}}(t)$ is assigned to $\Gamma_{a,s}^{(0)}(l_{1s})$-transformation  (\ref{zgama}), and
\begin{eqnarray}
\Delta_{fL}^{(-s)}(z,t';\{q\})=2\varepsilon_s(z)\partial_z
\acute S_\sigma^{(b,s-)}(z,t';\{q\}) +2\varepsilon_s\varepsilon_s'\acute S_\sigma^{(f,s-)}(z,t';\{q\})\,,
\nonumber\\
\Delta_{bL}^{(s-)}(z,t';\{q\})=2\varepsilon_s(z)\acute S_\sigma^{(f,s-)}(z,t';\{q\})\,.
\label{strkbfaa}
\end{eqnarray}
Eq. (\ref{ssiggfy}) at zeroth $(\mu_s,\nu_s)$-parameters and above-listed relations being employed,
it can be verified that eqs. (\ref{oins}) are equivalent to the equations
\begin{eqnarray}
S_\sigma^{(b)}(z,t_1;\{q\})=G_b(z,z';\{q\}_0)\vartheta'
+\sum_j\int_{C_j}\frac{dz'}{2\pi i}
G_b(z,z';\{q\}_0)\varepsilon_j(z')S_\sigma^{(f)}(z',t_1;\{q\})\,,
\nonumber\\
S_\sigma^{(f)}(z,t_1;\{q\})=S_\sigma^{(f)}(z,z';\{q\}_0)
- \sum_j\int_{C_j}\frac{dz'}{2\pi i}
S_\sigma^{(f)}(z,z';\{q\}_0)
\nonumber\\
\times
[-2\varepsilon_j'+\varepsilon_j(z')\partial_{z'}]
S_\sigma^{(b)}(z',t_1;\{q\})\,.
\label{seq}
\end{eqnarray}
To obtain 3/2-supertensors in (\ref{ssigtr})) the
the $z\to g_m(z)$
transformation of both sides of (\ref{seq}) is performed.
In this case $z$-point must be moved  inside the $C_m$ contour.
Due to the pole at $z=z_1$, the integrated term in (\ref{eqj})
arises, cf. the deriving of (\ref{ome}). The desired 3/2-supertensors are found
as follows
\begin{eqnarray}
(1+\varepsilon_s\varepsilon_s')\Psi_{\sigma;n_s}(t;;\{q\})=\chi_{n_s}(z;\{q\}_0)
\vartheta+\sum_j\int_{C_j}\frac{dz'}{2\pi i}
\chi_{n_s}(z';\{q\}_0)\varepsilon_j(z')S_\sigma^{(f)}(z',t;\{q\})\,,
\nonumber\\
(1-\frac{3}{2}\varepsilon_s\varepsilon_s')\Psi_{\sigma;\alpha_s}(t;\{q\})=
\biggl[\Psi_{\sigma;\alpha_s}(z;\{q\}_0)-
\sum_j\int_{C_j}\frac{dz'}{2\pi i}
\Psi_{\sigma;\alpha_s}(z';\{q\}_0)
\nonumber\\
\times
[-2\varepsilon_j'+\varepsilon_j(z')\partial_{z'}]
S_\sigma^{(b)}(z',t;\{q\})\biggl]
\label{zseq}
\end{eqnarray}
where $n_s=(k_s,u_s,v_s)$, $\alpha_s=(\mu_s,\nu_s)$ and
$\chi_{\alpha_n}(z';\{q\}_0)$ is defined by (\ref{gcor})  taken
at zero Grassmann parameters.

\section{Integration measure}

The integration measures
(partition functions) $Z_{L,L'}(\{q,\overline q\})$ in (\ref{ampl}) are
obtained from equations \cite{danphr,dan90} which provide the independence
of the superstring amplitudes from local variations of
the vierbein and of the gravitino field.  The equations relate
derivatives of $Z_{L,L'}(\{q,\overline q\})$
via moduli  with the superfield
vacuum correlators.
Due to a separation in right and left movers,
the  integration measure in (\ref{ampl}) is
represented as
\begin{equation}
Z_{L,L'}(\{q,\overline
q\})=(4\pi)^{5n}[\det\Omega_{L,L'}(\{q,\overline q
\})]^{-5} Z_L(\{q\}) \overline {Z_{L'}(\{q\})}
\label{hol}
\end{equation}
where $Z_L(\{q\})$ is a holomorphic function of
$q$ and $\Omega_{L,L'}(\{q,\overline q \})$
is given by (\ref{grom}).

The holomorphic measure  $Z_L(\{q\})$  is a product
\begin{equation}
Z_L(\{q)=Z_{m,L}(\{q\})Z_{gh,L}(\{q\})
\label{holme}
\end{equation}
where $Z_{m,L}(\{q\})$ is due to the matter superfields
and $Z_{gh,L}(\{q\})$ is due to  the ghosts ones.
The results of \cite{danphr} being employed,  equations for  $Z_{m,L}(\{q)$ and $Z_{gh,L}(\{q)$
can   be given as follows\footnote{Throughout the paper
the derivatives with respect to modular parameters are treated
as the derivatives on the right.}
\begin{eqnarray}
\partial_{q_s}\ln Z_{m,L}(\{q\})=-5\sum_{p=a,b}\int_{C_p^{(s)}}
\frac{dz\,d\vartheta}{2\pi i}
\partial_zK_L(t,t';\{q\})|_{t=t'}Y_{p,q_s}(t)\,,
\label{mat}\\
\partial_{q_s}\ln Z_{gh,L}(\{q\})=\sum_{p=a,b}\int_{C_p^{(s)}}
\frac{dz\,d\vartheta}{2\pi i}\biggl[
E^{(-2)}_{p,q_s}
(G(t,t;\{q\}))
\nonumber\\
-\sum_{N_s}\chi_{N_s}(t)\partial_{N_s}Y_{p,q_s}(t)
(-1)^{\eta(N_s)\eta(q_s)}\biggl]
\label{ghost}
\end{eqnarray}
where $C_p^{(s)}$-contour is defined just as in (\ref{gfsgsig}).
Furthermore, $G(t,t;\{q\})=G(t,t';\{q\})-(\vartheta-\vartheta')
(z-z')^{-1}$ at $t=t'$ and
$\partial_zK_L(t,t';\{q\})$ at
$t=t'$ is defined as
$\partial_z[K_L(t,t';\{q\})-(\vartheta-\vartheta')/(z-z')]$
at $t=t'$.  In the last term on the right side of (\ref{ghost}) the
summation is performed over the set $N_s=(k_s,u_s,v_s,\mu_s,\nu_s)$. In
addition, $\eta(N_s)=1$ for $N_s=(\mu_s,\nu_s)$, and  $\eta(N_s)=0$ for
$N_s=(k_s,u_s,v_s)$.  For any function $F(t)$, the $E_{p,N_s}^{(d)}(
F(t))$ expression is given \cite{danphr} by
\begin{eqnarray}
E_{p,N_s}^{(d)}(F(t))=\frac{d}{2}F(t)\partial_zY_{p,N_s}(t)+
\frac{\hat\epsilon(F)}{2}[D(t) F(t)]D(t)Y_{p,N_s}(t)
\nonumber\\
+
[\partial_zF(t)]Y_{p,N_s}(t)
\label{epf}
\end{eqnarray}
where $\hat\epsilon(F)=1$, if $F$ obeys the fermion statistics
and $\hat\epsilon(F)=-1$ in the opposite case.
By definition,
\begin{equation}
E_{p,N_s}^{(d)}( F(t,t))=E_{p,N_s}^{(d)}( F(t,t'))|_{t'=t}\,.
\label{ett}
\end{equation}
Further, if
\begin{equation}
F_d(t)\biggl|_{t\to\Gamma_{p,s}(t)}=
Q_{\Gamma_{p,s}}^d(t)[F_d(t)+\Delta_F(t)]\,,
\label{fchtg}
\end{equation}
$\Delta_F(t)$ being a certain function,
then $\partial_{N_s}F_d(t)$ is changed  as
\begin{equation}
\partial_{N_s}F_d(t)\biggl|_{t\to\Gamma_{p,s}(t)}
=Q_{\Gamma_{p,s}}^d(t)\biggl[\partial_{N_s}\biggl(F_d(t)+\Delta_F(t)\biggl)
-E_{p,N_s}^{(d)}(F_d(t))-E_{p,N_s}^{(d)}(\Delta_F(t))\biggl]\,.
\label{pardt}
\end{equation}
Eq. (\ref{pardt}) is an extension of the corresponding relation
given in \cite{danphr} for the case when
$\Delta_F(t)\equiv0$.
Due to (\ref{gfsgs}), the $G^{(r)}(t,t;\{q\})$ function
in (\ref{ghost}) is expressed through
$S_\sigma^{(r)}(t,t';\{q\})$. Then the first term on the right side
of (\ref{ghost}) appears to be
\begin{eqnarray}
E^{(-2)}_{p,q_s}(G(t,t;\{q\}))=E^{(-2)}_{p,q_s}(S_\sigma
(t,t;\{q\}))
+\sum_{r=1}^n\sum_{N_r}
\int_{C_r} E^{(-2)}_{p,q_s}\biggl(S_\sigma(t,t_1;\{q\})
\frac{dz_1d\vartheta_1}{2\pi i}
\nonumber\\
\times
Y_{N_r}(t_1)\chi_{N_r}(t;\{q\})\biggl).
\label{sgtwo}
\end{eqnarray}
Eq.(\ref{sgtwo}) being substituted in (\ref{ghost}), the order of
integration over $t$ and $t_1$ being interchanged, the first term
in (\ref{ghost}) is found to be as follows
\begin{eqnarray}
\sum_{p=a,b}\int_{C_p^{(s)}}
\frac{dz\,d\vartheta}{2\pi i}E^{(-2)}_{p,q_s}(S_\sigma(t,t;\{q\}))
+\sum_{r=1}^n\sum_{N_r}\sum_{p=a,b}\int_{C_p^{(s)}}
\frac{dz\,d\vartheta}{2\pi i}\chi_{N_r}(t;\{q\})E^{(-2)}_{p,q_s}(Y_{N_r}(t))
\nonumber\\
-
\sum_{r=1}^n\sum_{N_r}
\int_{C_r}Y_{N_r}(t_1)\frac{dz_1d\vartheta_1}{2\pi i}
\sum_{p=a,b}\int_{C_p^{(s)}}
\frac{dz\,d\vartheta}{2\pi i}\chi_{N_r}(t;\{q\})
E^{(-2)}_{p,q_s}(S_\sigma(t,t_1;\{q\}))\,.
\label{ghcont}
\end{eqnarray}
The second term in (\ref{ghcont}) is due to the pole at  $z=z_1$
in $S_\sigma(t,t_1;\{q\})$.

To calculate the first term in (\ref{ghcont}) we  represent
$S_\sigma(t,t_1;\{q\})$  through  the genus-1 functions
$S_{\sigma,s}^{(1)}(t,t')$ where $S_{\sigma,s}^{(1)}(t,t')$ is
associated with the handle $s$. We denote $S_{\sigma,s}^{(1)}(t,t')$ as
$S_{\sigma,sr}^{(1)}$, if $t$ is inside $C_s$ contour and $t'$ is
inside the $C_r$ one.  Further, we introduce an integration matrix
operator $\hat S_\sigma^{(1)} =\{\hat S_{\sigma,sr}^{(1)}\}$.  For
$s\neq r$, the $\hat S_{\sigma,sr}^{(1)}$ operator performs the
integration over $t'$ along $C_r$-contour, the kernel being
$S_{\sigma,sr}^{(1)}(t,t')$.  And $\hat S_{\sigma,ss}^{(1)}=0$.  For
$z$ and  $z'$ both being inside $C_s$-contour,  $S_{\sigma}(t,t')\equiv
S_{\sigma,ss}$ is written down as follows \cite{danphr}
\begin{equation}
S_{\sigma,ss}=S_{\sigma,ss}^{(1)}+
\sum_{n}\hat S_{\sigma,sn}^{(1)}(I-\hat
S_\sigma^{(1)})_{ns}^{-1} S_{\sigma,ss}^{(1)}
\label{sgrfss}
\end{equation}
If $z'$
is inside $C_s$,
then
3/2-supertensor $\Psi_{\sigma,N_s}(t';\{q\})\equiv \Psi_{\sigma,N_s}$
in (\ref{ssigtr})
can be written down as follows \cite{danphr}
\begin{equation}
\Psi_{\sigma,N_s}=\Psi_{\sigma,N_s}^{(1)}+
\sum_{n\neq s}\hat\Psi_{\sigma,N_sn} ^{(1)}(I-\hat
S_\sigma^{(1)})_{ns}^{-1} S_{\sigma,ss}^{(1)}
\label{pssi}
\end{equation}
where  the $\hat\Psi_{\sigma,N_sn}^{(1)}$ operator,
its kernel being
$\Psi_{\sigma,N_s}^{(1)}(t_1)dt_1$,
performs the
integration over $t_1$ along $C_n$-contour, and
$\Psi_{\sigma,N_s}^{(1)}(t_1)$ is the $3/2$-supertensor of the genus-1.

Contribution to (\ref{ghcont}) of the second term on the right side of
(\ref{sgrfss}) contains an expression
\begin{equation}
\sum_{p=a,b}\int_{C_p^{(s)}}S_{\sigma,s}^{(1)}(t_s,t)dt
E_{p,q_s}^{(-2)}(S_{\sigma,s}^{(1)}(t,t_n))
\label{fsect}
\end{equation}
which is calculated through $\partial_{q_s}
S_{\sigma,s}^{(1)}(t_s,t_n)$ by the relation
\begin{eqnarray}
\partial_{q_s} S_\sigma(t,t';\{q\})=
\sum_{p=a,b}\int_{C_p^{(s)}}S_\sigma(t,t_1;\{q\})dt_1
E_{p,q_s}^{(-2)}(S_\sigma(t_1,t';\{q\}))
\nonumber\\
+\int_{C_b^{(s)}}S_\sigma(t,t_1;\{q\})dt_1
\sum_{M_s}\biggl[
E_{p,q_s}^{(-2)}(Y_{M_s}^{(1)}(t_1)\Psi_{\sigma;M_s}(t';\{q\}))
\nonumber\\
-\partial_{N_s}
\biggl(Y_{M_s}^{(1)}(t_1)\Psi_{\sigma;M_s}(t';\{q\})\biggl)\biggl]
\label{dssig}
\end{eqnarray}
Eq.(\ref{dssig}) follows from the identity
\begin{equation}
\partial_{N_s} S_\sigma(t,t';\{q\})=
\int S_\sigma(t,t_1;\{q\})\,dt_1\partial_{N_s}S_\sigma(t_1,t';\{q\})
\label{doin}
\end{equation}
where the integration is performed along a contour surrounding
$z$-point. Deforming the contour on the boundary of the fundamental
region, eqs.(\ref{ssigtr}) and (\ref{pardt}) being used,
one obtains eq. (\ref{dssig}). Employing eq. (\ref{dssig}) for
$S_{\sigma,s}^{(1)}(t,t')$, one obtains the
first term in (\ref{ghcont}) as follows
\begin{eqnarray}
\sum_{p=a,b}\int_{C_p^{(s)}}
\frac{dz\,d\vartheta}{2\pi i}
E_{p,q_s}^{(-2)}(S_\sigma(t,t;\{q\})=\sum_{p=a,b}\int_{C_p^{(s)}}
\frac{dz\,d\vartheta}{2\pi i}
E_{p,q_s}^{(-2)}(S_{\sigma,s}^{(1)}(t,t))
\nonumber\\
+\int_{C_b^{(s)}}\sum_{M_s}(-1)^{\eta(M_s)+1}\biggl(
\Psi_{\sigma;M_s}(t;\{q\})-\Psi_{\sigma;M_s}^{(1)}(t)\biggl)dt
\nonumber\\
\times
\biggl(\partial_{q_s} Y_{M_s}^{(1)}(t)-E^{(-2)}_{p,q_s}
(Y_{M_s}^{(1)}(t))\biggl)
+\partial_{q_s} trace\ln(1-\hat S_\sigma^{(1)})\,.
\label{scont}
\end{eqnarray}
To simplify the last term in (\ref{ghcont}),
one employs the relation
\begin{eqnarray}
\sum_{p=a,b}\int_{C_p^{(s)}}
\chi_{N_r}(t;\{q\})\frac{dz\,d\vartheta}{2\pi i}
E^{(-2)}_{p,q_s}(S_\sigma(t,t_1;\{q\}))+
\int_{C_b^{(s)}}\chi_{N_r}(t;\{q\})\frac{dz\,d\vartheta}{2\pi i}
\nonumber\\
\times
\sum_{N_s}E^{(-2)}_{b,q_s}(Y_{\sigma,N_s}^{(1)}(t)
\Psi_{\sigma,N_s}(t_1;\{q\}))
\nonumber\\
-\sum_n\sum_{N_n}\int_{C_b^{(n)}}\chi_{N_r}(t;\{q\})
\frac{dz\,d\vartheta}{2\pi i}
\partial_{q_s}\biggl(Y_{\sigma,N_n}^{(1)}(t)\Psi_{\sigma,N_n}(t_1;\{q\})
\biggl)=0
\label{ssigrel}
\end{eqnarray}
Eq.(\ref{ssigrel}) follows from the identity
\begin{equation}
0=
\int\chi_{N_r}(t;\{q\})\frac{dz\,d\vartheta}{2\pi i}\partial_{N_s}
S_\sigma(t,t_1;\{q\})
\label{sdoin}
\end{equation}
where the integration is performed along any closed contour laying in
the fundamental region. Deforming the contour on the boundary
of the fundamental region, eqs.(\ref{ssigtr}) and (\ref{pardt}) being
used, one obtains eq. (\ref{ssigrel}).
Then
two last terms in (\ref{ghcont}) can be represented as follows
\begin{eqnarray}
\sum_{p=a,b}\int_{C_p^{(s)}}
\sum_{N_s}(-1)^{\eta(N_s)+1}\chi_{N_s}(t;\{q\})dt
\biggl(\partial_{q_s}Y_{p,N_s}(t)-
E^{(-2)}_{p,q_s}(Y_{p,N_s}(t))\biggl)
\nonumber\\
+\partial_{q_s}
\ln sdet M(\{\sigma\};\{q\})
\nonumber\\
+\sum_{N_s}(-1)^{\eta(N_s)}
\int_{C_b^{(s)}}\Psi_{\sigma;N_s}(t;\{q\})dt\biggl(
\partial_{q_s}Y_{N_s}^{(1)}(t)-E^{(-2)}_{b,q_s}(Y_{N_s}^{(1)}(t))\biggl)
\label{addgh}
\end{eqnarray}
where $sdet M(\{\sigma\};\{q\})$ is the superdeterminant \cite{danphr}
of the $M(\{\sigma\};\{q\})$
matrix that is given by (\ref{mma}).
Eq. (\ref{scont}) being added, the result is as follows
\begin{eqnarray}
\sum_{p=a,b}\int_{C_p^{(s)}}
\frac{dz\,d\vartheta}{2\pi i}
E_{p,q_s}^{(-2)}(S_{\sigma,s}^{(1)}(t,t))
+\partial_{q_s} trace\ln(1-\hat S_\sigma^{(1)})
+\partial_{q_s}\ln sdet M(\{\sigma\};\{q\})
\nonumber\\
-
\int_{C_b^{(s)}}\sum_{M_s}(-1)^{\eta(M_s)+1}\Psi_{\sigma;M_s}^{(1)}(t)dt
\biggl(\partial_{q_s} Y_{M_s}^{(1)}(t)
-E^{(-2)}_{p,q_s}(Y_{M_s}^{(1)}(t))\biggl)
\nonumber\\
+
\sum_{p=a,b}\int_{C_p^{(s)}}
\sum_{N_s}(-1)^{\eta(N_s)+1}\chi_{N_s}(t;\{q\})dt
\biggl(\partial_{q_s}Y_{p,N_s}(t)-
E^{(-2)}_{p,q_s}(Y_{p,N_s}(t))\biggl)\,.
\label{whole}
\end{eqnarray}
Once eq.(\ref{whole}) is substituted in (\ref{ghost}),
the direct calculation shows that
the last term in (\ref{whole}) is canceled
by the last term in (\ref{ghost}). In so doing eq. (\ref{chiy}) from
Appendix A is used.
As the result, eq.(\ref{ghost}) appears to be
\begin{eqnarray}
\partial_{q_s}\ln Z_{gh,L}(\{q\})=\sum_{p=a,b}\int_{C_p^{(s)}}
\frac{dz\,d\vartheta}{2\pi i}
E_{p,q_s}^{(-2)}(S_{\sigma,s}^{(1)}(t,t))
+\partial_{q_s} trace\ln(1-\hat S_\sigma^{(1)})
\nonumber\\
+\partial_{q_s}\ln sdet M(\{\sigma\};\{q\})
\nonumber\\
+\int_{C_b^{(s)}}\sum_{M_s}(-1)^{\eta(M_s)}
\Psi_{\sigma;M_s}^{(1)}(t)dt
\biggl(\partial_{q_s} Y_{M_s}^{(1)}(t)-E^{(-2)}_{p,q_s}(Y_{M_s}^{(1)}(t))
\biggl)\,.
\label{ghostf}
\end{eqnarray}
The calculation of the right side of (\ref{mat}) is similar to the
calculation of the first term on the right side of (\ref{ghcont}).
In so doing $K_L(t,t_1;\{q\})$ is represented through
the genus-1 functions
$K_{L,s}^{(1)}(t,t')$ each being assigned to $s$-handle, see
\cite{danphr}.
The integration matrix  operator $\hat K_L^{(1)} =\{\hat
K_{L,sr}^{(1)}\}$ is defined as follows.  For $s\neq r$, the $\hat K_{L,sr}^{(1)}$
operator performs the integration over $t'$ along $C_r$-contour, the
kernel being $K_{L,sr}^{(1)}(t,t')dt'$.  And $\hat
K_{L,ss}^{(1)}=0$. If the handle has the odd spin structure,
then a scalar zero mode $f(t)$ and a spinor zero mode $\varphi=D(t)f(t)$
exist. Correspondingly to this, the diagonal matrix $\hat\varphi=
\{\hat\varphi_{sr}\}$
and the integration operator $\hat f=\{\hat f_{sr}\}$ are defined.
The matrix elements of this matrix are
$\hat\varphi_{ss}=\varphi_s(t)$ and $\hat\varphi_{sr}=0$
for $s\neq r$. The kernel of
the $\hat f_{sr}$ operator is unequal to zero only for $s=r$.
The $\hat f_{ss}$ operator performs the integration
over $t'$ along $C_b^{(s)}$-contour, its kernel being
$f_s(t')dt'$. It has been shown
\cite{danphr} that (\ref{mat}) is equivalent to the equation
\begin{equation}
\partial_{q_s}\ln Z_{m,L}(\{q\})=\sum_{p=a,b}\int_{C_p^{(s)}}
\frac{dz\,d\vartheta}{2\pi i}\partial_zK_{L,s}^{(1)}(t,t')|_{t=t'}
Y_{p,q_s}(t)-5\partial_{N_r}trace\ln(I-\hat K^{(1)}+\hat\varphi
\hat f)
\label{zmat}
\end{equation}
The right side of (\ref{ghostf}) does not depend on the
choice of the $\{\sigma\}$ set and, therefore, it can be symmetrize
with respect to the $\{\sigma\}\leftrightarrow\{-\sigma\}$ replacement
that simplifies the solution of eq. (\ref{ghostf}). In the calculation
of the right side in (\ref{ghostf}) and in (\ref{zmat})  Appendix A is employed.
The holomorphic partition
function  is represented as follows
\begin{equation}
Z_L(\{q\})=\widetilde Z_L(\{q\})\prod_s(u_s-v_s-\mu_s\nu_s)^{-1}
\label{zht}
\end{equation}
where $\widetilde Z_L(\{q\})$ is invariant under the
$SL(2)$ transformations because
\begin{equation}
du_sdv_sd\mu_s d\nu_s/(u_s-v_s-\mu_s\nu_s)
\label{sliv}
\end{equation}
is $SL(2)$ invariant.
Employing the results of  \cite{danphr}, one can conclude that
\begin{eqnarray}
\widetilde Z_L(\{q\})=[\det\grave M(\sigma,\{q\})\det
\grave M(-\sigma,\{q\})]^{-1/2}\Biggl[\prod_s
\frac{2^{10l_{1s}}k_s^{l_{1s}}[1-(-1)^{2l_{2s}-1}k_s]^{2l_{1s}}}{
k_s^{3/2}[1-(-1)^{2l_{2s}-1}\sqrt{k_s}]^{4l_{1s}}}
\nonumber\\
\times
(-1)^{2l_{1s}+2l_{2_s}-1}\prod_{m=1}\biggl[\frac{1-(-1)^{2l_{2s}-1}
k_s^mk_s^{(2l_{1s}-1)/2}}
{1-k_s^m}\biggl]^8\Biggl]
\nonumber\\
\times
\exp[\frac{1}{2}
trace\ln(1-\hat S_\sigma^{(1)})+\frac{1}{2}
trace\ln(1-\hat S_{-\sigma}^{(1)})-5
trace\ln(1-\hat K^{(1)}+\hat\varphi\hat f)]\,,
\label{almes}
\end{eqnarray}
entries of $\grave M(\sigma,\{q\})$-matrix being
$M^{(r)}_{\alpha_s\beta_p}(\sigma,\{q\})$ where $\alpha_s$ runs the
$(\mu_s,\nu_s)$ set and $\beta_p$ runs the $(\mu_p,\nu_p)$ set. So $\grave M(\sigma,\{q\})$ is submatrix of the
$M^{(r)}(\sigma,\{q\})$ matrix
(\ref{mmr}). Eqs.
(\ref{mat}) and (\ref{ghost}) calculate every $\widetilde Z_L(\{q\})$ apart from a
numerical factor which is established from the condition of the
cancelation of the singularities at $k_s\to0$ in $\widetilde
Z_L(\{q\})$ after the summing over spin structures.

If Grassmann $(\mu,\nu)$ parameters vanish, then
$\widetilde Z_L(\{q\})\equiv\widetilde Z_L(\{q\}_0)$
is alternatively calculated using explicit
vacuum correlators that allows to simplify expression (\ref{almes}).
Taking this in mind, we represent $\widetilde Z_L(\{q\})$ as
follows
\begin{equation}
\widetilde Z_L(\{q\})=\biggl(\prod_s(-1)^{2l_{1s}+2l_{2_s}-1}\biggl)
\widetilde Z_{m,L}(\{q\}_0)\widetilde Z_{gh,L}(\{q\}_0)
\Upsilon_L(\{q\})
\label{expfz}
\end{equation}
where $\ln\Upsilon_L(\{q\})$ is proportional to
Grassmann parameters,
$\widetilde Z_{m,L}(\{q\}_0)$ is due to the matter
fields and $\widetilde Z_{gh,L}(\{q\}_0)$ is due to the ghost ones.
Both they have been given in \cite{danphr}.
The $Z_{m,L}(\{q\}_0)$
function can also be represented as\footnote{In
this paper $\widetilde Z_{m,L}(\{q\}_0)$ and
$\widetilde Z_{gh,L}(\{q\}_0)$ are given in a more
convenient form than in
\cite{danphr}}
\begin{equation}
\widetilde Z_{m,L}(\{q\}_0)=\Theta^5[L](0|\omega^{(0)})\prod_{(k)}
\prod_{m =1}^{\infty}\frac{1}{(1-k^m)^{15}}
\label{zmate}
\end{equation}
where $\omega^{(0)}\equiv\omega(\{q\}_0)$ and
$\Theta[L](0|\omega^{(0)})$ is the theta-function whose characteristics
are $L=(l_1,l_2)$. The product in (\ref{zmate}) is calculated over
those Schottky group multipliers $k$ which
are not powers of the other ones.

In the calculation \cite{danphr} of $\widetilde Z_L(\{q\}_0)$ the
$G_f(z,z';\{q\}_0)$ fermion Green function in (\ref{gzgm}) is expressed
by (\ref{gsig}) through $G_\sigma(z,z';\{q\}_0)$.
The $\widetilde Z_{gh,L}(\{q\}_0)$ function can be represented as follows
\begin{eqnarray}
\widetilde Z_{gh,L}(\{q\}_0)=e^{-\pi i
\sum_{r,s}l_{1s}l_{1r}\sigma_s\sigma_r\omega^{(0)}_{rs}}
[\det\widetilde M_L(\sigma)
\det\widetilde M_L(-\sigma)]^{-1/2}
\nonumber\\
\times
\biggl(\prod_s\frac{(1-k_s)^2(-1)^{(2l_{2s}-1+2l_{1s})}}
{(1+(-1)^{2l_{2s}}\sqrt{k_s})^2k_s^{3/2}}\biggl)
\biggl(\prod_{(k)}
\prod_{m =2}^{\infty}{\cal P}_m(\sigma,k){\cal
P}_m(-\sigma,k)\biggl)^{-1}
\label{zzgh}
\end{eqnarray}
where the product is calculated over
those Schottky group multipliers $k$ which
are not powers of the other ones,
$\widetilde M_L(\sigma)$ is given by
(\ref{tilmmac}), and
\begin{equation}
{\cal P}_m(\sigma,k)=\frac{1-k^{m-1/2}e^{\Omega(\sigma,k)}}{1-k^m}\,.
\label{Ome}
\end{equation}
In this case $\Omega(\sigma,k)\equiv\Omega_{\Gamma_{(k)}}(\{\sigma_p\})$
where $\Omega_{\Gamma_{(k)}}(\{\sigma_p\})$
is given by (\ref{omgs}) for a Schottky
group product $\Gamma_{(k)}$ whose
multiplier is $k$.

Eq. (\ref{zmate})
could be   re-written down as
\begin{equation}
\widetilde Z^{(m)}_n(\sigma,q_b)=e^{5\pi i
\sum_{r,s}l_{1s}l_{1r}\sigma_s\sigma_r\omega^{(0)}_{rs}}\prod_{(k)}
\prod_{m =1}^{\infty}{\cal P}_m^5(\sigma,k){\cal P}_m^5(-\sigma,k)\,,
\label{zmatee}
\end{equation}
but the product over $k$ for $m=1$  may be
divergent when  the Ramond handles are presented in the spin structure
under consideration.

Further, $\det\widetilde M_L(\sigma)$ in (\ref{zzgh}) can be reduced
to the determinant of the matrix whose entries are obtained from
(\ref{tilmmac}) through a replacement $Y_{N_r}^{(0)}(z)$ by either $z$,
or the unity.
It is achieved through a replacement each of
$Y^{(0)}_{N_s}(z)$ in (\ref{tilmmac}) by relevant linear combination of
them.
Meanwhile one  employs identities
\begin{eqnarray}
\sum_{s=1}^n\int_{C_s}\Phi_{\sigma,\mu_p}(z;\{q\}_0)\frac{zdz}{2\pi
i}=\grave N_{\mu_p}[\hat t
+u_p]t_{\mu_p}\,,
\sum_{s=1}^n\int_{C_s}\Phi_{\sigma,\nu_p}(z;\{q\}_0)\frac{zdz}{2\pi
i}=\grave N_{\nu_p}[\hat t
+v_p]t_{\nu_p}\,,
\nonumber\\
\sum_{s=1}^n\int_{C_s}\Phi_{\sigma,\mu_p}(z;\{q\}_0)\frac{dz}{2\pi
i}=\grave N_{\mu_p}t_{\mu_p}\,,\quad
\sum_{s=1}^n\int_{C_s}\Phi_{\sigma,\nu_p}(z;\{q\}_0)\frac{dz}{2\pi
i}=\grave N_{\nu_p}t_{\nu_p}
\label{suiph}
\end{eqnarray}
where $\grave N_{\mu_p}$ and $\grave N_{\nu_p}$ are given by (\ref{nors}), and
\begin{equation}
\hat t=-z\sum_r2\pi il_{1r}\sigma_rJ_r(z)\biggl|_{z\to\infty}\,,\quad
t_{\mu_p}=1-\frac{e^{-\Omega_p(\{\sigma_s\})}}{\sqrt k_p}\,,
\quad t_{\nu_p}=1-\sqrt k_pe^{-\Omega_p(\{\sigma_s\})}\,.
\label{tht}
\end{equation}
In this case $\Omega_p(\{\sigma_s\}\equiv\Omega_\Gamma(\{\sigma_s\})$
at $\Gamma=g_p$ and $\Omega_\Gamma(\{\sigma_s\})$ is given by
(\ref{omgs}).  Identifies (\ref{suiph}) follow from the fact that each
of integrands  in (\ref{suiph}) has no singularities out of the
integration contour. Hence the integral is reduced to integral along
the infinity large circle. This integral is calculated through the
asymptotic at $z\to\infty$ of the 3/2-tensors (\ref{phi}).

It is convenient to extract $\grave N_{\mu_p}$ and $\grave N_{\nu_p}$
from 3/2-tensors (\ref{phi}) defining new 3/2-tensors
$\Phi_{\sigma,\mu_p}(z;\{q\}_0)$
and $\Phi_{\sigma,\mu_p}(z;\{q\}_0)$ by relations
\begin{equation}
\Phi_{\sigma,\mu_p}(z;\{q\}_0)=\grave N_{\mu_p}
\hat\Phi_{\sigma,\mu_p}(z;\{q\}_0)\,,\quad
\Phi_{\sigma,\nu_p}(z;\{q\}_0)=
\grave N_{\nu_p}\hat\Phi_{\sigma,\nu_p}(z;\{q\}_0)\,.
\label{phhphi}
\end{equation}
Employing eqs. (\ref{phhphi}) and choosing
relevant linear combinations of columns in
$\widetilde M_L(\sigma)$ one can
represent $\det\widetilde M_L(\sigma)$ in (\ref{zzgh}) as
\begin{equation}
\det\widetilde M_L(\sigma)=\prod_p\biggl(\frac{4\grave N_{\mu_p}
\grave N_{\nu_p}}{u_p-v_p}\biggl)\,\det\acute M(L;\sigma)
\label{phdet}
\end{equation}
where entries $\acute M_{r,s}(L;\sigma)$ of $\acute M(L;\sigma)$ are as
follows
\begin{eqnarray}
\acute M_{2r-1,2s-1}(L;\sigma)=\int_{C_s}
\hat\Phi_{\sigma,\mu_r}(z;\{q\}_0)z\frac{dz}{2\pi i}\,,\quad
\acute M_{2r,2s-1}(L;\sigma)=\int_{C_s}\hat\Phi_{\sigma,\nu_r}(z;\{q\}_0)z
\frac{dz}{2\pi i}\,,
\nonumber\\
\acute M_{2r-1,2s}(L;\sigma)=\int_{C_s}
\hat\Phi_{\sigma,\mu_r}(z;\{q\}_0)\frac{dz}{2\pi i}\,,\quad
\acute M_{2r,2s}(L;\sigma)=\int_{C_s}
\hat\Phi_{\sigma,\nu_r}(z;\{q\}_0)\frac{dz}{2\pi i}\,,
\nonumber\\
\acute M_{2r-1,2n}(L;\sigma)=t_{\mu_p}\,,\quad
\acute M_{2r,2n}(L;\sigma)=t_{\nu_p}\,,\quad
\acute M_{2r-1,2n-1}(L;\sigma)=u_pt_{\mu_p}\,,
\nonumber\\
\acute M_{2r,2n-1}(L;\sigma)=v_pt_{\nu_p}
\label{hmentr}
\end{eqnarray}
where $1\leq r \leq n$ and $1\leq s \leq( n-1)$.

The $\Upsilon_L(\{q\})$ factor in (\ref{expfz}) is
obtained from (\ref{almes}) as follows
\begin{eqnarray}
\Upsilon_L(\{q\})=det\frac{\grave M(\sigma;\{q\}_0)
\grave M(-\sigma;\{q\}_0)}{
\grave M(\sigma;\{q\})
\grave M(-\sigma;\{q\})}
\exp[\frac{1}{2}trace\ln(1-\widehat\Delta_\sigma)
\nonumber\\
+\frac{1}{2}trace\ln(1-\widehat\Delta_{-\sigma})
-5trace\ln(1-\widehat\Delta^{(m)})]
\label{munucor}
\end{eqnarray}
where
$\widehat\Delta_\sigma$ and $\widehat\Delta^{(m)}$ are integral
operators. They are formed by the $\{\widehat\Delta_{\sigma,p}\}$ set
and, respectively, by the $\widehat\Delta_p^{(m)}$ set
of integral operators. The kernel of the $\widehat\Delta_{\sigma,p}$
operator is $\Delta_{\sigma,p}(t,t')dt'$,
and the kernel of $\widehat\Delta_p^{(m)}$ is
$\Delta_p^{(m)}(t,t')dt'$.  We again define the
kernel together with the differential $dt'$. These operators perform
integrating over $t'$ along the $C_p$ contour. In this case
\begin{eqnarray}
\Delta_p^{(m)}(t,t')=\sum_{r\neq p}\int_{C_r}K_L(t,t_1;\{q\}_0)
dt_1\delta K_{L,p}^{(1)}(t_1,t')
-\int_{C_p}K_L(t;\{q\}_0)dt_1\delta[\varphi_p(t_1)f_p(t')]\,,
\label{kernm}\\
\Delta_{\sigma,p}(t,t')=\sum_{r\neq p}\int_{C_r}S_\sigma(t;\{q\}_0)
dt_1\delta S_{\sigma,p}^{(1)}(t_1,t')
\label{kerngh}
\end{eqnarray}
where $\delta S_{\sigma,p}^{(1)}(t_1,t')$, $\delta
K_{L,p}^{(1)}(t_1,t')$ and $\delta[\varphi_p(t_1)f_p(t')]$ is the
proportional to $(\mu_p,\nu_p)$ part of the corresponding function
$S_{\sigma,p}^{(1)}(t_1,t')$,
$K_{L,p}^{(1)}(t_1,t')$ or $\varphi_p(t_1)f_p(t')$. The last term on
the right side of (\ref{kernm}) appears, if $l_{1p}=l_{2p}=1/2$.
From (\ref{kernm}) and (\ref{kerngh}) the  exponent in (\ref{munucor}) contains genus-1 functions.
Apart from notations, eq.(\ref{munucor})
has been obtained in \cite{danphr}. Now we transform  (\ref{munucor}) to represent the
the exponent in (\ref{munucor})  through
genus-$n$ functions only.

As an example, we consider
$trace\ln(1-\widehat\Delta^{(m)})$. For the sake of simplicity,
we assume no to be the handles with $l_1=l_2=1/2$.
In this case
\begin{eqnarray}
\Delta_p^{(m)}(t,t')=\sum_{r\neq p}\int_{C_r}\frac{dz_1}{2\pi i}
\Biggl[-\vartheta R_{f,L}(z,z_1;\{q\}_0)\varepsilon_r(z_1)
\biggl(\partial_{z_1}\partial_{z'}R^{(1)}_{b,p}(z_1,z')\biggl)
[\vartheta'-\varepsilon_r(z')]
\nonumber\\
+\biggl(\partial_{z_1}R_b(z,z_1;\{q\}_0)\biggl)\varepsilon_r(z_1)
\biggl[R_{f,r}^{(1)}(z_1,z')+
\partial_{z'}\biggl(R_{f,p}^{(1)}(z_1,z')
\varepsilon_r(z')\biggl)\vartheta'\biggl]\Biggl]\,.
\label{expdm}
\end{eqnarray}
Therefore,
\begin{eqnarray}
trace\ln(1-\widehat\Delta^{(m)})=\sum_r\int_{C_r}\Biggl[
\sum_{p\neq r}\int_{C_p}\biggl[
\biggl(\partial_{z_1}R_b(z_3,z_1;\{q\}_0)\biggl)\varepsilon_r(z_1)
\partial_{z_3}\biggl(R_{f,r}^{(1)}(z_1,z_3)
\varepsilon_r(z_3)\biggl)
\nonumber\\
+R_{f,L}(z_3,z_1;\{q\}_0)\varepsilon_r(z_1)
\biggl(\partial_{z_1}\partial_zR^{(1)}_{b,r}(z_1,z_3)\biggl)
\varepsilon_r(z_3)\biggl]\frac{dz_3}{2\pi i}\Biggl]\frac{dz_1}{2\pi i}
\nonumber\\
-\sum_r\int_{C_r}\frac{dz_1}{2\pi
i}\sum_{{r'}}\int_{C_{r'}}\frac{dz_2}{2\pi i}
\int_{C[r']}
\frac{dz}{2\pi i}\int_{C[r]}\frac{dz'}{2\pi i}
\biggl(\partial_{z_1}R_b(z,z_1;\{q\}_0)\biggl)\varepsilon_r(z_1)
R_{f,r}^{(1)}(z_1,z')
\nonumber\\
\times
R_{f,L}(z',z_2;\{q\}_0)\varepsilon_{r'}(z_2)
\biggl(\partial_{z_2}\partial_{z}R^{(1)}_{b,r'}(z_2,z)\biggl)+\dots
\label{splnm}
\end{eqnarray}
where the dots denote the terms of the higher order in the Grassmann
parameters and the $C[r]$ contour (the $C[r']$ contour) is the sum over
$p\neq r$ (the sum over $p\neq r'$) of the $C_p$ contours.
In this case the points $z'=z_1$ and $z'=z_2$
are situated inside the $C[r]$ contour. Correspondingly,
$z=z_1$ and $z=z_2$ lay inside  $C[r']$.
Furthermore,
if $z_1$ and $z_2$ both lay outside  the $C_r$ contour, then
\begin{equation}
\int_{C_r}R_{f,r}^{(1)}(z_1,z')
R_{f,L}(z',z_2;\{q\}_0)dz'=0\,,\quad
\int_{C_r}\partial_{z_1}R_b(z,z_1;\{q\}_0)
\partial_{z}R^{(1)}_{b,r}(z_2,z)dz=0\,.
\label{mino}
\end{equation}
So the integral over $z'$ in  (\ref{splnm}) is not changed,
if the integration contour $C[r]$ is added by the $C_{r}$ contour
drawing in such a way that $z'=z_1$ and $z'=z_2$ are situated outside
$C_{r}$. Hence the result of the integration with respect to $z'$ is
given by  residues at those poles of the integrand that lay outside the
$C[r]+C_r$  contour.  Accordingly, the result of the integration over
$z$  is given by the sum over residues at the  poles laying outside the
$C[r']+C_{r'}$  contour.

In the $r=r'$ case the poles at $z'=z_1$ and at $z'=z_2$ both lay
outside the $C[r']+C_{r'}$ contour. In addition, the poles at $z=z_1$
and at $z=z_2$ lay outside the $C[r']+C_{r'}$ contour. The integration
over $z'$ and over $z$ being performed, the result is the sum of the
terms corresponding to  pairs of the poles at $z'=z_j$ and $z=z_l$
where $j$ and $l$ indices  each  run numbers 1 and 2. In this case the
terms due to ($z'=z_1$, $z=z_1$) and ($z'=z_2$, $z=z_2$)are canceled by
the first term on the right side of (\ref{splnm}).  The term due to
($z'=z_2$, $z=z_1$) disappears after  integrations over $z_1$ and over
$z_2$ to be performed. Indeed, in this case all the singularities in
($z_1,z_2$) of the integrand are to be inside the integration contours
$C_r$ and $C_{r'}$. Hence only the term due to ($z'=z_1$, $z=z_2$)
contributes to the right side of (\ref{splnm}) and only this term
contributes to (\ref{splnm}), if $r\neq r'$. As the result,
the right side of
(\ref{splnm}) becomes as follows
\begin{equation}
\sum_s\int_{C_s}\frac{dz}{2\pi i}
\sum_r\int_{C_r}\frac{dz_1}{2\pi i}
\partial_z\partial_{z_1} R_b(z,z_1;\{q\}_0)\varepsilon_r(z_1)
R_f(z_1,z;\{q\}_0)\varepsilon_s(z)+\dots
\label{rights}
\end{equation}
where the dots code the terms of a higher order in Grassmann parameters.
These terms can be rearranged in the same manner as the two-order terms
above. The resulted expression is found to be
\begin{equation}
trace\ln(1-\widehat\Delta^{(m)})=
trace\ln(1+\widehat K)
\label{mup}
\end{equation}
where $\widehat K$ is formed by the set of the $\widehat K_s$
operator. The $\widehat K_s$ operator performs the
integration over $z'$ along the $C_s$ contour. Its kernel  is $\widehat
K_s(z,z')dz'/(2\pi i)$ where
\begin{equation}
\widehat K_s(z,z')= \sum_r\int_{C_r}\frac{dz_1}{2\pi i}
\partial_z\partial_{z_1} R_b(z,z_1;\{q\}_0)\varepsilon_r(z_1)
R_f(z_1,z';\{q\}_0)\varepsilon_s(z')\,.
\label{mups}
\end{equation}
The expression
$trace\ln(1-\widehat\Delta_\sigma)$
in (\ref{munucor}) is rearranged in very kindred manner,
the result being as follows
\begin{equation}
trace\ln(1-\widehat\Delta_\sigma)=
trace\ln(1+\widehat S_\sigma)
\label{mghup}
\end{equation}
where $\widehat S_\sigma$ is formed by the set of the $\widehat S_{\sigma,s}$
operators. The $\widehat S_{\sigma,s}$ operator performs the
integration over $z'$ along the $C_s$ contour, its  kernel  is $\widehat
S_{\sigma,s}(z,z')dz'/(2\pi i)$ where
\begin{equation}
\widehat S_{\sigma,s}(z,z')= \sum_r\int_{C_r}\frac{dz_1}{2\pi i}
G_b(z,z_1;\{q\}_0)\varepsilon_r(z_1)
S_\sigma(z_1,z';\{q\}_0)[-2\varepsilon_s'+\varepsilon_s(z')\partial_{z'}]\,.
\label{ghups}
\end{equation}
The determinant of   $\grave M(\sigma;\{q\})/\grave M(\sigma;\{q\}_0)$
matrix in (\ref{munucor})  is calculated using
(\ref{mmrr}), (\ref{zseq}) and (\ref{tilzer}) to be as follows
\begin{equation}
\det\frac{\grave M(\sigma;\{q\})}
{\grave M(\sigma;\{q\}_0)}=\det[1+{\cal M}]
\label{sde}
\end{equation}
and the ${\cal M}_{nm}$ entry of the ${\cal M}$ matrix is given by
\begin{equation}
{\cal M}_{nm}=
\sum_j\int_{C_j}\frac{dz'}{2\pi i}
\chi_{\alpha_n}(z';\{q\}_0)
[-2\varepsilon_j'+\varepsilon_j(z')\partial_{z'}]
\int_{C^{(m)}} S_\sigma^{(b)}(z',t_1;\{q\})dt_1
Y_{\beta_m}^{(r)}(t_1)
\label{calm}
\end{equation}
where $\alpha_n$ runs the $(\mu_n,\nu_n)$ set and
$\beta_m$ runs the set $(\mu_m,\nu_m)$. To derive  (\ref{sde}), eq.
(\ref{psichi}) has been employed,
$\chi_{\alpha_n}(z';\{q\}_0)$ being defined by (\ref{gcor})  taken
at zero Grassmann parameters. If $j=m$, then the integration contour
over $z'$ is situated inside the integration contour over $z_1$.

In configurations where Schottky circles overlap each other
the integration measures are obtained by the analytic continuation of
the foregoing formulae like as it was discussed in the end of Section 4.

\section{Uncertainties of the amplitude}

The calculation of the amplitude (\ref{ampl}) includes
Grassmann integrations as well as integrations over local variables, and the
integrand has singularities at that. Such
integrations are ambiguous with respect to a non-split replacement of
the integration variables. Indeed, in this case
the result of the integration is either finite or
divergent, depending on a choice of the integration variables as  it is
seen for an easy integral
\begin{equation}
I_{(ex)}=\int\frac{dxdyd\alpha
d\beta d\bar\alpha d\bar\beta}
{|z-\alpha\beta|^p}\theta(1-|z|^2)
\label{examp}
\end{equation}
where $z=x+iy$, $(\alpha,\beta)$ are Grassmann
variables and
$p$ characterizes the
strength of the singularity. For the sake of simplicity we bound
the integration region by $|z|^2\leq1$. Eq. (\ref{examp})
is re-written as follows
\begin{eqnarray}
I_{(ex)}=\int\frac{dxdy}
{|z|^p}\theta(1-|z|^2)\,d\alpha\,
d\beta\,d\bar\alpha\,d\bar\beta+
\nonumber\\
+p^2\int\frac{dxdy} {4|z|^{p+2}}\theta(1-|z|^2)
\alpha
\beta\bar\alpha\bar\beta\,
d\alpha\,
d\beta\,d\bar\alpha\,d\bar\beta
\,.
\label{exam}
\end{eqnarray}
The first integral on the right side is equal to
zero since its integrand does not contain Grassmann
variables.  The second integral is divergent at $z=0$, if
$Re\,p>0$. On the other side, in (\ref{examp}) one can introduce
the $\tilde z=z-\alpha\beta$ variable instead of $z$. Then the Grassmann
variables will be present only in the step function $\theta(|\tilde
z+\alpha\beta|^2)$. Grassmann integrations being
performed, the integral is reduced to the integral along the
circle $|z|^2=1$. Thus for any $p$ the result is
finite as follows
\begin{equation}
I_{(ex)}=-\int\frac{d\tilde
xd\tilde yd\alpha d\beta d\bar\alpha d\bar\beta} {|\tilde
z|^p}\alpha\beta\bar\alpha\bar\beta \left[\delta(|\tilde
z|^2-1)+|\tilde z|^2\frac{d\delta(|\tilde z|^2-1)}{d|\tilde
z|^2}\right]
=-\frac{\pi p}{2}\,.
\label{ex}
\end{equation}
So (\ref{examp}) depends on the integration
variables. This ambiguity  arises because the
integrand is expanded in a series over the Grassmann variables
in the  singular point $z=0$.
If the integral is convergent, then it does not depend on the choice of
the integration variables provided that a transition to news
integration variables remains it to be convergent.  As an example,
the integral (\ref{examp}) is equal to (\ref{ex}) at $Re p<0$   as employing
$z$ to be the integration variable, so when the integration variable is
$\tilde z$.  One could, however, replace the integration variable $z$
by $z+\sum_{i=1}^N\delta_i\delta_i^{(1)}$ where $\delta_i$ and
$\delta_i^{(1)}$ are arbitrary Grassmann numbers. When $p$ is
not an negative even number, then the resulted integrand has the
singularity $\sim|z|^{-(p+2+2N)}$, and  the integral is divergent,
if $p+2N>0$.

The ambiguity in the superstring calculations is resolved by the
requirement to preserve local symmetries of the superstring. Then
the
amplitude must be finite. Otherwise the conformal symmetry is broken
because a cutoff parameter appears. In this case the amplitude becomes
to depend on $\{N_0\}$ set (\ref{fixm})  that falls the theory.
Besides,  the 10-dim. space-time
supersymmetry requires that the vacuum amplitude is equal to zero along
with 1-, 2- and 3- point massless boson amplitudes. The proposed
amplitude satisfies these conditions.

Due to the singularity (\ref{lim}) at $z=z'$ in the Green
function, the vacuum expectation of the interaction vertices in
(\ref{ampl}) is singular in nodal regions where some number $m_1>1$ of
the interaction vertices go to the same point $z_0$. In addition, a
certain number of the Schottky group limiting points may be also moved
to $z_0$ that corresponds to a degeneration of the Riemann surface.  If
$1<m_1<(m-1)$ and $m>3$, then the strength of the singularity depends
on the 10-energy invariant corresponding to the given reaction channel.
In this case the integral is calculated \cite{gsw} for those energies
(laying below the reaction threshold), where it is convergent.  The
result is analytically continued to energies that are above the
threshold.  In this case the amplitude receives singularities required
by the unitarity equations.  Due to the energy-momentum conservation,
there is no a domain  of the 10-energy invariants where the integral
over all the nodal regions is convergent.  Hence the amplitude is
obtained by the summing of the pieces obtained by the analytical
continuation in the distinct  the 10-invariants. For instance, the
calculation of the scattering amplitude includes an analytical
continuation in $s$, $t$ and $u$-Mandelstam's invariants of the
integral over a relevant nodal region.  Every  region  contains a pair
of the vertices going to each other.  Each  integral gives rise to the
cut in $s$, $t$ or $u$. The analytical continuation  under
consideration is evidently consistent with the local symmetries of the
amplitude. Therefore, only configurations where $m_1=0$,
$m_1=1$, $m_1=(m-1)$ and $m_1=m$  might lead to divergences in the
amplitude.

From (\ref{almes}), the integration measure has a singularity at $k_s\to0$.
As it has been discussed just below eq.(\ref{ddig}), the $k_s\to1$
multipliers are not contained in the integration region. Hence the
singularity at $k_s\to0$ is an only singularity in $k_s$ of the
amplitude (\ref{ampl}).  The leading singularity is
$1/(|k_s|^3\ln^5|k_s|)$ when $l_{1s}=0$, but it is canceled in the sum
over $l_{2s}=1$ and $l_{2s}=0$ spin structures. The resulted
singularity is $1/(|k_s|^2\ln^5|k_s|)$ that is the integrable
singularity. The same singularity arises at $l_{1s}=1/2$.  So, the
singularity at $k_s\to0$ becomes to be integrable due to the summation
over $l_{2s}=1$ and $l_{2s}=0$ spin structures.

The
holomorphic partition function  (\ref{zht}) has the
$(u_s-v_s-\mu_s\nu_s)^{-1}$ singularity at $u_s\to v_s$.
From
(\ref{zmate}) and from (\ref{zzgh}), the leading at  $u_s\to v_s$ term
in the integration measure (\ref{hol}) at zeroth
$\{\mu_p,\nu_p\}$-parameters is  proportional to the genus-1
integration measure $Z_{L_s,L'_s} (k_s,\bar k_s)$ where
\begin{eqnarray}
Z_{L_s,L'_s} (k_s,\bar k_s)=\biggl(-\frac{2\pi}{\ln|k_s|}\biggl)^5
(-1)^{2l_{1s}+2l_{1s}+2l'_{1s}+2l'_{1s}}\Theta^4[l_{1s},l_{2s}]
\biggl(0\biggl|\frac{\ln k}{2\pi i}\biggl)
\overline{\Theta^4[l'_{1s},l'_{2s}]\biggl(0\biggl|
\frac{\ln k}{2\pi i}\biggl)}
\nonumber\\
\times
\biggl(\prod_m^\infty|1-k_s^m|\biggl)^{-24}.
\label{g1mes}
\end{eqnarray}
If no interaction vertices are nearby $v_s\to u_s$, then
the local amplitude $\breve A^{(n)}_m(\{q,\bar q\};\{t,\bar t\})$ (the
integrand) in (\ref{ampl}) is
\begin{eqnarray}
\breve
A^{(n)}_m(\{q,\bar q\};\{z,\bar z\})=\frac{1}{|u_s-v_s-\mu_s\nu_s|^2}
\sum_{L_s,L_s'}{\cal O}^{(1)}\biggl(\frac{\ln k_s}{2\pi i},
-\frac{\ln\overline{k_s}}{2\pi i}\biggl)Z_{L_s,L'_s} (k_s,\bar k_s)
\nonumber\\
\times
\biggl[
\breve A^{(n-1)}_m(\{q,\bar q\};\{t,\bar
t\}) +\dots\biggl]
\label{degs}
\end{eqnarray}
where $\{z,\bar z\}$ is the set of the interaction vertex coordinates,
$\breve A^{(n-1)}_m(\{q,\bar q\};\{t,\bar
t\})$ is the integrand for the $(n-1)$-loop, $m$-point  amplitude
and the step functions ${\cal O}^{(1)}\biggl(\frac{\ln k_s}{2\pi i},
-\frac{\ln\overline{k_s}}{2\pi i}\biggl)$
bound the  fundamental region of the genus-1 modular group
\cite{siegal}.  The dots crypt the terms which  depend on
$(u_s|\mu_s)$ and  on $(v_s|\nu_s)$ as well as on their complex
conjugative. Being summed  either over $L_s$ or over $L'_s$ genus-1
even spin structures, the $Z_{L_s,L'_s}(k_s,\bar k_s)$ function
(\ref{g1mes}) is nullified due to the Riemann identities. It is the
well known result \cite{gsw} for the  vacuum function of the torus.  If
either $L_s$ or $L'_s$ the genus-1 spin structure is odd, then
(\ref{g1mes}) is equal to zero by itself.  Therefore, the singularity
in (\ref{degs}) at $u_s\to v_s$ might appear only because of those
`dots" terms  in (\ref{degs}) which depend on both $L_s$ and $L'_s$
genus-1 spin structures.

Questioned terms appear due to the theta-constant in
(\ref{zmate}) and due to the theta-function in the fermion Green
function (\ref{fgrfr}). They  are of $\sim(u_s-v_s)^2$ and,
therefore, they do not origin the divergence in the amplitude. The
questioned terms  in (\ref{zzgh}) are of  $\sim(u_s-v_s)^4$,
if $l_{1s}=0$.  And they are of $\sim(u_s-v_s)^2$ for  the
$l_{1s}=1/2$, $l_{2s}=0$ spin structure. Linear in $(u_s-v_s)$ terms
are absent due to the symmetry with respect to $\sigma_s\to-\sigma_s$.
The  discussed terms cancel the $\mu_s\nu_s(u_s-v_s)^{-2}$ singularity
at $u_s\to v_s$ in (\ref{degs}) and, therefore, they do not origin the
divergence in the amplitude.  The calculation of  $\det\widetilde
M_L(\sigma)$ in (\ref{zzgh}) can be performed employing  eqs.
(\ref{phdet}) and (\ref{hmentr}).  The $C_s$ contour in (\ref{hmentr})
can be moved through a distance of $\rho>\rho_s>>|u_s-v_s|$ where
$\rho$ is the distance to the nearest singularity of the integrand.
Therefore, the asymptotic in $z$ of the integrand at
$|u_s-v_s|<<z<<\rho$ can be employed in the calculation.  The same
trick is useful for the calculation of proportional to $(\mu_s,\nu_s)$
terms in (\ref{degs}). The $L_s$- and $L_s'$-dependent terms are found
to be of $\sim|u_s-v_s|^2$. Therefore, they do not origin the
divergence in the amplitude.

If  a single  interaction vertex coordinate $z$ also moves to $v_s$,
then  additional terms might contribute to the discussed singularity due to
the dependent on $z$ vacuum correlator (\ref{corr}). In this case
$|z-v_s|\sim|u_s-v_s|$. Indeed, if $|z-v_s|>>|u_s-v_s|$, then by
aforesaid, the discussed terms are small as $|u_s-v_s|/|z-v_s|$.
Under the discussed conditions
the superholomorphic function
$R_L(t,t';\{q\})$ in  (\ref{corr}) is given as follows
\begin{eqnarray}
R_L(t,t';\{q\})=R_{L_{(n-1})}(\tilde t,t';\{q\}_{(n-1)})+
\vartheta^{(s)}\Xi(z^{(s)};\{q_s\}_0)
\biggl[1-\frac{\mu_s\nu_s}{2(u_s-v_s)}\biggl]
\nonumber\\
\times
\biggl[R_{fL_{(n-1)}}(v_s,t';\{q\}_{(n-1)})+
\varepsilon_s(z^{(s)})
\partial_{v_s}R_{bL_{(n-1)}}(v_s,t';\{q\}_{(n-1)})\biggl]
+\dots
\label{ap1gr}
\end{eqnarray}
where $\Xi(z^{(s)};\{q_s\}_0)$ is defined by (\ref{xi}) and $\tilde
t=(v_s|\vartheta)$.  In this case
$R_{fL_{(n-1)}}(v_s,t';\{q\}_{(n-1)})$ and
$R_{bL_{(n-1)}}(v_s,t';\{q\}_{(n-1)})$ are given by (\ref{rlbf}) on the
genus-$(n-1)$ supermanifold which is obtained by removing the handle
`$s$' from the  genus-$n$ supermanifold under consideration.  Eq.
(\ref{ap1gr}) corresponds to eq. (\ref{apgr}) from Appendix B for the
genus-$n_1=1$ case where the Green genus-1 function is given by
(\ref{gir}).  The dots code the terms that appear due to the boson part
of (\ref{gir}) and, therefore, they are independent of $L_s$.

To integrate over $t$, it is
convenient to introduce
$t^{(s)}=(z^{(s)}|\vartheta^{(s)})$ as the integration variable.
The product of the leading term in the dilaton type pairing
(\ref{ngcor}) by the $\sim \Xi(z^{(s)};\{q_s\}_0)$ term in
(\ref{ap1gr}) does not appear in the amplitude because both multipliers
are $\sim\vartheta_1^{(s)}$.  The bilinear in $\Xi(z^{(s)};\{q_s\}_0)$
term in the amplitude contains the $|u_s-v_s-\mu_s\nu_s|^2/|u_s-v_s|^2$
factor which reduces the $1/|u_s-v_s-\mu_s\nu_s|^2$ singularity to
$1/|u_s-v_s|^2$. The singularity of the remaining contributions from
(\ref{ap1gr}) to the amplitude is also  not stronger than
$1/|u_s-v_s|^2$.  The integration over $t$ being performed, the
singularity is compensated  by the $\sim|u_s-v_s|^2$ smallness of the
integration volume.  The singularity also disappears at $z^{(s)}$ to be
fixed. In this case the singularity disappears  due to the Riemann
identity that agrees with the nullification of the torus two-point
functions \cite{gsw}.
By kindred reasons, the divergence does not appear when $m_1=(m-1)$
where $m_1$ is a number of the vertices moving to $v_s$.
If $m_1=m$, then the resulted expression is proportional to the
genus-$(n-1)$ vacuum amplitude. As is verified in the next Section, the
arbitrary genus vacuum amplitude is nullified. So, at $m_1=m$ the
considered divergence does not  appear, too.

The divergence does not also  appear when
$|u_s-v_s|<<\rho\to\infty$ where $\rho$ is the
minimal distance between $u_s$ or $v_s$ and every  other limiting
point of the Schottky group. Indeed, this  configuration is reduced
by a relevant $L(2)$ transformation to the  above-considered case
$u_s\to v_s$.

So the integrand in (\ref{ampl}) has no non-integrable singularities.
In the next Section, there is verified that  divergences in
(\ref{ampl}) do not also appear owing to the integration over
degenerated configurations  of higher genuses.

\section{Finiteness  of the superstring amplitudes}

In the case of interest  the
super-Schottky genus-$n$ group is degenerated into a group product of
the lower genus sub-groups.  It appears when all the
distances between the local limiting points of the given sub-group  go
to zero. Another configuration is that where
distances between the local limiting points of the given sub-group all
are nothing like as much as  $\rho\to\infty$ where $\rho$ is the
minimal distance between the above-mentioned points and the rest
limiting points of the genus-$n$ group.  Only the first configuration
is discussed below because the second configuration is reduced  to the
first configuration by means of a relevant $L(2)$ transformation.  The
spin structure  of the discussed configuration  is supposed to be even.
Otherwise its contribution to the amplitude  disappears due to the
theta-constant factor in (\ref{zmate}).

We begin with the case when
the genus-$n$ surface is degenerated into a sum of
the genus-2 supermanifold and of the genus-$(n-2)$ one.
The genus-2 supermanifold is given by transformations (\ref{gamab})
corresponding to  $s$- and $p$-handles. For definiteness,  $u_s$,
$v_s$ and $u_p$ limiting points are supposed moving to
$v_p$.
We show that the discussed configuration does not origin
divergences in the amplitude (\ref{ampl}) and that the genus-2 vacuum
amplitude is nullified.

For this purpose
we integrate in (\ref{ampl}) over limiting points assigned to any one
from the $s$ and $p$ handles, for example, over $(u_s|\mu_s)$,
$(v_s|\nu_s)$ and their complex conjugated. The integrand
of the resulted integral is denoted
as ${\cal A}_m^n
(U_p,\overline U_p,V_p,\overline
V_p;\{q',\bar q'\},\{t_j,\bar t_j\})$.
In this case $\{q'\bar q'\}$ is the set of super-Schottky group parameters
of the genus-$(n-2)$ supermanifold and
$\{t_j,\bar t_j\}$ is the same set as in (\ref{ampl}).
As above,
$U_i=(u_i|\mu_i)$ and $V_i=(v_p|\nu_p)$.
For the sake of brevity, there is omitted an explicit dependence on
the Schottky group multipliers.
No interaction vertices being nearby  $v_p$, then
\begin{equation}
{\cal A}_m^n
(U_p,\overline U_p,V_p,\overline
V_p;\{q',\bar q'\},\{t_j,\bar t_j\})=
{\cal A}(U_p,\overline U_p,V_p,\overline{V_p})
{\cal A}_m^{(n-2)}
(\{q,\bar q\},\{t_j,\bar t_j\})+\dots
\label{gen2deg}
\end{equation}
where ${\cal A}(U_p,\overline U_p,V_p,\overline{V_p})$ depends only on the
super-Schottky group variables of the genus-2  supermanifold
while ${\cal A}_m^{(n-2} (\{q,\bar q\},\{t_j,\bar t_j\})$ is the
integrand of the integral (\ref{ampl}) for
the $(n-2)$-loop, $m$-point amplitude.
The `dots" crypt the correction terms.
In this case
\begin{equation}
{\cal A}(U_p,\overline U_p,V_p,\overline{V_p})
=\int d^2u_s\,d\mu_s\,d\bar\mu_s \theta(\Lambda-|u_s-v_p|)
\widetilde{\cal A}(U_p, V_p,\overline{U_p},
\overline{V_p},U_s,\overline{U_s})\,.
\label{calatca}
\end{equation}
where
$\Lambda<<\rho$ and $\rho$ is the minimal
distance from the genus-2 configuration to the remaining
genus-$(n-2)$ part of the genus-n supermanifold.
In turn,
$\widetilde{\cal A}(U_p, V_p,\overline{U_p},\overline{V_p},
U_s,\overline{U_s})$ is
as follows
\begin{equation}
\widetilde{\cal A}(U_p, V_p,\overline{U_p},\overline{V_p},
U_s,\overline{U_s})=\int\widetilde{\cal Z}_2(U_s,U_p,\overline U_s,
\overline U_p,V_s,V_p,\overline V_s,\overline V_p)
d^2v_sd\nu_sd\bar\nu_s
\label{inz}
\end{equation}
where $\widetilde{\cal
Z}_2(U_s,U_p,\overline U_s,\overline U_p,V_s,V_p,\overline
V_s,\overline V_p)$ is obtained from (\ref{ampl}) at $n=2$ by removing of
the
factor (\ref{fixm})
and of the interaction vertex product.
By
the previous Section, the integral (\ref{inz}) is convergent at $v_s\to
u_s$.
Since the genus-1 function is nullified,
only the  $|v_s-v_p|\sim |u_p-v_p|$ region contributes to (\ref{inz}).
So there is no necessity to introduce a cutoff
restricting the size of the integrated configuration.
Hence (\ref{inz}) has
an explicit   symmetry   under those $SL(2)$ transformations
which are independent of the spin structure.
In particular, (\ref{inz}) is invariant under the super-boost
transformation
of its arguments. The super-boost transformation
$(z|\vartheta)\to(\tilde z|\tilde\vartheta)$ is defined as follows
\begin{equation}
z=\tilde z+z_0+\tilde\vartheta\vartheta_0\,,\quad
\vartheta= \tilde\vartheta+\vartheta_0\,,
\label{trco}
\end{equation}
where $z_0$ and $\vartheta_0$ are  transformation parameters.
Therefore, (\ref{inz})
depends on differences between
$u_s$, $u_p$ and $v_p$.
Then
$\widetilde{\cal A}
(U_p, V_p,\overline{U_p},\overline{V_p},U_s,\overline{U_s})$
can be written down
through certain functions $\psi_{mn}(r_p,w_p,
\overline r_p,\overline w_p)$ of
$r_p=(u_p-u_s)$ and of $w_p=(u_p-v_p-\mu_p\nu_p)$, and through
Grassmann variables as follows
\begin{equation}
\widetilde{\cal A}
(U_p, V_p,\overline{U_p},\overline{V_p},U_s,\overline{U_s})
=\sum_{m=1}^3\sum_{n=1}^3{\cal V}_m\overline{{\cal V}}_n\psi_{mn}(r_p,w_p,
\overline r_p,\overline w_p)
\label{wtca}
\end{equation}
where
\begin{equation}
{\cal V}_1=\mu_s\mu_p\nu_p,\quad{\cal V}_2=(\mu_p-\mu_s),
\quad{\cal V}_3=(\nu_p-\mu_s)\,.
\label{caln}
\end{equation}
The right part of (\ref{wtca}) to be invariant under  transformation
(\ref{trco}) with $\vartheta_0\neq0$,
it requires that
\begin{eqnarray}
\psi_{11}(r_p,w_p,
\overline r_p,\overline w_p)=\partial_{u_s}\partial_{\bar
u_s}\psi_{33}(r_p,w_p,
\overline r_p,\overline w_p)\,,
\nonumber\\
\psi_{13}(r_p,w_p,
\overline r_p,\overline w_p)=\partial_{u_s}\psi_{33}(r_p,w_p,
\overline r_p,\overline w_p)\,,
\nonumber\\
\psi_{31}(r_p,w_p,
\overline r_p,\overline w_p)=\partial_{\bar u_s}\psi_{33}(r_p,w_p,
\overline r_p,\overline w_p)\,,
\nonumber\\
\psi_{12}(r_p,w_p,
\overline r_p,\overline w_p)=\partial_{u_s}\psi_{32}(r_p,w_p,
\overline r_p,\overline w_p)\,,
\nonumber\\
\psi_{21}(r_p,w_p,
\overline r_p,\overline w_p)=\partial_{\bar u_s}\psi_{23}(r_p,w_p,
\overline r_p,\overline w_p)
\label{calmn}
\end{eqnarray}
where the derivatives are calculated keeping $w_p$
and $\overline w_p$ to be fixed.
The integration over $\mu_s$ and over $\bar\mu_s$ in
(\ref{calatca}) being performed,
${\cal A}(U_p,\overline U_p,V_p,\overline{V_p})$
is represented  as follows
\begin{equation}
{\cal A}(U_p,\overline U_p,V_p,\overline{V_p})
=\int d^2u_s\theta(\Lambda-|u_s-v_p|)
\grave{{\cal A}}
(U_p,\overline{U_p},w_p,\overline{w_p},u_s,\overline{u_s},
\nu_p,\overline\nu_p)
\label{calatcaf}
\end{equation}
where
\begin{equation}
\grave{{\cal A}}
(U_p,w_p,\overline{U_p},\overline{w_p},u_s,\overline{u_s},
\nu_p,\overline\nu_p)=\int\widetilde{\cal Z}_2(U_s,U_p,\overline U_s,
\overline U_p,V_s,V_p,\overline V_s,\overline V_p)
d^2v_sd\nu_sd\bar\nu_sd\mu_sd\bar\mu_s
\label{calgr}
\end{equation}
which can be  also represented as follows
\begin{equation}
\grave{{\cal A}}
(U_p,w_p,\overline{U_p},\overline{w_p},u_s,\overline{u_s},
\nu_p,\overline\nu_p)
=-\sum_{m=1}^3\sum_{n=1}^3{\cal U}_m\overline{{\cal U}}_n\psi_{mn}(r_p,w_p,
\overline r_p,\overline w_p)\,,
\label{wtcaf}
\end{equation}
$\psi_{mn}(r_p,w_p,
\overline r_p,\overline w_p)$ being
identical that in (\ref{wtca}), and
\begin{equation}
{\cal U}_1=\mu_p\nu_p,\quad {\cal U}_2={\cal U}_3=-1\,.
\label{calnf}
\end{equation}
Firstly, we discuss $\grave{{\cal A}}
(U_p,w_p,\overline{U_p},\overline{w_p},u_s,\overline{u_s},
\nu_p,\overline\nu_p)$ in the $u_p\to v_p$ limit keeping $u_s$ to be fixed.
Since the genus-1 function is
nullified, only the region
$|v_s-u_p|^2\sim|v_p-u_p|^2$ could contribute to (\ref{calgr}). Due to
the small size of the integration region, the function
\begin{equation}
\grave{{\cal A}}_0
(u_p,v_p,\overline u_p,\overline v_p,u_s,\overline u_s)\equiv
\grave{{\cal A}}
(U_p,w_p,\overline{U_p},\overline{w_p},u_s,\overline{u_s},
\nu_p,\overline\nu_p)\biggl|_{\mu_p=\nu_p=\overline\mu_p=\overline\nu_p=0}
\label{grclaz}
\end{equation}
has no a singularity at $u_p=v_p$.
In this case $\psi_{mn}(r_p,w_p,
\overline r_p,\overline w_p)$ with both $m\neq1$ and $n\neq1$  has no a
singularity  at $u_p=v_p$ and at $w_p=0$. Then, as it follows from
(\ref{calmn}), the $\psi_{1n}(r_p,w_p, \overline r_p,\overline w_p)$
and $\psi_{m1}(r_p,w_p, \overline r_p,\overline w_p)$ functions  also
have no a singularity at $u_p=v_p$ and at $w_p=0$.
Therefore, $u_s$ to be fixed, $\grave{{\cal A}}
(U_p,w_p,\overline{U_p},\overline{w_p},u_s,\overline{u_s},
\nu_p,\overline\nu_p)$ has no singularity
at $u_p=v_p$ and at $w_p=0$.
Furthermore, $\grave{{\cal A}}
(U_p,w_p,\overline{U_p},\overline{w_p},u_s,\overline{u_s},
\nu_p,\overline\nu_p)$
is expressed through the same
function at $\mu_p=\nu_p=
\overline\mu_p=\overline\nu_p=0$ as follows (see Appendix B for details)
\begin{equation}
\grave{{\cal A}}
(U_p,w_p,\overline{U_p},\overline{w_p},u_s,\overline{u_s},
\nu_p,\overline\nu_p)=
\biggl|1-\frac{\mu_p\nu_p}{v_p-u_p}\biggl|^2
\grave{\cal A}_0(u_p,v_p,\overline u_p,
\overline v_p,u_s,\overline u_s)
\label{mnz}
\end{equation}
where
$\grave{\cal A}_0(u_p,v_p,\overline u_p,
\overline v_p,u_s,\overline u_s)$ is defined by (\ref{grclaz})
By aforesaid, the left side of
(\ref{mnz}) is finite at $u_p\to v_p$. Therefore, $\grave{\cal
A}_0(u_p,v_p,\overline u_p, \overline v_p,u_s,\overline u_s)$ is equal
to zero at $u_p\to v_p$.

Due to $L(2)$ symmetry,
$|(u_p-u_s)(u_s-v_p)|^2\grave{\cal A}_0(u_p, v_p,\overline u_p,
\overline v_p,u_s,\overline u_s)$
does not depend on $u_p$ ,$v_p$ and $u_s$. Therefore
$\grave{\cal A}_0(u_p,v_p,\overline u_p,
\overline v_p,u_s,\overline u_s)$
is equal to
zero identically in its arguments. Then  the vacuum genus-2
amplitude is nullified. Indeed, this amplitude is represented
by the integral (\ref{ampl}) for $m=0$ in the genus-2 case. Further,
$\mu_1$ and $\nu_1$ can be taken to be $\mu_1=\nu_2=0$ (as it has
been discussed in Section 3, the integral does not depend on $u_1$,
$v_1$, $u_2$, $\mu_1$ and $\nu_1$). After the integration over $v_2$,
$\mu_2$ and $\nu_2$ (and over their complex conjugated)
the integral over the Schottky group multipliers appears, the
integrated function being proportional to $\grave{\cal A}_0(u_1,
v_1,\overline u_1, \overline v_1,u_2,\overline u_2)\equiv0$.
Therefore, the vacuum amplitude is nullified.
The nullification of the vacuum amplitude is achieved due
to the integration over the limiting points of the super-Schottky
group.  The local in the moduli space vacuum function is, generally,
not equal to zero.  Due to the nullification of the vacuum amplitude,
the contribution to (\ref{gen2deg}) from the first term on the right
side is nullified.  The `dots" terms (\ref{gen2deg}) are finite for
reasons which are quite similar to those  for the genus-1 case. So
the considered configuration give the finite contribution to the
interaction amplitude (\ref{ampl}).

If, in addition,  a single  interaction vertex coordinate $z$ moves to
$v_p$, then only the region
$|z-v_p|\sim|u_s-v_p|\sim|v_s-v_p|\sim|u_s-v_s|\sim|u_p-v_p|$
might additionally contribute to the $u_p\to v_p$ singularity.
In this case the superholomorphic
function $R_L(t,t';\{q\})$  is given
by (\ref{apgr}) where  the genus-2 supermanifoldis formed by
`$s$' and `$p$'-handles.
It is useful to represent this genus-2 function
$R(t_1,t_2;\{q\}_{sp};L_{sp})$  through the variables $\hat t_1$ and
$\hat t_2$ which are related with
$t_1$ and $t_2$ by transformation (\ref{itrn})
given in Appendix C. In this case
\begin{eqnarray}
R(t_1,t_2;\{q\}_{sp};L_{sp})=\ln(z_1-z_2-\vartheta_1\vartheta_2)+
\widetilde R_{bb}(\hat z_1,\hat z_2;\{q\}_{sp})
-\hat\vartheta_1
\hat\vartheta_2\widetilde R_{ff}(\hat z_1,\hat z_2;\{q\}_{sp};L_{sp})
\nonumber\\
-\hat\vartheta_1R_{fb}(\hat z_1,\hat z_2;\{q\}_{sp};L_{sp})
\nonumber\\
+
R_{bf}(\hat z_1,\hat z_2;\{q\}_{sp};L_{sp})\hat\vartheta_2
+\kappa_{sp}[\Xi^{(b)}_{sp}(\hat t_2;\{q\}_{sp};L_{sp})
+\Xi^{(b)}_{sp}(\hat t_1;\{q\}_{sp};L_{sp})]
\nonumber\\
-l_{sp}\varepsilon_p'[\Xi^{(f)}_{sp}(\hat t_2;\{q\}_{sp};L_{sp})
+\Xi^{(f)}_{sp}(\hat t_1;\{q\}_{sp};L_{sp})]
\label{rsp}
\end{eqnarray}
where $R_{fb}(\hat z_1,\hat z_2;\{q\}_{sp};L_{sp})=
R_{bf}(\hat z_2,\hat z_1;\{q\}_{sp};L_{sp})$ and
\begin{eqnarray}
\Xi^{(b)}_{sp}(\hat t_2;\{q\}_{sp};L_{sp})=
\widetilde\Xi^{(bb)}_{sp}(\hat z_2;\{q\}_{sp};L_{sp})
+\Xi^{(bf)}_{sp}(\hat z_2;\{q\}_{sp};L_{sp})
\hat\vartheta_2\,,
\nonumber\\
\Xi^{(f)}_{sp}(\hat t_2;\{q\}_{sp};L_{sp})
=\widetilde\Xi^{(ff)}_{sp}(\hat z_2;\{q\}_{sp};L_{sp})\hat\vartheta_2
+\Xi^{(fb)}_{sp}(\hat z_2;\{q\}_{sp};L_{sp})\,,
\nonumber\\
l_{sp}=1+\frac{1}{2}(u_p-v_p)
\kappa_{sp}\,,\quad
\kappa_{sp}=\frac{\mu_s\varepsilon_p(u_s)}{(u_s-u_p)(u_s-v_p)}\,.
\label{xisp}
\end{eqnarray}
The terms on the right side in (\ref{xisp}) determine the asymptotic
at $z_1\to\infty$   as follows
\begin{eqnarray}
\widetilde R_{bb}(\hat z_1,
\hat z_2;\{q\}_{sp};L_{sp})\to\frac{\widetilde\Xi^{(bb)}_{sp}(\hat z_2;
\{q\}_{sp};L_{sp})}{(\hat z_1-v_p)}\,,
\nonumber\\
\widetilde R_{ff}(\hat z_1,\hat z_2;
\{q\}_{sp};L_{sp})\to\frac{\widetilde\Xi^{(ff)}_{sp}(\hat z_2;
\{q\}_{sp};L_{sp})}{(\hat z_1-v_p)}\,,
\nonumber\\
R_{fb}(\hat z_1,\hat z_2;\{q\}_{sp};L_{sp})\to\frac{\Xi^{(fb)}_{sp}(t_2;
\{q\}_{sp};L_{sp})}{(\hat z_1-v_p)}\,,
\nonumber\\
R_{bf}(\hat z_1,\hat z_2;\{q\}_{sp};L_{sp})\to\frac{\Xi^{(bf)}_{sp}(\hat z_2;
\{q\}_{sp};L_{sp})}{(\hat z_1-v_p)}\,.
\label{spxi}
\end{eqnarray}
In this case (see Appendix B)
\begin{eqnarray}
R_L(t,t';\{q\})=\Xi^{(b)}_{sp}(\hat t_1;\{q\}_{sp};
L_{sp})[(u_p-v_p)\kappa_{sp}\partial_{v_p}R_{bL_2}(v_p,t';\{q\}_2)
\nonumber\\
-l_{sp}\varepsilon_p'
R_{fL_2}(v_p,t';\{q\}_2)]
+\Xi^{(f)}_{sp}(\hat t_1;\{q\}_{sp};L_{sp})l_{sp}(1+\frac{1}{2}
\varepsilon_p\varepsilon_p')
\nonumber\\
\times
[\nu_p
\partial_{v_p}R_{bL_2}(v_p,t';\{q\}_2)-R_{fL_2}(v_p,t';\{q\}_2)]
+R_{bL_2}(v_p,t';\{q\}_2)
\nonumber\\
-\vartheta_1(\hat t_1)R_{fL_2}(v_p,t';\{q\}_2)+\dots
\label{ap2gr}
\end{eqnarray}
where $\vartheta_1(\hat t_1)$ is calculated through $\hat t_1$ using
eqs. (\ref{trn}) and (\ref{trnsf}) of Appendix C and `dots' code terms
of   higher orders in $(u_p-v_p)$. The $R_{bL_2}(v_p,t';\{q\}_2)$ and
$R_{fL_2}(v_p,t';\{q\}_2)$ functions are the same as
$R_{bL}(z,t';\{q\})$ and $R_{fL}(z,t';\{q\})$ in (\ref{rlbf}) given on
the genus-$(n-2)$ supermanifold which is obtained  omitting $s$- and
$p$-handles from the former genus-$n$ supermanifold. Other definitions
are the same as in (\ref{rsp}) and in (\ref{xisp}).  Eq.(\ref{rsp}) is
an extension  to the higher-genus supermanifolds of eq. (\ref{xi}).
Like the genus-1 case,  (\ref{ap2gr}) leads to the singularity that is
not stronger than  $1/|u_s-v_s|^2$.  The integration over $t$ being
performed, the singularity is compensated  by the $\sim|u_s-v_s|^2$
smallness of the integration volume.  So the contribution to the
amplitude of the considered configuration is finite.

Two- and three-point amplitudes
are nullified like
the vacuum amplitude. The questioned amplitudes
are nominally given by (\ref{ampl}) for $m=2$ and $m=3$.
The vacuum expectation of the interaction vertex product is
as follows
\begin{equation}
<\prod_{r=1}^mV(t_r,\overline t_r;p_r;\zeta^{(r)})>_{L,L'}=
\sum_s{\cal P}^{(s)}{\cal L}_m^{(s)}(\{t_j,\overline t_j;{L,L'})
\label{vex}
\end{equation}
where ${\cal P}^{(s)}$ depend only on
the 10-dim momenta and polarizations of the bosons.
As above,
$<...>$ is the vacuum expectation of the
interaction vertex product.
In this case
two- and three-point integrals $\widetilde{\cal A}_m(U_p,
V_p,\overline{U_p},\overline{V_p}, U_s,\overline{U_s})$
are
constructed like (\ref{inz}) as follows
\begin{eqnarray}
\widetilde{\cal A}_m(U_p,
V_p,\overline{U_p},\overline{V_p},
U_s,\overline{U_s})=\int\widetilde{\cal Z}_2^{(j)}(U_s,U_p,\overline U_s,
\overline U_p,V_s,V_p,\overline V_s,\overline V_p)
d^2v_sd\nu_sd\bar\nu_s\,,
\label{inzm}\\
\widetilde{\cal Z}_2^{(j)}(U_s,U_p,\overline U_s,
\overline U_p,V_s,V_p,\overline V_s,\overline V_p)
\nonumber\\
=
\sum_{L,L'}\widetilde{\cal Z}_2(U_s,U_p,\overline U_s,
\overline U_p,V_s,V_p,\overline V_s,\overline V_p;L,L'){\cal L}_m^{(s)}(L,L')
\label{inzmll}
\end{eqnarray}
where $m=2$ or $m=3$ and ${\cal L}_m^{(s)}$ is obtained by the
integration of ${\cal L}_m^{(s)}(\{t_j\},\overline t_j\};L,L')$ over
interaction vertex coordinates.  Integrating the function (\ref{inzm})
over $\mu_s$ and $\overline\mu_s$,
one obtains the
$\grave{\cal A}_m
(U_p,w_p,\overline{U_p},\overline{w_p},u_s,\overline{u_s},
\nu_p,\overline\nu_p)$ function.
The questioned ${\cal L}_m^{(s)}$ is invariant under $SL(2)$-transformations.
So, like (\ref{inz}), the integral (\ref{inzm}) is invariant
under (\ref{trco}) and (\ref{trn}) transformations. As the result,
$\grave{\cal A}_m
(U_p,w_p,\overline{U_p},\overline{w_p},u_s,\overline{u_s},
\nu_p,\overline\nu_p)$ is equal to zero for the same reasons as
$\grave{\cal A}
(U_p,w_p,\overline{U_p},\overline{w_p},u_s,\overline{u_s},
\nu_p,\overline\nu_p)$.

The nullification of the vacuum amplitude with  more then two loops as
well as the nullification of the 2- and 3-point ones is proved by the
induction method.  Simultaneously the absence of divergences  is
established in the case when the $n$-loop, $m$-point amplitude is
degenerated into a product  of the vacuum $n_1>2$-loop amplitude and of
the $(n-n_1)$-loop, $m$-point one.  In doing so, the desired
properties are assumed for the $(n_1-1)$-loop amplitude and the
genus-$(n_1-1)$ function (\ref{calatca}) is constructed.  Then the
consideration is performed in the manner given above for the $n_1=2$
case.
An absence of the divergences in the higher-genus amplitude is
similarity    verified for configurations where a number $m_1$ of the
vertices  moving to limiting points of the Schottky group
is $m_1=1$, $m_1=(m-1)$ and $m_1=m$ where $m$ is number of the vertices
in the considered amplitude.

The one-point amplitude is nullified because by (\ref{ngcor}), it is proportionle to the vacuum amplitude. In this case
the nullification appears after the integration over the vertex coordinate being performed.

The nullification of the vacuum amplitude arises once
the integration has been performed over  limiting points
of the genus-$n_1$ super-Schottky group.  The local in the moduli space
vacuum function is not nullified.
The nullification of the 2- and 3-point amplitudes requires, in addition, that the integration over  the interaction vertex coordinates
to be performed. Indeed, the vacuum
correlator (\ref{corr}) is, generally, not invariant under
$SL(2)$ transformations. Really, the above correlator receives
two additional terms, every term being dependent on only one of
the points. As the result,
the ${\cal L}_m^{(s)}(\{t_j\},\overline t_j\};L,L')$ function has not
$SL(2)$-symmetry. At the same time the right side of (\ref{vex})
possesses $SL(2)$-symmetry because the vacuum expectation of the
interaction vertex product has the symmetry in question.  The  ${\cal
P}^{(s)}$ factor being taken into account, the 2- and 3-point amplitude
is nullified independently from which $(3|2)$  variables are fixed
among the super-Shottky group limiting points and the interaction
vertex coordinates.

Though  local spinning string amplitudes are covariant under
the supermodular transformations \cite{dannph}, the modular
invariance of the whole amplitude (\ref{ampl}) is not seemingly
obvious. Indeed, the supermodular transformations depend on the
superspin structure \cite{dannph}, but
the integral of a single superspin structure is ill-defined
due to the sigularities of the integration measure (\ref{hol}).
For the same reason, it is not obvious the invariance of the amplitude
(\ref{ampl})
with respect to
the superspin dependent $SL(2)$-transformations, in particular, with
respect to the super-Schottky group transformations.
To verify the invariance of (\ref{ampl}) under considered
transformations, it is useful to remaind that
any part $\Omega(L,L')$ of the
integration region in (\ref{ampl}) can replaced its congruent part
$\Omega_c(L,L')$. If the
amplitude has no singularities in the
$\Omega_c(L,L')-\Omega(L,L')$ domain, then
the integration of the local amplitude over
$\Omega_c(L,L')-\Omega(L,L')$ can be performed separately for each a
superspin structure, the result being  equal to zero. So
(\ref{ampl}) is invariant under questioned transformations.
To verify  the symmetry, one can also regularize the
integral for every superspin structure by relevant restrictions on the
period matrix entries $\omega_{jl}(\{q\},L)$. In doing so, a
regularization at $k_j\to0$ is achieved by atop  cutoff
$1/\epsilon_{jj}$  on the diagonal entry $\omega_{jj}(\{q\},L)$ of the
period matrix.  The regularization at $v_j\to u_j$ and  at $v_l\to u_l$
is achieved by a bottom cutoff $\epsilon_{jl}$  on
$\omega_{jl}(\{q\},L)$. The regularization is realized introducing
relevant step-function
multipliers in (\ref{ampl}) which in line with previous Sections are
treated as Taylor series over Grassmann variables.  In the $k_j\to0$
limit or in the $v_j\to u_j$ one, the period matrix entries cease to
depend on the spin structure of the $j$-th handle.  Therefore, the
discussed regularization corresponds to such a  cutoff on the Schottky
group parameters which is independent of the spin structure of the
degenerated handle. In this case the cutoff sum over superspin
structures in the $\epsilon_{jl}$ limit coincides with the sum in
(\ref{ampl}).  At the same time,  the desired 2D-transformation can be
performed in every  cutoff integral corresponding to the given
superspin structure. Under $SL(2)$-transformation the period matrix
is not changed, and is changed by (\ref{modtr}) under the supermodular
transformation. Therefore, like the former period matrix, the resulted
period matrix in the given nodal region does not depend on the spin
structure of the relevant degenerated handle.  Therefore,
the  cutoff on the resulted Schottky
group parameters does not depend on the spin structure of the
degenerated handle.  In this case
the cutoff resulted sum over superspin
structures in the $\epsilon_{jl}$ limit coincides with the former sum
that is the amplitude (\ref{ampl}) is invariant under the questioned
2D-transformations. Therefore, the obtained amplitude (\ref{ampl}) is
consistent with the symmetries of the spinning string and with the
space-time supersymmetry.
\\ \\
The paper is  supported by RSF grant
No. 14-22-00281.

\appendix
\def\thesection{Appendix \Alph{section}}
\def\theequation{\Alph{section}.\arabic{equation}}
\setcounter{equation}{0}

\section{Polynomials}
The column $Y_p^{(s)}$ of the $Y_{p,N_s}$ polynomials \cite{danphr} for $p$ being $p=a$ and $p=b$,
can be represented as follows
\begin{equation}
Y_p^{(s)}=\hat Y_p^{(s)T}\widehat T
\label{ypol}
\end{equation}
where "$T$" index denotes transposition and
\begin{equation}
\widehat T=
\left( \begin{array}{ccccc}
                  1 & 0 & 0 & 0 & 0 \\
                  0 & 1+\frac{\mu\nu}{u-v} & 0 & \mu & 0 \\
                  0 & 0 & 1+\frac{\mu\nu}{u-v} & 0 & \nu \\
                  0 & \frac{\mu-\nu}{u-v} & 0 & 1-\frac{\mu\nu}{2(u-v)} & 0 \\
                  0 & 0 & \frac{\mu-\nu}{u-v} & 0 & 1-\frac{\mu\nu}{2(u-v)} \\
                \end{array}
              \right)
\label{Mmatr}
\end{equation}
Therefore,
\begin{equation}
\widehat T^{-1}=
\left( \begin{array}{ccccc}
                  1 & 0 & 0 & 0 & 0 \\
                  0 & 1-2\frac{\mu\nu}{u-v} & 0 & -\mu & 0 \\
                  0 & 0 & 1-2\frac{\mu\nu}{u-v} & 0 & -\nu \\
                  0 & -\frac{\mu-\nu}{u-v} & 0 & 1+\frac{3\mu\nu}{2(u-v)} & 0 \\
                  0 & 0 &- \frac{\mu-\nu}{u-v} & 0 & 1+\frac{3\mu\nu}{2(u-v)} \\
                \end{array}
              \right)\,.
\label{Mmatrr}
\end{equation}
Entries  $Y_{p,N_s}^{(r)}$ of the $\hat Y_p^{(s)}$ column are
\begin{equation}
Y_{p,N_s}^{(r)}(t)=\frac{\grave Y_{p,N_s}^{(0)}(t_s)}{Q^2_{\tilde
\Gamma_s}(t_s)}
\label{tilzer}
\end{equation}
where, in turn,  $\grave
Y_{p,N_s}^{(0)}(t_s)=\vartheta_s Y_{p,N_s}^{(0)}(z_s)$ when
$N_s=(\mu_s,\nu_s)$ and $\grave Y_{p,N_s}^{(0)}(t_s)=
Y_{N_s}^{(0)}(z_s)$ when  $N_s=(k_s,u_s,v_s)$. Further (index "s"
is omitted),
\begin{eqnarray}
Y_{b,k}^{(0)}(z)=\frac{(z-u)(z-v)}{k(u-v)}\,,\quad Y_{b,u}^{(0)}(z)
=\frac{(1-k)(z-v)^2}{k(u-v)^2}\,,\quad Y_{b,v}^{(0)}(z)
=\frac{(k-1)(z-u)^2}{k(v-u)^2}\,,
\nonumber\\
Y_{b,\mu}^{(0)}(z)=2\frac{(1-\sqrt k)(z-v)}{\sqrt
k(u-v)}\,,\quad Y_{b,\nu}^{(0)}(z)=2\frac{(\sqrt
k-1)(z-u)}{(v-u)}\,,\quad
Y_{a,N}^{(0)}(z)=Y_{b,N}^{(0)}(z)\biggl|_{\sqrt{k}=-1}\,.
\label{yb}
\end{eqnarray}
Furthermore,  $Y_{p,N_s}(t)$ for $N_s\neq k_s$
are expressed through some $Y_{N_s}(t)$
polynomials as follows
\begin{equation}
Y_{p,N_s}(t)=Q^2_{\Gamma_{p,s}}(t)Y_{N_s}(\Gamma_{p,s}(t))
-Y_{N_s}(t)\,.
\label{ypy}
\end{equation}
The $Y_{N_s}$ set forms the $\hat Y^{(s)}$ column, and
\begin{equation}
Y^{(s)}=\hat Y^{(s)T}\widehat T
\label{ypl}
\end{equation}
where "$T$" index denotes transposition, $\widehat T$ matrix is
given by (\ref{Mmatr}) and
entries  $Y_{N_s}^{(r)}$ of the $\hat Y^{(s)}$ column are
represented as follows
\begin{equation}
Y_{N_s}^{(r)}(t)=
\frac{\grave Y_{N_s}^{(0)}(t_s)}{Q^2_{\tilde \Gamma_s}(t_s)}\,.
\label{tlzer}
\end{equation}
In turn,  $\grave Y_{N_s}^{(0)}(t_s)=\vartheta_s Y_{N_s}^{(0)}(z_s)$
when $N_s=(\mu_s,\nu_s)$
and $\grave Y_{N_s}^{(0)}(t_s)= Y_{N_s}^{(0)}(z_s)$, if
$N_s=(k_s,u_s,v_s)$. In this case
(index "s" is omitted)
\begin{equation}
Y_u^{(0)}(z)
=\frac{(z-v)^2}{(u-v)^2}\,,\quad Y_v^{(0)}(z)
=\frac{(z-u)^2}{(v-u)^2}\,,\quad
Y_\mu^{(0)}(z)=\frac{2(z-v)}{(u-v)}\,,\quad
Y_\nu^{(0)}(z)=\frac{2(z-u)}{(v-u)}\,.
\label{yu}
\end{equation}
It is useful to note also that
\begin{eqnarray}
\sum_{p=a,b}\int_{C_p^{(s)}}
\chi_{N_s}(t;\{q\})\frac{ dzd\vartheta}{2\pi i}
Y_{p,N_r}(t)=\delta_{N_sN_r}\,,
\label{chiy}\\
\int_{C_b^{(s)}}
\Psi_{\sigma,N_s}(t;\{q\})\frac{ dzd\vartheta}{2\pi i}
\hat
Y_{\sigma,N_r}^{(1)}(t)=\delta_{N_sN_r}
\label{psihy}
\end{eqnarray}
and that
\begin{eqnarray}
\sum_{p=a,b}\int_{C_p^{(s)}}
G(t,t';\{q\})\frac{ dz'd\vartheta'}{2\pi i}
Y_{p,N_s}(t')
=0\,,
\label{gfy}\\
\int_{C_b^{(s)}}
S_\sigma(t,t';\{q\})\frac{ dz'd\vartheta'}{2\pi i}
\hat
Y_{\sigma,N_s}^{(1)}(t')
=0\,.
\label{ssiggfy}
\end{eqnarray}
In this case $t=(z|\vartheta)$ and  $z$ lays
in the fundamental region.

\section{Correlators on degenerated surfaces}
\setcounter{equation}{0}

It is supposed that the genus-$n$ supermanifold is degenerated into a
sum of the genus-$n_1$ supermanifold and of the genus-$n_2$ one,
$n_2=n-n_1$. In doing so, it is assumed that the local limiting points
of the genus-$n_1$ supermanifold are  moved to $v_p$ which is one among
them.  Further, if  $t=(z|\vartheta)$ and $z$ is moved to $v_p$, then
the super-holomorphic Green function $R_L(t,t';\{q\})$
in (\ref{scorr}) is approximated as follows
\begin{equation}
R_L(t,t';\{q\})\approx \int d\vartheta\widetilde R_{L_1}(t,t_2;\{q\}_1)(z-z_2)\biggl|_{z_2\to\infty}D(\tilde t_2)
R_{L_2}(\tilde t_2,t';\{q\}_2)
+R_{L_2}(t,t';\{q\}_2)
\label{apgr}
\end{equation}
where  $\widetilde R_{L_1}(t,t_2;\{q\}_1)=R_{L_1}(t,t_2;\{q\}_1)-\ln(z-z_2-\vartheta\vartheta_2)$
is given on the genus-$n_1$ supermanifold,
$R_{L_2}(\tilde t_2,t';\{q\}_2)$ is given on the genus-$n_2$ supermanifold,
$t_2=(z_2|\vartheta_2)$, $\tilde t_2=(v_p|\vartheta_2)$  and $\tilde
t=(v_p|\vartheta)$.  Eq.(\ref{apgr}) follows directly from
(\ref{rbos}), (\ref{fgrfr}) and (\ref{eqsr}).

If the $m_1$ handle is assigned to the genus-$n_1$ supermanifold
under consideration and a certain point $z'$ is not nearby
$v_p$, then in accordance with (\ref{apgr}),
the superscalar function $J_{m_1}(t';\{q\};L)$ is approximated as follows
\begin{equation}
J_{m_1}(t';\{q\};L)\approx \int d\vartheta_2\,J_{m_1}(t_2;\{q\}_1;L_1)
(z-z_2)\biggl|_{z_2\to\infty}D(\tilde t_2)
R_{L_2}(\tilde t_2,t';\{q\}_2)+R_{L_2}(t,t';\{q\}_2)\,.
\label{apscrf}
\end{equation}
If the $m_2$ handle is assigned to the genus-$n_2$ supermanifold and  $z$ is  nearby
$v_p$, then  the superscalar function $J_{m_2}(t;\{q\};L)$ is approximated as follows
\begin{equation}
J_{m_2}(t;\{q\};L)\approx \int d\vartheta_2\,R_{L_1}(t,t_2;\{q\}_1)
(z-z_2)\biggl|_{z_2\to\infty}D(\tilde t_2)
J_{m_2}(\tilde t_2;\{q\}_2;L_2)+J_{m_2}(t;\{q\}_2;L_2)
\label{apsc2f}
\end{equation}
In accordance  with both (\ref{apscrf}) and (\ref{apsc2f}),
the $\omega_{m_1m_2}(\{q\};L)$ entry of the period matrix is approximated by
\begin{equation}
\omega_{m_1m_2}(\{q\};L)\approx \int d\vartheta_2\,J_{m_1}(t_2;\{q\}_1;L_1)
(z-z_2)\biggl|_{z_2\to\infty}D(\tilde t_2)
J_{m_2}(\tilde t_2;\{q\}_2;L_2)\,.
\label{degperm}
\end{equation}

\section{Deriving of eq.(\ref{mnz})}
\setcounter{equation}{0}

To derive eq.(\ref{mnz}) we employ the $SL(2)$ transformation that
reducing $\mu_p$ and $\nu_p$ to zero,
preserves $u_p$, $v_p$ and $u_s$. The transformation is
\begin{equation}
z={\it f}(\hat z)+{\it f}'(\hat z)\hat\vartheta\xi(\hat z)\,,
\quad
\vartheta=\sqrt{{\it f}'(\hat z)}[(1+\frac{1}{2}\xi\xi')
\hat\vartheta+\xi(\hat z)]
\label{trn}
\end{equation}
where
\begin{eqnarray}
{\it f}(\hat z)=\hat z-\frac{(\hat z-u_p)(\hat z-v_p)}
{(u_s-u_p)(u_s-v_p)}\hat\mu_s\xi(u_s)\,,
\nonumber\\
\xi(\hat z)=\frac{\mu_p(\hat z-v_p)}{(u_p-v_p)\sqrt{{\it
f}'(u_p)}} -\frac{\nu_p(\hat z-u_p)}{(u_p-v_p)\sqrt{{\it
f}'(v_p)}}\,.
\label{trnsf}
\end{eqnarray}
Evidently, ${\it f}'(u_p){\it f}'(v_p)=1$. The inverse transformation is as follows
\begin{equation}
\hat z=\widetilde{\it f}(z)-\widetilde{\it f}'(z)\vartheta\varepsilon_p(z)\,,
\quad
\hat\vartheta=\sqrt{\widetilde{\it f}'(z)}[(1+\frac{1}{2}\varepsilon_p\varepsilon_p')
\vartheta-\varepsilon(z)]
\label{itrn}
\end{equation}
where
\begin{equation}
\widetilde{\it f}(z)=z+\frac{(z-u_p)(z-v_p)}
{(u_s-u_p)(u_s-v_p)}\mu_s\varepsilon_p(u_s)\,.
\label{itrnsf}
\end{equation}

To obtain the desired equation we
represent $d\mu_sd\bar\mu_s\widehat{\cal A}(U_p,V_p,\overline
U_p,\overline V_p,U_s,\overline U_s)$ in (\ref{calatca}) as
$[d\mu_sd\bar\mu_s\widehat{\cal A}(U_p,V_p,\overline U_p,\overline
V_p,U_s,\overline U_s)|\widehat H(U_p,V_p,U_s)|^2]$ $\times$ $|\widehat
H(U_p,V_p,U_s)|^{-2}$ where the $\widehat H(U_p,V_p,U_s)$  factor is
given by (\ref{fixm}).   Then we  transform $\mu_s\to\hat\mu_s$
by the $SL(2)$ transformation (\ref{trnsf}).
The expression inside the square brackets is transformed into
the same expression at $\mu_p=\nu_p=0$ where $\mu_s$ is replaced by
$\hat\mu_s$. By dimension reasons,this expression is proportional to
$\hat\mu_s\overline{\hat\mu_s}$. So $H^{-1}(U_2,V_2,U_1)$ in front of
the square brackets can be replaced by  $[1+\mu_p\nu_p/(u_p-v_p)]$.
Thereupon  the integration over $\mu_s$ and $\overline\mu_s$
being performed, a relation for
$\widetilde{\cal A}(U_p,V_p,\overline U_p,\overline V_p,u_s,\overline
u_s)$ arises as follows
\begin{equation}
\widetilde{\cal A}(U_p,V_p,\overline U_p,\overline V_p,u_s,\overline
u_s)= \biggl|1-\frac{\mu_p\nu_p}{v_p-u_p} \biggl|^2
\widetilde{\cal A}_0(u_p,v_p,\overline u_p,
\overline v_p,u_s,\overline u_s)
\label{linvc}
\end{equation}
where
$\widetilde{\cal A}_0(u_p,v_p,\overline u_p,
\overline v_p,u_s,\overline u_s)$ is
$\widetilde{\cal A}(U_p,V_p,\overline
U_p,\overline V_p,u_s,\overline u_s)$ at $\mu_p=\nu_p=0$.
Eq.(\ref{mnz}) is obtained by
substituting $v_p=u_p-w_p-\mu_p\nu_p$ into (\ref{linvc}).


\begin{thebibliography}{99}
\bibitem{rnshw}
P.  Ramond,   Phys.Rev. D3
(1971) 2415.\\ A. Neveu  and J.H. Schwarz, Nucl.Phys. B31 (1971) 86.
\bibitem{ver}
E. Verlinde and H.  Verlinde, Phys.  Lett.  B 192 (1987)
95; Nucl. Phys B 288 (1987) 357.
\bibitem{swit}
N.  Seiberg, and E.
Witten, Nucl.  Phys. B 276 (1986) 272.
\bibitem{momor} G. Moore and A.
Morozov, Nucl.  Phys.  B 306 (1988) 387.
\bibitem{as}
J. Atick, J. Rabin, and A.  Sen, Nucl. Phys.  B 299 (1988) 279.
\bibitem{hok}
Eric
D'Hoker and D.H. Phong, Phys. Lett. B529 (2002) 241; Nucl.  Phys.  B
636 (2002) 3, 61; B 639 (2002) 129; B 715 (2005) 3, 91.
\bibitem{dancqg}
G.S. Danilov,
Class. Quantum Grav. 29 (2012) 235009; arXiv:1211.3280 [hep-th].
\bibitem{bsh}
M.A.  Baranov and A.S.  Schwarz, Pis'ma ZhETF 42 (1985)
340 [JETP Lett.  49 (1986) 419]; D.  Friedan, Proc.  Santa Barbara
Workshop on Unified String theories, eds.  D.  Gross and M.  Green
(World Scientific, Singapure, 1986).
\bibitem{fried}
D. Friedan, E. Martinec and S. Shenker,
Nucl. Phys. B 271 (1986) 93.
\bibitem{vec}
P.  Di Vecchia, K.
Hornfeck, M. Frau, A. Ledra and S.  Sciuto, Phys. Lett.  B 211 (1988)
301; J.L. Petersen, J.R.  Sidenius and A.K. Tollst{\'e}n, Phys Lett. B
213 (1988) 30; Nucl.  Phys.  B 317 (1989) 109; B.E.W.  Nilsson, A.K.
Tollst{\'e}n and A.  W{\"a}tterstam, Phys Lett. B 222 (1989) 399;
\bibitem{dan93}
G.S.  Danilov, JETP Lett.  58 (1993) 796 [ Pis'ma JhETF
58 (1993)]; G.S.  Danilov, Class. Quantum Grav. 11 (1994) 2155.
\bibitem{danphr}
G. S.  Danilov, Phys. Rev. D 51 (1995) 4359,
52 (1995) 6201 (Erratum).
\bibitem{dannph}
G.S.  Danilov, Nucl. Phys. B 463 (1996) 443.
\bibitem{siegal}
C.L.  Siegal, Topics in Complex Function Theory, (New
York, Wiley, 1973), vol. 3.
\bibitem{dan90}
G.S.  Danilov, Phys. Lett. B 257 (1991) 285;
Sov.J.Nucl.Phys. 52 (1990) 727 [Yad.Fiz. 52 (1990) 1143].
\bibitem{dan04}
G. S. Danilov, Phys.Atom.Nucl. 67 (2004) 1035 [Yad.Fiz.
67 (2004) 1059].
\bibitem{capige}
Sergio L. Cacciatori, Francesco dalla Piazza, and Bert Van Geemen,
Nucl. Phys.  B 800 (2008) 565;
Sergio L.  Cacciatori, Francesco dalla Piazza, and Bert Van Geemen,
Lett. Math.  Phys. 85 (2008) 185;
G.  Grushevsky, Com. Math. Phys.  287 (2009) 749;
M. Oura, C.  Poor, R. Salvati Manni, and
D. S.  Yuen, Math. Ann.  346 (2010) 447;
Marco Matone and Roberto Volpato, Nucl. Phys. 732 (2006) 321.
\bibitem{dunmorsl}
P. Dunin-Barkowski, A.  Morozov and A. Sleptsov, JHEP 10 (2009) 072.
\bibitem{ma2vo}
Marco Matone and Roberto Volpato, Nucl. Phys. B 839 (2010) 21.
\bibitem{ford}
L. Ford, Automorphic Functions (New York, Chelsea 1951).
\bibitem{gsw}
M.B. Green, J.H. Schwarz and E. Witten, Superstring Theory,
vols.I and II  ( Cambridge Univesity Press, England,
1987).
\bibitem{dan89}
G.S. Danilov,  Sov. J. Nucl. Phys. 49 (1989) 1106
[ Jadernaja Fizika 49 (1989) 1787 ].
\bibitem{mart}
E.  Martinec, Nucl.  Phys. B 281 (1986) 157.
\end{thebibliography}
\end{document}